\documentstyle[12pt]{article}   

\makeatletter
\@addtoreset{equation}{section}
\makeatother

\newtheorem{Definition}{Definition}[section]

\def\be{\begin{equation}}
\def\ee{\end{equation}}
\def\ba{\begin{eqnarray}}
\def\ea{\end{eqnarray}}
\def\a{{\cal A}}
\def\ab{\overline{\a}}

\def\g{{\cal G}}

\def\ag{\a/\g}
\def\agb{\overline{\ag}}

\def\Nl{{\mathchoice
{\setbox0=\hbox{$\displaystyle\rm N$}\hbox{\hbox to0pt
{\kern0.4\wd0\vrule height0.9\ht0\hss}\box0}}
{\setbox0=\hbox{$\textstyle\rm N$}\hbox{\hbox to0pt
{\kern0.4\wd0\vrule height0.9\ht0\hss}\box0}}
{\setbox0=\hbox{$\scriptstyle\rm N$}\hbox{\hbox to0pt
{\kern0.4\wd0\vrule height0.9\ht0\hss}\box0}}
{\setbox0=\hbox{$\scriptscriptstyle\rm N$}\hbox{\hbox to0pt
{\kern0.4\wd0\vrule height0.9\ht0\hss}\box0}}}}
\def\Zl{{\mathchoice
{\setbox0=\hbox{$\displaystyle\rm Z$}\hbox{\hbox to0pt
{\kern0.4\wd0\vrule height0.9\ht0\hss}\box0}}
{\setbox0=\hbox{$\textstyle\rm Z$}\hbox{\hbox to0pt
{\kern0.4\wd0\vrule height0.9\ht0\hss}\box0}}
{\setbox0=\hbox{$\scriptstyle\rm Z$}\hbox{\hbox to0pt
{\kern0.4\wd0\vrule height0.9\ht0\hss}\box0}}
{\setbox0=\hbox{$\scriptscriptstyle\rm Z$}\hbox{\hbox to0pt
{\kern0.4\wd0\vrule height0.9\ht0\hss}\box0}}}}
\def\Ql{{\mathchoice
{\setbox0=\hbox{$\displaystyle\rm Q$}\hbox{\hbox to0pt
{\kern0.4\wd0\vrule height0.9\ht0\hss}\box0}}
{\setbox0=\hbox{$\textstyle\rm Q$}\hbox{\hbox to0pt
{\kern0.4\wd0\vrule height0.9\ht0\hss}\box0}}
{\setbox0=\hbox{$\scriptstyle\rm Q$}\hbox{\hbox to0pt
{\kern0.4\wd0\vrule height0.9\ht0\hss}\box0}}
{\setbox0=\hbox{$\scriptscriptstyle\rm Q$}\hbox{\hbox to0pt
{\kern0.4\wd0\vrule height0.9\ht0\hss}\box0}}}}
\def\Rl{{\mathchoice
{\setbox0=\hbox{$\displaystyle\rm R$}\hbox{\hbox to0pt
{\kern0.4\wd0\vrule height0.9\ht0\hss}\box0}}
{\setbox0=\hbox{$\textstyle\rm R$}\hbox{\hbox to0pt
{\kern0.4\wd0\vrule height0.9\ht0\hss}\box0}}
{\setbox0=\hbox{$\scriptstyle\rm R$}\hbox{\hbox to0pt
{\kern0.4\wd0\vrule height0.9\ht0\hss}\box0}}
{\setbox0=\hbox{$\scriptscriptstyle\rm R$}\hbox{\hbox to0pt
{\kern0.4\wd0\vrule height0.9\ht0\hss}\box0}}}}
\def\Cl{{\mathchoice
{\setbox0=\hbox{$\displaystyle\rm C$}\hbox{\hbox to0pt
{\kern0.4\wd0\vrule height0.9\ht0\hss}\box0}}
{\setbox0=\hbox{$\textstyle\rm C$}\hbox{\hbox to0pt
{\kern0.4\wd0\vrule height0.9\ht0\hss}\box0}}
{\setbox0=\hbox{$\scriptstyle\rm C$}\hbox{\hbox to0pt
{\kern0.4\wd0\vrule height0.9\ht0\hss}\box0}}
{\setbox0=\hbox{$\scriptscriptstyle\rm C$}\hbox{\hbox to0pt
{\kern0.4\wd0\vrule height0.9\ht0\hss}\box0}}}}
\def\Hl{{\mathchoice
{\setbox0=\hbox{$\displaystyle\rm H$}\hbox{\hbox to0pt
{\kern0.4\wd0\vrule height0.9\ht0\hss}\box0}}
{\setbox0=\hbox{$\textstyle\rm H$}\hbox{\hbox to0pt
{\kern0.4\wd0\vrule height0.9\ht0\hss}\box0}}
{\setbox0=\hbox{$\scriptstyle\rm H$}\hbox{\hbox to0pt
{\kern0.4\wd0\vrule height0.9\ht0\hss}\box0}}
{\setbox0=\hbox{$\scriptscriptstyle\rm H$}\hbox{\hbox to0pt
{\kern0.4\wd0\vrule height0.9\ht0\hss}\box0}}}}
\def\Ol{{\mathchoice
{\setbox0=\hbox{$\displaystyle\rm O$}\hbox{\hbox to0pt
{\kern0.4\wd0\vrule height0.9\ht0\hss}\box0}}
{\setbox0=\hbox{$\textstyle\rm O$}\hbox{\hbox to0pt
{\kern0.4\wd0\vrule height0.9\ht0\hss}\box0}}
{\setbox0=\hbox{$\scriptstyle\rm O$}\hbox{\hbox to0pt
{\kern0.4\wd0\vrule height0.9\ht0\hss}\box0}}
{\setbox0=\hbox{$\scriptscriptstyle\rm O$}\hbox{\hbox to0pt
{\kern0.4\wd0\vrule height0.9\ht0\hss}\box0}}}}

\title{Complexifier Coherent States for Quantum General Relativity}
\author{T. Thiemann\thanks{thiemann@aei-potsdam.mpg.de} \\
       MPI f. Gravitationsphysik, Albert-Einstein-Institut, \\
           Am M\"uhlenberg 1, 14476 Golm near Potsdam, Germany}
\date{{\small Preprint AEI-2002-045}}

\begin{document}

\maketitle

\begin{abstract}
Recently, substantial amount of activity in Quantum General Relativity
(QGR) has focussed on the semiclassical analysis of the theory. In this 
paper we want to comment on two such developments: 
1) Polymer-like states for Maxwell theory and 
linearized gravity constructed by Varadarajan which use much of the 
Hilbert space machinery that has proved useful in QGR and 2) coherent 
states for QGR, based on the general complexifier method, with built -- 
in semiclassical properties. We show the following:

A) Varadarajan's states {\it are} complexifier coherent states. This 
unifies all states constructed so far under the general complexifier
principle.

B) Ashtekar and Lewandowski suggested a non-Abelean generalization of 
Varadarajan's states to QGR which, however, are no longer of the 
complexifier type. We construct a new class of non-Abelean complexifiers 
which come close to the one underlying Varadarajan's construction.

C) Non-Abelean complexifiers close to Varadarajan's induce new types of 
Hilbert spaces which do not support the operator algebra of QGR. 
The analysis suggests that if one sticks to the 
present kinematical framework of QGR and if kinematical coherent states
are at all useful, then normalizable, graph dependent 
states must be used which are produced by the complexifier method as well. 

D) Present proposals for states with mildened graph dependence, 
obtained by performing a graph average, do not approximate well 
coordinate dependent observables. However, graph dependent states, whether
averaged or not, seem to be well suited for the semiclassical analysis 
of QGR with respect to coordinate independent operators.
\end{abstract}

\section{Introduction}
\label{s1}

A mathematically well-defined candidate 
Wheeler-DeWitt (or Hamiltonian constraint) operator for canonical Quantum 
General Relativity (QGR) has been proposed in \cite{0,1}
on the kinematical Hilbert space ${\cal H}_0$ defined by Ashtekar, Isham
and Lewandowski \cite{21}. This Hilbert space presently underlies 
literally all the constructions within 
QGR. (See \cite{6} for an up-dated, detailed introduction to QGR and 
\cite{7,8} for non-technical overviews). Thus, there is  
modest hope 
that ${\cal H}_0$ indeed supports a (dual) representation of all constraint 
operators and Dirac observables of the theory and that
there exists a well-defined quantum field theory for Lorentzian metrics 
(plus, possibly supersymmetric, matter) in four spacetime dimensions.
It is possible to explicitly state (and to 
some extent even solve) the, necessarily discrete, quantum time evolution.
Open problems are the construction of an inner product on the space of 
solutions and of a suitable set of gauge invariant observables.

However, before tackling these problems, the more crucial question is, 
whether that
quantum field theory can possibly have general relativity as its 
classical limit. The 
answer to that question is far from obvious because the theory is background
independent, that is, non-perturbatively defined: Usual quantization 
procedures, of which we know that they guarantee the correct classical limit
such as for Maxwell theory, always make crucially use of the  
(Minkowski) background metric. However, general relativity does 
not distinguish any background, it is a backgound independent theory, in 
other words, the background metric becomes dynamical, a quantum field 
operator. Therefore these usual quantization techniques, usually based on 
Fock spaces (notice that the Minkowski metric explicitly slips into the 
definition of a Fock space through, e.g. the d'Alembert operator, the 
positive and negative frequency one-particle wave functions etc.) cannot 
be used. This fact manifests itself in the regularization and 
renormalization of the operator in \cite{0} on the background independent 
Hilbert space ${\cal H}_0$ which employs completely new mathematical 
techniques  
which we have little experience with and it is therefore indeed not 
manifest 
whether the quantum Einstein equations proposed reduce to the classical
ones in the classical limit.

To settle this question one has several possibilities. As a minimum test
one should verify whether the new quantization technique works in 
situations where the theory can be solved explicitly, say in dimensionally
or Killing reduced models of general relativity. This has been done in 
\cite{1} for 2+1 gravity and in \cite{8a} for Bianchi cosmologies, in both
cases with success. The analysis in \cite{8a} is based on a quantum symmetry
reduction, that is, one works on the Hilbert space of the full theory and 
imposes the Killing constraints there, in contrast to the usual procedure 
of imposing the symmetry before quantization. In that sense the calculations
performed in \cite{8a} are almost fully 3+1 dimensional, just that one 
neglects the
excitations of all but a finite number of degrees of freedom, which is why
this provides a really firm test of the proposal made in \cite{0}. The 
agreement of the quantum theory with the classical theory demonstrated 
in \cite{8a} is indeed spectacularly sharp, the discrete spectrum of 
quantum metric operators constructed in \cite{0} approaches the classical
curve much faster (for very low quantum numbers) than one could have 
hoped for (correspondence principle). What is more, the operator proposed 
in \cite{0} certainly suffers from quantization ambiguities but in \cite{8a}
these were shown to be irrelevant for the classical limit. Finally, 
for the same reason that the quantization technique in in \cite{0} 
made the Wheeler-DeWitt operator finite, the deeper reason being background 
independence, it avoids certain classical curvature singularities which 
could be checked explicitly in \cite{8a} where it was shown that the
classical big bang singularity is avoided in the quantum theory. These 
discrepancies between classical and quantum predictions actually are 
large only in the deep Planckian regime.

These results are very promising but they must be generalized to the full 
theory (all excitations). To that end, a general construction principle,
the so-called complexifier method, for coherent states (in principle for 
any canonical theory based on a cotangent bundle) has been 
formulated in \cite{9} which guarantees that many
semiclassical properties are built in. 
These states are {\it kinematical}, that is, they are annihilated 
neither by the spatial diffeomorphism constraint nor by the Hamiltonian
constraint for well motivated reasons:\\ 
1) In order to construct physical
coherent states we would need to spend a lot of work in order to find the 
entire space of solutions to all constraints which we do not want to do
unless we are sure that the Hamiltonian constraint operator has the 
correct classical limit.\\ 
2) Whether or not the Hamiltonian conatraint 
operator has the correct classical limit cannot be checked on its kernel
where it vanishes by definition.\\ 
3) We cannot make our coherent states at 
least diffeomorphism invariant because the Hamiltonian constraint does 
not preserve the space of diffeomorphism invariant states. We will
report on the classical limit of the Hamiltonian constraint with respect 
to these states elsewehere \cite{16} and focus on another issue in the 
present paper.

When applying the complexifier method to QGR at the kinematical level, 
then it does not give rise to a normalizable (with respect to the 
kinematical inner product of QGR) state but rather a complex probability 
measure $\psi$ which also
can be considered as a {\it distribution} over a dense subset of the 
kinematical Hilbert space. The underlying reason is that the kinematical
Hilbert space is not seperable, a situation that might change once we go 
to the spatially diffeomorphism invariant level. For instance,
we showed that if the volume operator is used 
as the complexifier then one obtains a distributional, gauge invariant 
and diffeomorphism covariant coherent state. Thus, in order to 
obtain normalizable states one must restrict these distributions to
(countable) graphs, resulting in ``cut-off" states $\psi_\gamma$.
This we did in \cite{10,11} with respect to a natural (modification 
of a) Laplacian operator as the complexifier considered already in 
\cite{30} and 
obtained a complete family of gauge invariant and spatially 
diffeomorphism covariant coherent states for quantum general relativity 
These are based on coherent states that have 
been constructed for the phase space of a cotangent bundle over a compact 
gauge group (in this case SU(2)) in \cite{15} (see also references 
therein) and whose semiclassical 
properties (such as overcompleteness, peakedness, Ehrenfest properties,
minimal uncertainty, annihilation operator eigenstate property, small 
fluctuations) have been explicitly proved in the just mentioned works.\\
\\
To shed light on the issue of the classical analysis from a 
seemingly completely 
independent perspective, Varadarajan has constructed a polymer 
like representation of Maxwell theory and linearized general relativity
in \cite{17} by using GNS technology in an ingenious way. More in detail, by 
making explicit use of the Minkowski 
background metric, he was able to construct an image of 
the usual Fock states on {\it a distributional extension of} the type of 
background independent Hilbert space ${\cal H}_0$ on which quantum general 
relativity currently is based. Varadarajan's work is 
ground breaking for several reasons: First of all, this is the first 
time that a clear link between the usual Fock Hilbert spaces of free Maxwell
theory or linearized general relativity on the one hand and the 
polymer-like Hilbert spaces of the type in use for QGR on the other hand has 
been achieved in a mathematically rigorous way. Secondly, his work shows
that the space of distributional connections $\ab$ on which the 
background {\it independent} measure underlying ${\cal H}_0$ is supported
is also a natural support for the background {\it dependend} measure 
\cite{18} which underlies Varadarajan's GNS Hilbert space. Finally, at least
on the level of linearized general relativity we now have, for the first 
time, a very good idea of how to think about gravitons in the framework
of QGR and a first glimpse of how to extend his results to full 
non-perturbative QGR.

There are many more applications of Varadarajan's technique, in particular
it goes much beyond the mere purpose of constructing semiclassical or, more
specifically, coherent states. In 
this paper we confine ourselves to a limited set of questions, namely which
implications his work has for full Quantum General Relativity. More 
specifically, we are interested in whether the coherent states that are 
implicit in his construction, suitably generalized to non-Abelean gauge 
theories, are good candidate semiclassical states for QGR. Recalling that
his states, from the point of view of ${\cal H}_0$, are distributions,
Varadarajan's analysis suggests two conclusions:\\
i) There is an
appropriate substitute for usual Fock states (photons, gravitons) and the 
usual coherent states (realized, e.g. in lasers) in full QGR.\\
ii) However, these states are not normalizable with respect to the 
kinematical 
inner product used in quantum general relativity, rather, Fock 
states and semiclassical states define a completely new representation of 
the canonical commutation and adjointness relations based on a new type
of background dependent measure \cite{18} on the quantum configuration 
space $\ab$ of QGR. 

These speculations rest on the assumption that the answers to the 
following two questions are affirmative:\\ 
A) Can these results, obtained for the gauge 
groups $U(1)$ for Maxwell theory and $U(1)^3$ for linearized gravity, 
be transferred to $SU(2)$, the gauge group of fully non-linear general
relativity ?\\
B) Do the basic operators of quantum general 
relativity, the one-dimensionally smeared holonomy operators and the 
two-dimensionally smeared electric field operators, continue to be 
well-defined in the new representation ?

To answer A), in \cite{19} Ashtekar and Lewandowski already 
proposed such an extension to $SU(2)$ using the {\it key} observation
that the formulas derived by Varadarajan allow for a natural 
generalization if one replaces the operators 
associated with the Lie algebra $u(1)$ by their analogs for $su(2)$.
However, issue B) is more difficult to settle because Varadarajan's 
construction explicitly employs {\it three-dimensionally smeared electric
fields} (while holonomies are by construction well-defined). This is quite 
critical since all the operators so far 
constructed in quantum general relativity are based on electric fields 
smeared in two dimensions only. In particular, in the non-Abelean case 
the holonomy and 3d smeared electric field operators no longer 
form a closed subalgebra of the Poisson algebra. Notice that we do not 
insist that electric field and holonomy operators are physically interesting
operators {\it per se}, however, they serve to build more complicated,
physically interesting, composite operators such as constraint operators 
(in the limit of small loops or surfaces).

Thus, if B), as we will show, does not hold, then one can have the 
following attitudes:\\ 
I) We want to keep the present, background independent and non-perturbative
kinematical set-up of QGR which has proved so successful in supporting 
constraint operators, area operators, volume operators etc. Then the 
distributional states constructed by Varadarajan or those which 
are produced by the complexifier machinery are not good semiclassical states
for QGR because the corresponding representations that they define does not 
support the operator algebra of QGR.\\
II) We do not insist on using the current formulation of QGR but wish to 
use the new representations of the Varadarajan type. Then the regularization
and quantization of all operators of interest has to be repeated for
the new Varadarajan type of representation, probably in a background 
dependent way.\\ 
III) It does not make sense to discuss the semiclassical analysis at 
the kinematical level at all. That is, one should first solve all the
constraints, leading to new Hilbert spaces on the space of solutions
which from the point of view of ${\cal H}_0$ are expected to be 
distributional as well, before diving into the semiclassical analysis.

Option II) would not only 
make all current results within quantum general relativity obsolete, it 
would also raise serious conceptual
difficulties because the representations constructed by Varadarajan make
explicit use of a background metric while we wanted to build a manifestly
background independent quantum gravity theory and thus our working 
principle would be lost ! What would be our fundamental theory if for each 
background we have to construct a new representation ? Also option III)
would be desastrous because we would have no possibility to check on the 
correctness of the Hamiltonian constraint before actually solving it.

Being optimistic that the current formulation of QGR has at least some
bearing for the final theory, we 
thus adopt viewpont I) and the question arises what
one can do with the distributional states of the Varadarajan type or the 
complexifier coherent states within the present stage of affairs anyway.
The strategy used in \cite{10,11} was to cut off these 
distributional coherent states at given (countable) graphs resulting in 
families of normalizable coherent states indexed by graphs. A similar 
procedure has also been applied to the Varadarajan states in \cite{19} where 
the analog of cut-off states were coined ``shadows". (The difference 
between a 
cut-off state and a shadow is that the former form a cylindrically 
consistent family of complex probability measures while shadows are their
corresponding spin-network projections). Since such states are embedded 
into the full kinematical Hilbert space, there is hope is that they 
display suitable semiclassical properties. This has been demonstrated to 
some extent for the states \cite{11} as already mentioned. Notice that 
this strategy {\it does not make use of the new representation of the 
Varadarajan type} but that
either the complexifier method or Varadarajan's method just serve to 
produce those cut-off states or shadows respectively, the kinematical 
framework of QGR remains untouched.

Cut-off states have been criticized due to their graph dependence which
leads to ``staircase problems" with the semiclassical properties of 
certain types of operators such as the area operator, see the last
reference in \cite{11}. In 
order to remove this graph dependence, in \cite{19} graph -- averaging 
techniques have been advocated which are based on the Dirichlet-Voronoi 
construction \cite{20}.\\
\\
The aim of this paper is to review these developments, to show 
how they are interrelated and to communicate some observations which 
hopefully add some conceptual and technical clarity to the subject.
The architecture of this article is as follows:\\
\\
In order to make this paper close to being self-contained we review
the complexifier construction and the Varadarajan construction in section   
two, using a position space language which will help us to compare them 
later on. The main results then are contained in the subsequent sections: 
\begin{itemize}
\item[i)] {\it Complexifier Coherent States Versus Varadarajan States}\\
In section three we begin by observing that the usual coherent 
states on Fock spaces are 
also obtained via the complexifier method where for Maxwell theory the 
complexifier is related to
the electric field energy. Using an operator correspondence 
$\cal I$ between pairs of 3d 
smeared holonomy and non-smeared elecric field operators on Fock space on 
the one hand and pairs of non-smeared holonomy and 3d smeared elecric 
fields on the QGR 
Hilbert space ${\cal H}_0$ due to Ashtekar, Isham and and Lewandowski 
\cite{21}
on the other hand, on which Varadarajan's beautiful construction rests, we 
obtain 
as a simple corollary that also the Varadarajan states are obtained via 
the complexifier method from the QGR Hilbert space. This explains the 
miracolous identities found in \cite{17} which are responsible for the fact 
that the coefficients of the formal linear combinations of charge network 
states, defining the distributional image of the Fock vacuum states, can be 
obtained by asking that the Varadarajan states 
are generalized eigenstates of the image, under the just mentioned 
operator correspondence, of the exponentiated Fock space annihilation 
operator.

Once we know that the Varadarajan states come from a complexifier we can 
employ the techniques developed in \cite{9} in order to examine their 
semiclassical properties. For instance, we find the 
point in phase space at which these states are peaked which implies 
corresponding Ehrenfest properties.

Finally we show that the Varadarajan representation can indeed be 
obtained just by using the QGR inner product, without recourse to the 
Fock representation and the isomorphism $\cal I$, 
although these states are 
not normalizable with respect to the QGR inner product. The mechanism 
responsible for this is very similar to the way that one obtains 
rigorously defined Gaussian measures $d\mu$ from formal measures 
of the kind $[DA] e^{-S}/Z$. This last point is conceptually interesting
because it shows that despite the appearance of distributional states one 
can still use the original inner product in order to obtain the new 
representation so that it is still the ``fundamental representation"
${\cal H}_0$ from which all others arise through usual limiting procedures
of constructive quantum field theory. Thus, also the complexifier method 
induces new $L_2$ Hilbert spaces on $\ab$.
\item[ii)] {\it Electric Flux Operator}\\
Since the isomorphism $\cal I$ explicitly uses 3d smeared electric fields 
it is not surprising that the 2d smeared electric field 
operator is not a well-defined operator in Varadarajan's representation
We then ask  
whether it is possible to construct a different complexifier, for the 
$U(1)$ theory say,
which {\it does} give rise to a well-defined electric flux operator.
We find that for complexifiers that are bilinear in the electric field 
(so that on the Fock side the measure is Gaussian and we can make practical
computations) it is not possible to have the fluctuations of both the 
holonomy and electric flux operator finite, no matter how one chooses the 
smearing of the complexifier. This indicates an obstruction to using 
this type of new representations produced by Varadarajan's or the 
complexifier method, for the holonomy and electric 
flux operator algebra crucial for QGR, at least for those that come from 
Gaussian measures.
\item[iii)] {\it Generalization to $SU(2)$}\\
In section five we examine whether we can use the complexifier method in 
order to arrive 
at Varadarajan type of states directly for $SU(2)$ as anticipated in 
\cite{19}. We find that the obvious gauge invariant $SU(2)$ 
generalization of the Varadarajan complexifier
underlying the states considered in \cite{17} for $U(1),U(1)^3$ is not a 
well-defined operator any longer due to the fact that $SU(2)$ is not 
Abelean, roughly speaking the corresponding operator is no longer densely 
(cylindrically) defined. 

Next we consider the ``shadows" of \cite{19} and find that these states 
do not come from a complexifier, the corresponding would-be complexifier 
is not cylindrically consistent. This means that the methods of \cite{9} 
cannot be used in order to spell out their semiclassical properties and 
that more work is required in order to determine them. 

To improve on this, we construct a new type of well-defined complexifier 
on the QGR Hilbert space with the following properties:\\
1.\\ 
It is almost bilinear in the electric fields, more precisely it is 
bilinear in certain area operators which by themselves are square roots 
of quadratic polynomials in the electric field operators.\\
2.\\
There are several variants of this operator. One variant does not use 
background metric but requires three, everywhere
linearly independent, families of foliations of the spatial manifold as a 
background structure together with a {\it parquet} (partition into 
disjoint 2d surfaces) of each leaf of these foliations, which is 
quite similar to the polyhedronal decompositions that have been used in 
\cite{10}. There is also a family of background dependent variants of this
operator where the freedom consists in choosing a covariance 
kernel (propagator) of a Gaussian measure. Notice that since 
we want to approximate a given classical metric we may use it naturally 
as a background mewtric in order to construct our coherent state.\\
3.\\
The resulting distributional state is gauge invariant
and diffeomorphism covariant. It can be considered as the closest analog to 
the Abelean case. One can also define these new complexifier states for the
Abelean theory and if one would use the isomorphism $\cal I$ to map the 
resulting representations of this type
to the ``Fock side", then the resulting measure would no longer be 
free (Gaussian) but interacting.\\ 
4.\\
The cut-off states resulting from this complexifier are very similar
to those of \cite{10,11} (and also to the Abelean ones corresponding to 
Varadarajan's complexifier) but in 
contrast to those the complexifier now does have a classical limit.

Despite of these positive features, the electric flux still does not have 
finite 
fluctuations in the associated new representation induced by these 
distributional states. This indicates an obstruction to using these 
representations for the semiclassical analysis of QGR and suggests that
we are forced to work with cut-off states (or maybe with a completely 
different type of complexifier). This is unfortunate because it would
be much more natural to work with graph independent objects.
\item[iv)] {\it Dirichlet-Voronoi Construction}\\
An idea of how to milden the graph dependence of cut-off states is to 
perform an average over an ensemble of graphs. An extremely elegant way of 
how to do that is based on the Dirichlet-Voronoi \cite{19,20} construction 
which we describe in some detail. As 
observed by many, the resulting states or density matrices are strictly 
speaking zero states and zero operators respectively on the QGR Hilbert 
space, however, if one interchanges two limits then the density 
matrix turns from a trace class operator on the QGR Hilbert space
into an expectation value functional which defines a new representation 
through the GNS construction. 

We show that in this
representation the holonomy operator necessarily has zero expectation value
(far off the classically allowed range of being a unitary matrix) whether 
or not one thickens the path on which the holonomy depends. The only
way of how to obtain non-zero expectation values is by averaging 
coherent states over a countable (necessarily finite, if the spatial 
manifold is compact and our graphs are piecewise analytic) number of 
graphs in which case we may 
as well work with the coherent state on the countable graph that results 
by taking the union of all sample graphs in the ensemble. If one does not 
do that but rather works with a countable averaging then expectation 
values for holonomy operators are non-zero but still way off their 
classical values. 

This seems to indicate that the Dirichlet-Voronoi procedure produces nice 
weaves
(states that approximate the intrinsic three metric, defined in terms of 
area operators, for instance) but definitely not
semiclassical states (states approximating also the extrinsic curvature) 
since the holonomy expectation values come out grossly wrong. The underlying 
reason is the discrete structure of the QGR Hilbert space which is 
related to background independence. Thus, it seems that better averaging 
procedures have to be designed.
\item[v)] {\it A Resolution: Diffeomorphism Invariant Observables}\\
The upshot of the analysis presented in sections three trough six 
is as follows:\\
1) If the fundamental theory of quantum general relativity,
based on the holonomy -- electric flux operator algebra, is not to be changed
then the best coherent states available at this point must be states in
the kinematical QGR Hilbert space, that is, normalizable and therefore  
graph dependent.\\
2) If we do not average the graph dependent states, then they suffer from 
the staircase problem and electric flux and area operators do not have the 
correct expectation values. If we average them, then holonomy operators 
have the wrong expectation values.\\
\\
This seems to be a desastrous situation. In section seven we propose the 
following way out of the dilemma:

We must remember that 
a given set of states may not be adapted to all possible operators of the 
theory. In particular, it is an {\it additional physical input} what 
the basic observables of the theory should be with respect to which a 
given set of states is to behave semiclassically. Our first guess 
was that these should be the electric flux and holonomy operators since all
others can be constructed from them. However, the above analysis shows 
that the coherent states constructed so far do not approximate them well. 
Are there better suited quantum observables ?

The analysis in \cite{31} has revealed that while gauge invariant but 
coordinate dependent operators such as Wilson loop and area operators
are simply not well approximated, the present coherent states  
are very well geared to approximating 
{\it spatially diffeomorphism invariant operators} (and more general ones
which arise from volume integrals of spatial scalar densities of weight 
one), specifically those that use 
matter (or those of the type of the Hamiltonian constraint). Intuitively
this happens because, physically {\bf matter can only be where geometry is 
excited}.
In other words, while a coordinate surface or loop can be chosen quite 
independently of a given graph underlying a coherent state, if the 
surface or loop is defined by matter 
then they {\bf automatically lie within the graph}. This is, in a nutshell, 
the simple mechanism which avoids both the staircase problem and the 
vanishing of holonomy expectation values.

Notice that since the classical counterparts of such operators suffice to 
separate the points of the (spatially diffeomorphism invariant) classical
phase space, that only those are well approximated is quite sufficient 
for our purposes in view of the fact that they are more physical anyway. 

Thus, it might be that graph dependent coherent states, especially those 
based on {\it random graphs} \cite{31} and those that are averaged over a 
countable ensemble, are maybe not
so bad for the modest main application that we have in mind, namely to 
test whether the quantum dynamics of the theory as presently proposed 
has the correct classical limit (including matter). 
\end{itemize}
The paper ends with a summary in section eight and in the appendices we 
prove some technical results.

\section{Review of Complexifier Coherent States and Varadarajan States}
\label{s2}

This section reviews the complexifier coherent states construction and the 
Vardarajan construction in order to prepare the stage for the next section 
where we compare them. Readers familiar with these concepts can skip this 
section, which is fairly self-contained. There are two new twists in 
our presentation however: 1) We 
use entirely a position space language since we are interested later on 
in complexifiers which do not use a flat background metric so that Fourier
transformation is not available. 2) We perform a small change of viewpoint
in that we consider the distributions that come out of both constructions 
formally as (non-normalizable) states of the original Hilbert space at the 
aid of which one can pass to a new representation through a limiting 
procedure familiar from constructive quantum field theory. 

Appendix \ref{s2.2} contains a review of the kinematical structure of 
diffeomorphism invariant theories of connections and an explanation why
kinematical representations of the type ${\cal H}_0$ are a natural 
starting point.

\subsection{The Complexifier Coherent State Concept}
\label{s2.1}

The purpose of the present subsection is to introduce the concept of 
a complexifier which can be used to {\it generate} coherent states.
As it will become obvious, this method is quite powerful and proviedes
a clean {\it construction mechanism} for coherent states. The method has 
been introduced for the first time in \cite{23} and is by now also 
appreciated by mathematicians (see the second and third reference of 
\cite{15}).\\
\\

Let $({\cal M},\Omega)$ be a symplectic manifold with strong symplectic 
structure $\Omega$ (notice that $\cal M$ is allowed to be infinite 
dimensional). We will assume that ${\cal M}=T^\ast {\cal C}$
is a cotangent bundle. Let us then choose a real polarization of $\cal M$, 
that is, a real Lagrangean submanifold $\cal C$ which will play the role of
our configuration space. Then a lose definition of a complexifier is as 
follows:
\begin{Definition} \label{def2.1}  
A complexifier is a positive definite function $C$ on $\cal M$ with the 
dimension of an action, which is smooth
a.e. (with respect to the Liouville measure induced from $\Omega$) and 
whose Hamiltonian vector field is everywhere non-vanishing on $\cal C$.
Moreover, for each point $q\in {\cal C}$ the function $p\mapsto 
C_q(p)=C(q,p)$ grows stronger than linearly with $||p||_q$
where $p$ is a local momentum coordinate and $||.||_q$ is a suitable norm 
on $T^\ast_q({\cal C})$.
\end{Definition}
In the course of our discussion we will motivate all of these 
requirements.

The reason for the name {\it complexifier} is that $C$ enables us to
generate a {\it complex polarization} of $\cal M$ from $\cal C$ as 
follows: If we denote by $q$ local coordinates of $\cal C$ (we do not 
display any discrete or continuous labels but we assume that local fields
have been properly smeared with test functions) 
then 
\be \label{2.1}
z(m):=\sum_{n=0}^\infty \frac{i^n}{n !} \{q,C\}_{(n)}(m)
\ee
define local complex coordinates of $\cal M$ provided we can invert 
$z,\bar{z}$ for $m:=(q,p)$ where $p$ are the fibre (momentum) 
coordinates of $\cal M$. This is granted at least locally by 
definition (\ref{def2.1}). Here the multiple Poisson bracket
is inductively defined by $\{C,q\}_{(0)}=q,\;
\{C,q\}_{(n+1)}=\{C,\{C,q\}_{(n)}\}$ and makes sense due to the required 
smoothness. What is interesting about 
(\ref{2.1}) is that it implies the following bracket structure
\be \label{2.2}
\{z,z\}=\{\bar{z},\bar{z}\}=0
\ee
while $\{z,\bar{z}\}$ is necessarily non-vanishing. The reason for this
is that (\ref{2.1}) may be written in the more compact form
\be \label{2.3}
z=e^{-i{\cal L}_{\chi_C}} q=([\varphi_{\chi_C}^t]^\ast q)_{t=-i}
\ee
where $\chi_C$ denotes the Hamiltonian vector field of $C$, unamibiguously
defined by $i_{\chi_C}\Omega+dC=0$, $\cal L$ denotes the Lie 
derivative and $\varphi^t_{\chi_C}$ is the one -- parameter family of 
symplectomorphisms generated by $\chi_C$. 
Formula (\ref{2.3}) displays the transformation (\ref{2.1})
as the analytic extension to imaginary values of the one parameter family
of diffeomorphisms generated by $\chi_C$ and since the flow generated by 
Hamiltonian vector fields leaves Poisson brackets invariant, (\ref{2.2})
follows from the definition of a Lagrangean submanifold. The fact that
we have to continued to the negative imaginary axis rather than the 
positive one is important in what follows and has to do with the required
positivity of $C$.

The importance of 
this observation is that either of $z,\bar{z}$ are coordinates of a 
Lagrangean submanifold of the complexification ${\cal M}^\Cl$, i.e. a 
complex polarization and thus 
may serve to define a Bargmann-Segal representation of the quantum theory
(wave functions are holomorphic functions of $z$). The diffeomorphism 
${\cal M}\to {\cal C}^\Cl;\;m\mapsto z(m)$ shows that we may think of 
$\cal M$ either as a symplectic manifold or as a complex manifold 
(complexification of the configuration space). Indeed, the 
polarization is usually a positive K\"ahler polarization with respect
to the natural $\Omega$-compatible complex structure on a cotangent bundle 
defined by local Darboux coordinates, if we choose the complexifier to be 
a function of $p$ only.  These facts make the associated Segal-Bargmann 
representation especially attractive.\\
\\
We now apply the rules of canonical quantization: a suitable 
Poisson algebra $\cal O$ of functions $O$ on $\cal M$
is promoted to an algebra $\hat{{\cal O}}$ of operators $\hat{O}$ on a 
Hilbert 
space $\cal H$ subject to the condition that Poisson brackets turn
into commutators divided by $i\hbar$ and that reality conditions 
are reflected as adjointness relations, that is,
\be \label{2.4}
[\hat{O},\hat{O}']=i\hbar \widehat{\{O,O'\}}+o(\hbar),\;\;
\hat{O}^\dagger=\hat{\bar{O}}+o(\hbar)
\ee
where quantum corrections are allowed (and in principle unavoidable 
except if we restrict $\cal O$, say to functions linear in momenta).
We will assume that the Hilbert space can be represented as a space of 
square integrable 
functions on (a distributional extension $\overline{{\cal C}}$ of) 
$\cal C$ with respect a positive, faithful probability measure
$\mu$, that is, ${\cal H}=L_2(\overline{{\cal C}},d\mu)$ as it is 
motivated by the real polarization.   

The fact that $C$ is positive motivates to quantize it in such a way 
that it becomes a self-adjoint, positive definite operator. We will assume 
this to 
be the case in what follows.
Applying then the quantization rules to the functions $z$ in (\ref{2.1}) 
we arrive at
\be \label{2.5}
\hat{z}=\sum_{n=0}^\infty \frac{i^n}{n^!} 
\frac{[\hat{q},\hat{C}]_{(n)}}{(i\hbar)^n}=e^{-\hat{C}/\hbar}\hat{q}
e^{\hat{C}/\hbar}
\ee
The appearance of $1/\hbar$ in (\ref{2.5}) justifies the requirement
for $C/\hbar$ to be dimensionless in definition (\ref{2.1}).
We will call $\hat{z}$ {\it annihilation operator} for reasons that
will become obvious in a moment. 

Let now $q\mapsto \delta_{q'}(q)$ be the $\delta$-distribution with 
respect to $\mu$ with support at $q=q'$. (More in mathematical terms,
consider the complex probability measure, denoted as  
$\delta_{q'}d\mu$, which is defined by $\int \delta_{q'} d\mu f=f(q')$ 
for measurable $f$).
Notice that since $C$ is non-negative and necessarily depends 
non-trivially on momenta (which will turn into (functional) derivative 
operators in the quantum theory), the operator $e^{-\hat{C}/\hbar}$
is a {\it smoothening operator}. Therefore, although $\delta_{q'}$
is certainly not square integrable, the complex measure (which
is probability if $\hat{C}\cdot 1=0$) 
\be \label{2.6}
\psi_{q'}:=e^{-\hat{C}/\hbar}\delta_{q'}
\ee
has a chance to be an element of $\cal H$. Whether or not it does depends 
on the details of ${\cal M},\Omega,C$. For instance, if $C$ as a function
of $p$ at fixed $q$ has flat directions, then the smoothening effect
of $e^{-\hat{C}/\hbar}$ may be insufficient, so in order to avoid this
we required that $C$ is positive definite and not merely non-negative.
If $C$ would be indefinite, then (\ref{2.6}) has no chance to make sense 
as an $L_2$ function. 

We will see in a moment that (\ref{2.6}) qualifies as a candidate {\it
coherent state} if we are able to analytically extend (\ref{2.6}) to complex 
values $z$ of $q'$ where the label $z$ in $\psi_z$ will play the role
of the point in $\cal M$ at which the coherent state is peaked. In 
order that this is possible (and in order that the extended function is 
still square integrable), (\ref{2.6}) should be entire analytic.
Now $\delta_{q'}(q)$ roughly has an integral kernel of the form 
$e^{i(k,(q-q'))}$ (with some pairing $<.,.>$ between tangential and 
cotangential vectors) which is analytic in $q'$ but the integral over $k$,
after applying $e^{-\hat{C}/\hbar}$,
will produce an entire analytic function only if there is a damping factor 
which decreases faster than exponentially. This provides the intuitive 
explanation for the growth requirement in definition \ref{def2.1}.
Notice that the $\psi_z$ are not necessarily normalized.

Let us then assume that 
\be \label{2.7}
q\mapsto \psi_m(q):=[\psi_{q'}(q)]_{q'\to z(m)}=
[e^{-\hat{C}/\hbar}\delta_{q'}(q)]_{q'\to z(m)}
\ee
is an entire $L_2$ function. Then $\psi_m$ is automatically an 
{\it eigenfunction of the annihilation operator $\hat{z}$ with eigenvalue
$z$} since 
\be \label{2.8}
\hat{z}\psi_m=[e^{-\hat{C}/\hbar}\hat{q}\delta_{q'}]_{q'\to z(m)} 
=[q' e^{-\hat{C}/\hbar}\delta_{q'}]_{q'\to z(m)}=z(m)\psi_m
\ee
where in the second step we used that the delta distribution is a 
generalized eigenfunction of the operator $\hat{q}$. But to be an 
eigenfunction of an annihilation operator {\it is one of the accepted 
definitions of coherent states} ! 

Next, let us verify that $\psi_m$ indeed has a chance to be peaked at
$m$. To see this, let us consider the self-adjoint (modulo domain 
questions) combinations 
\be \label{2.9}
\hat{x}:=\frac{\hat{z}+\hat{z}^\dagger}{2},\;\; 
\hat{y}:=\frac{\hat{z}-\hat{z}^\dagger}{2i}
\ee
whose classical analogs provide real coordinates for $\cal M$.
Then we have automatically from (\ref{2.8})
\be \label{2.10}
<\hat{x}>_m:=\frac{<\psi_m,\hat{x}\psi_m>}{||\psi_m||^2}=
\frac{z(m)+\bar{z}(m)}{2}=:x(m)
\ee 
and similar for $y$. Equation (\ref{2.10}) tells us that the operator
$\hat{z}$ should really correspond to the function $m\mapsto z(m),\;m\in 
{\cal M}$.

Now we compute by similar methods that
\be \label{2.11}
<[\delta\hat{x}]^2>_m
:=\frac{<\psi_m,[\hat{x}-<\hat{x}>_m]^2\psi_m>}{||\psi_m||^2}
=<[\delta\hat{y}]^2>_m
=\frac{1}{2}|<[\hat{x},\hat{y}]>_m|
\ee
so that the $\psi_m$ are automatically {\it minimal uncertainty states for
$\hat{x},\hat{y}$}, moreover the fluctuations are unquenched. This is 
the second motivation for calling the $\psi_m$ coherent states. Certainly
one should not only check that the fluctuations are minimal but also that 
they are small as compared to the expectation value, at least at generic 
points of the phase space, in order that the quantum errors are small. 

The {\it infinitesimal Ehrenfest} property
\be \label{2.12}
\frac{<[\hat{x},\hat{y}]>_z}{i\hbar}=
\{x,y\}(m)+O(\hbar)
\ee
follows if we have properly implemented the canonical commutation 
relations and 
adjointness relations. The size of the correction, however, does not 
follow from these general considerations but the minimal uncertainty
property makes small corrections plausible. Condition (\ref{2.12}) 
supplies information about how well the symplectic structure is reproduced
in the quantum theory.  

For the same reason one 
expects that the peakedness property
\be \label{2.13}
|\frac{<\psi_m,\psi_{m'}>|^2}{||\psi_m||^2\;||\psi_{m'}||^2}
\approx \chi_{K_m}(m')
\ee
holds, where $K_m$ is a phase cell with center $m$ and Liouville volume  
$\approx\sqrt{<[\delta \hat{x}]^2>_m <[\delta \hat{y}]^2>_m}$ 
and $\chi$ denotes the characteristic function of a set.

Finally one wants coherent states to be overcomplete in order that every
state in $\cal H$ can be expanded in terms of them. This has to be checked
on a case by case analysis but the fact that our complexifier coherent 
states are for real $z$ nothing else than regularized $\delta$ 
distributions which in turn provide a (generalized) basis makes this 
property plausible to hold.

After these abstract considerations, the reader should work out a known 
example for himself in order to fill in the necessary intuition:\\
Phase space: ${\cal M}=T^\ast \Rl=\Rl^2$\\
Configuration space: ${\cal C}=\Rl$\\
Symplectic structure: $\{q,q\}=\{p,p\}=0,\;\{p,q\}=1$\\
Complexifier: $C=p^2/2$\\
This complexifier certainly meets all our requirements and the ambitious
reader will find out that the construction displayed above results in the 
usual coherent states for the harmonic oscillator as defined via 
energy eigenstates with Hamiltonian 
$H=(p^2+q^2)/2$ up to a phase
(we take $q,p,\hbar$ dimensionless for the sake of the example).\\
\\
Important Remark:\\
1.\\
It is very crucial to know the map $m\mapsto z(m)$. If we are just given
some states $\psi_z$ with $z\in {\cal C}^\Cl$ then we have no way
to find the point $m\in {\cal M}$ to which $z$ corresponds (there are 
certainly infinitely many diffeomorphisms between ${\cal M},{\cal C}^\Cl$)
and the connection with the classical phase space is lost. Without 
this knowledge we cannot check, for instance, whether the 
infinitesimal Ehrenfest property holds. This is one 
of the nice things that the complexifier method automatically does for 
us. In order to know the function $z(m)$ we must know what the classical
limit of $\hat{C}$ is, if we are just given some abstract operator 
without classical interpretation, then again we do not know $z(m)$.
Of course, if we are given just some set of states $\psi_z$ we could try 
to
construct an appropriate map $m\mapsto z(m)$ as follows: Find a 
(complete) set of basic operators $\hat{O}$ whose fluctuations are (close 
to) minimal and define a map $z\mapsto 
O'(z):=<\psi_z,\hat{O}\psi_z>/||\psi_z||^2$. Also define 
$\{O',\bar{O}'\}'(z):=\lim_{\hbar\to 0}
<\psi_z,\frac{[\hat{O},\hat{O}^\dagger]}{i\hbar}\psi_z>/||\psi_z||^2$.
Now construct $m\mapsto z(m)$ by asking that the pull-back functions 
$O(m):=O'(z(m))$ satisfy 
\be \label{2.13a}
\{O,\bar{O}\}(m)=\{O',\bar{O}'\}'(z(m))
\ee
in other words, that the symplectic structure defined by $\{.,.\}'$ is 
the symplectomorphic image of the original symplectic structure $\{.,.\}$
under the canonical transformation $m\mapsto z(m)$.
The reader will agree that this procedure is rather indirect and 
especially in field theory will be hard to carry out. Notice that by far 
not all symplectic structures are equivalent so that even to find 
appropriate operators for given $\psi_z$ such that at least one map
$m\mapsto z(m)$ exists will be a non-trivial task. The complexifier method
guarantees all of this to be the case from the outset since the 
transformation (\ref{2.1}) is a canonical transformation by construction.
\\ 
2.\\ 
Usually one calls a {\it complete} system of states $\psi_m,\;m\in {\cal 
M}$ semiclassical for an algebra $\hat{{\cal O}}$ if the expectation 
values
of the operators and their commutators in the states $\psi_m$ reproduce 
the values of the corresponding functions and their Poisson brackets 
at the point $m$ and if the relative errors (fluctuations) are small.
Coherent states have the additional properties of being annihilation 
operator eigenstates, to be minimal uncertainty states and to be peaked in 
phase space. The complexifier method thus provides a general 
construction guideline, but no algorithm, to arrive at a satisfactory 
candidate system of semiclassical, even coherent, states for a wide class 
of theories. It should be kept in mind, however, that while coherent states
are a natural choice for semiclassical states, they do not comprise the 
full set of semiclassical states and in some cases it maybe needed to 
employ more general classes of semiclassical states.

\subsection{Complexifier Coherent States for Diffeomorphism Invariant 
Theories of Connections}
\label{s2.3}

We assume the reader to be familiar with the basics of QGR. A short
introduction can be found in appendix \ref{s2.2} where we mainly introduce 
notation and explain why the Hilbert space ${\cal H}_0$ currently used 
in QGR is extremely natural from a representation theoretic point of view.
It therefore seems to be a suitable kinematical background independent 
starting point 
for further analysis, in particular, it should be considered as the 
``fundamental representation" from which all others can be derived.\\ 
\\
After having chosen a Hilbert space ${\cal H}_0=L_2(\ab,d\mu_0)$, the 
only input required in the complexifier construction is the choice of a 
complexifier itself. We will restrict our class of choices to functions
$C=C(E)$ which are gauge invariant but not necessarily diffeomorphism 
invariant (since we can use the $(D-)$metric to be approximated as a 
naturally available background metric) and only depend on the electric field
to make life simple.
We suppose that the associated operator $\hat{C}$ is a densely 
defined positive definite operator on ${\cal H}_0$ whose spectrum is pure 
point (discrete). The latter assumption is not really a restriction 
because operators which are constructed from (limits of) electric flux 
operators quite generically have this sort of spectrum, see e.g. \cite{6}.  
Let $T_s,\;s\in {\cal S}$ be the associated uncountably infinite orthonormal 
basis of eigenvectors. The labels
$s=(\gamma,\vec{\pi},\vec{I})$ are triples consisting of a piecewise 
analytic graph $\gamma$, an array of equivalence classes of 
non-trivial irreducible representations $\pi_e$, one for each edge $e$ of 
of $\gamma$ and an array of intertwiners $I_v$, one for each vertex $v$
of $\gamma$. The intertwiners are chosen in such a way that the $T_s$
are not only gauge invariant but also eigenfunctions of $\hat{C}$.
The space of possible $\vec{I}$ at given $\vec{\pi}$ is always finite 
dimensional and the operators of the form $\hat{C}$ can never change 
$\vec{\pi},\gamma$. Thus, the $T_s$ are just suitable 
linear combinations of the usual spin network functions \cite{29}.

Let $\lambda_s$ be the corresponding eigenvalues. Then 
\be \label{2.15}
\delta_{A'}=\sum_{s\in {\cal S}} T_s(A') \overline{T_s}
\ee
is a suitable representation of the $\delta$ distribution with respect to
$\mu_0$, i.e. 
\be \label{2.16}
\int_{\ab} d\mu_0(A) \delta_{A'}(A) f(A)=\sum_s T_s(A') <T_s,f>=f(A')
\ee
and our complexifier coherent states become explicitly
\be \label{2.17}
\psi_m=\sum_{s\in {\cal S}} e^{-\lambda_s/\hbar} T_s(Z(m)) \overline{T_s}
\ee
where we have made use of the fact that the expression for $\hat{C}$ is real,
$m=(A,E)$ is the point in $\cal M$ to be approximated and 
\be \label{2.18}
[z_a^j(m)](x):=[^{(\Cl)}A_a^j(m)](x):=A_a^j(x)-i\kappa\frac{\delta 
C}{E^a_j(x)}
=\sum_{n=0}^\infty \frac{i^n}{n!} \{A_a^j(x),C\}_{(n)}
\ee
is a {\it complex valued $G$-connection} since $C$ is supposed to be 
gauge invariant.

Since there are more than countably many terms different from zero in 
(\ref{2.17}) the states $\psi_m$ are not elements of ${\cal H}_0$. Rather,
they define algebraic distributions in Cyl$^\ast$ defined by 
\ba \label{2.18a}
\psi_m[f] &:=& <1,\psi_m \;f>
\nonumber\\
&=& <1,[e^{-\hat{C}/\hbar}\delta_{A'}] f>_{A'\to Z(m)}
=<\overline{f},[e^{-\hat{C}/\hbar}\delta_{A'}]>_{A'\to Z(m)}\nonumber\\
&=& <e^{-\hat{C}/\hbar}\overline{f},\delta_{A'}>_{A'\to Z(m)}
=<1,\delta_{A'}\overline{e^{-\hat{C}/\hbar}\overline{f}}>_{A'\to Z(m)}
\nonumber\\
&=&(\delta_{A'}[\overline{e^{-\hat{C}/\hbar}\overline{f}}])_{A'\to Z(m)}
\ea
For $f=T_s$ the right hand side of (\ref{2.18}) becomes
$e^{-\lambda_s/\hbar} T_s(Z(m))$ and since 
$C$ is supposed to depend on a sufficiently high power of $E$ and since
$|T_s(Z(m))|$ grows at most exponentially with $\vec{\pi}$, these numbers 
are actually bounded from above so that the distribution is well
defined. Equivalently, we can consider $\psi_m$ as a complex probability
measure (since the $\delta$ distribution is). 

Consider for each entire analytic path $e$ the 
{\it annihilation operators} 
\be \label{2.19}
\hat{g}_e:=e^{-\hat{C}/\hbar}\hat{A}(e) e^{\hat{C}/\hbar}
\ee
which are the quantum analogs of the classical functions 
$Z(m)(e)=h_e(^{(\Cl)}A(m))=g_e(m)$ 
where $h_e(A)=A(e)$ denotes the holonomy of $A$ along $e$. Thus 
$g_e(m)$ is the holonomy along $e$ of the complex connection $^{(\Cl)}A$.
The holonomy property can also be explicitly checked for the operators
$\hat{g}_e$ themselves since for a composition of paths $e=e_1\circ e_2$
we have from the holonomy property for $\hat{A}$ that
\be \label{2.19a}
\hat{g}_{e_1}\hat{g}_{e_2}=
e^{-\hat{C}/\hbar}\hat{A}(e_1)\hat{A}(e_2)e^{\hat{C}/\hbar}
=\hat{g}_{e} \mbox{ and }
\hat{g}_{e^{-1}}=(\hat{g}_e)^{-1}
\ee
where product and inversion is that within $G^{\Cl}$.

As one can explicitly check, $\psi_m$ is a simultaneous generalized 
eigenvector of all the $\hat{g}_e$, that is,
\ba \label{2.20}
(\hat{g}_e \psi_m)[f]&:=&<1,[\hat{g}_e\psi_m]\;f>
\nonumber\\
&=&<\hat{g}_e^\dagger\overline{f},\psi_m >
=<1,\psi_m\overline{\hat{g}_e^\dagger\overline{f}}>
=\psi_m[\overline{\hat{g}_e^\dagger\overline{f}}]
\nonumber\\
&=& h_e(Z(m)) \psi_m[f]
\ea
The crucial point is now that although the $\psi_m$ are not normalizable, 
we may be able to define a positive linear functional $\omega_m$ on 
our algebra of functions as expectation value functional
\be \label{2.21}
\omega_m(\hat{O}):=\frac{<\psi_m,\hat{O}\psi_m>}{||\psi_m||^2}
\ee
where we have {\it used the inner product on ${\cal H}_0$} and no other
additional inner product ! This is conceptually appealing because, if we 
can give meaning to (\ref{2.21}), then we arrive at a new 
representation of the canonical commutation relations {\it which is 
derived from ${\cal H}_0$} whence ${\cal H}_0$ plays the role of the 
fundamental representation, very much in the same way as temperature 
representations in ordinary quantum field theory can be derived from the 
Fock representation by limits of the kind performed in (\ref{2.21})  

Expression (\ref{2.21}) is very formal in the sense that it is the 
quotient of two uncountably infinite series. However, notice that we can 
easily give meaning to it at least for {\it normal ordered functions of 
annihilation and creation operators} as follows: For the numerator we 
write (the colons denote normal ordering of an operator 
$\hat{O}=O(\{\hat{g}_e,\hat{g}_e^\dagger\})$ which is an
analytic function of the 
$\hat{g}_e,\hat{g}_e^\dagger$) 
\ba \label{2.22}
<\psi_m,:\hat{O}:\psi_m> &:=&
\sum_s e^{-\lambda_s/\hbar} \overline{T_s(Z(m))} <\overline{T_s},
:\hat{O}:\psi_m>
\nonumber\\
&=& \sum_s e^{-\lambda_s/\hbar} \overline{T_s(Z(m))} 
[:\hat{O}:\psi_m](T_s)
\nonumber\\
&=& O(\{g_e(m),\overline{g_e(m)}\})
\sum_s e^{-\lambda_s/\hbar} \overline{T_s(Z(m))} <\overline{T_s}\psi_m>
\nonumber\\
&=& O(\{g_e(m),\overline{g_e(m)}\})
\sum_s e^{-2\lambda_s/\hbar} |T_s(Z(m)|^2
\nonumber\\
&=:& O(\{g_e(m),\overline{g_e(m)}\}) ||\psi_m||^2
\ea
The norm squared in the last line of (\ref{2.22}) is infinite but with 
proper regularization understood we arrive at (after taking the 
regulator away)
\be \label{2.23} \omega_m(:\hat{O}:)=O(m)=O(\{g_e(m),\overline{g_e(m)}\})
\ee
which has no quantum corrections at all. Thus, if the functions $m\mapsto
g_e(m)$ separate the points of $\cal M$ as $e$ varies, then we may use them
as the basic variables in the quantum theory and they, together with their 
adjoints, have the correct expectation values in the representation 
induced by $\omega_m$ via the GNS construction, moreover, that 
representation by construction also solves the adjointness and canonical 
commutation relations. Of course, (\ref{2.23}) will be 
an interesting functional only if the normal ordering corrections of 
interesting operators are finite. This can only be decided in a case by
case analysis.

As an illustrative example (see \cite{9} for more 
details) let $Q^{ab}:=E^a_j E^b_k \delta^{jk}$ and consider the 
diffeomorphism invariant complexifier 
(recall that $E$ is a density of weight one)
\be \label{2.24}
C:=\frac{1}{a\kappa}\int_\sigma d^Dx (\sqrt{\det(Q)})^{1/(D-1)}
\ee
where $a$ is a parameter with units of $(\hbar\kappa)^{1/(D-1)}$. Our 
convention is that $A$ has dimension of cm$^{-1}$, thus 
$\frac{1}{\kappa}\int_\Rl dt \int_\sigma d^Dx \dot{A}_a^j E^a_j$, the 
kinetic term in the canonical action, must have dimension of an action,
therefore $E/(\hbar\kappa)$ must have dimension cm$^{-(D-1)}$. Thus, in 
order that $C/\hbar$ be dimensionfree, $a$ must have the said dimension.
%
%
E.g. for general relativity in $D+1=4$ dimensions, 
$(\hbar\kappa)^{1/(D-1)}=\ell_p$ is the Planck length, (\ref{2.24}) is 
essentially the volume functional $V$ for $\sigma$ and if we are 
interested in 
cosmological questions or scales, then $a=1/\sqrt{\Lambda}$ would be 
a natural choice, where $\Lambda$ is the cosmological constant.
In that case the quantized complexifier would simply be given by
\be \label{2.25}
\hat{C}/\hbar=\frac{1}{a\ell_p^2}\hat{V}=\frac{\ell_p}{a}\hat{v}
\ee
where $\hat{v}=\hat{V}/\ell_p^3$ is the dimensionfree volume functional 
which has discrete spectrum with eigenvalues in multiples of $\ell_p^3$.
Thus $\hat{C}=t\hat{v}$ where the tiny {\it classicality parameter}
\be \label{2.26}
t=\frac{\ell_p}{a}=\sqrt{\hbar\kappa\Lambda}
\ee
has entered the stage. We easily compute the complexified connection in 
this case as 
\be \label{2.27}
^{(\Cl)}A=A-ie/(2a)
\ee
where $e$ is the dimensionfree co-triad. Thus, with the volume as the 
complexifier, the $g_e(m)$ indeed separate the points of $\cal M$ !

However, in order to qualify as a good 
semi-classical state, at the very least the fluctuations of our basic 
operators with respect to $\omega_m$ should be 
small as compared to the expectation values at generic points of the 
phase space, in particular, they should be finite. Whether or not this is 
the case has to be checked for the explicit choices for $C$ displayed in 
the subsequent sections.

It should be noted, however, that even if the fluctuations do not come out
finite, then we can still produce {\it graph dependent coherent states},
which in particular {\it are} elements of ${\cal H}_0$, as follows:\\
Given a graph $\gamma$, consider all of its subgraphs 
$\gamma'\subset\gamma$ obtained by removing edges in all possible ways.
Given a label $s$ we write $s=(\gamma(s),\vec{\pi}(s),\vec{I}(s))$ and 
define a graph dependent $\delta-$distribution
\be \label{2.28}
\delta_{A',\gamma}(A):=\sum_{\gamma'\subset\gamma}\;\;\;
\sum_{s;\;\gamma(s)=\gamma'}T_s(A') \overline{T_s(A)}
\ee
It is easy to check that (\ref{2.28}) is a $\delta$-distribution restricted
to those functions on $\ab$ which can be written in terms of the holonomies
$A(p)$ where $p\subset\gamma$. In fact, (\ref{2.28}) is the {\it cut-off}
of (\ref{2.15}) with the cut-off given by the graph $\gamma$ since
(\ref{2.28}) is the restriction of the uncountably infinite series in
(\ref{2.15}) to the countably infinite one in (\ref{2.28}) given by
restricting the sum over $s\in {\cal S}$ to $s\in {\cal S}_\gamma$ where
\be \label{2.29}
{\cal S}_\gamma=\{s\in{\cal S};\;\;\gamma(s)\subset\gamma\}
\ee
In fact, we can consider the Hilbert space ${\cal H}_{\gamma,0}=
L_2(\ab_\gamma,d\mu_{0,\gamma})$ where $\mu_{0,\gamma}$ is the pushforward
of $\mu_0$ to the space $\ab_\gamma$ which is the spectrum of holonomy
algebra restricted to paths within $\gamma$. Then $\delta_\gamma$ is in fact
the $\delta-$distribution with respect to $\mu_{0\gamma}$. In other words,
$\delta_\gamma$ is the cylindrical projection of the complex measure
$\delta$.

We now obtain normalizable graph, dependent coherent states 
\be \label{2.30}
\psi_{\gamma,m}(A)=
[e^{-\hat{C}/\hbar}\delta_{\gamma,A'}]_{A'\to\; ^{(\Cl)}A(m)} 
=\sum_{s\in {\cal S}_\gamma} e^{-\lambda_s/\hbar} T_s^{(\Cl)}A(m))
\overline{T_s(A)}
\ee
with norm
\be \label{2.30a}
||\psi_{\gamma,m}||^2
=\sum_{s\in {\cal S}_\gamma} e^{-2\lambda_s/\hbar} |T_s(^{(\Cl)}A(m))|^2
\ee
which converges due to our assumptions on the spectrum $\lambda_s$.
Notice that these assumptions might not hold for the volume complexifier
(the volume operator is only non-negative but not positive definite, the 
spectrum has flat directions and it would be crucial to know how generic
these are, a problem very similar in nature (but much simpler) to the 
convergence proof 
of Euclidean Yang-Mills theory). By arguments very similar to those 
from above it is easy to check that the $\psi_{\gamma,m}$ are still
eigenstates of the operators $\hat{g}_e$ provided that the path $e$
lies within $\gamma$. In other words, for normal ordered functions
of some set of operators $\hat{g}_e,\hat{g}_e^\dagger$ it is unimportant 
whether we work with the complete state $\psi_m$ or with the 
cut-off state $\psi_{\gamma,m}$, as far as expectation values are concerned,
as long as $\gamma$ contains all the paths $e$ under consideration. However,
the fluctuations will be significantly different in general since the 
square of a normal ordered operator is no longer normal ordered. As we will
see, it is the fluctuations which will force us to work with graph 
dependent coherent states.

Thus, we arrive at a {\it coherent state family
$\{\psi_{\gamma,m}\}_{\gamma\in \Gamma}$} for each $m\in {\cal M}$ where 
$\Gamma$ denotes the set of piecewise analytic, compactly supported 
graphs embedded into $\sigma$. They define a complex probability measure 
$\mu_m$ through the consistent family of measures 
$d\mu_{\gamma,m}:=\psi_{\gamma,m} d\mu_{0,\gamma}$.\\
To see that this 
family of measures is automatically consistent we consider for 
$\gamma'\subset\gamma$ the projections 
$p_{\gamma'\gamma}:\;\ab_\gamma\to
\ab_{\gamma'}$ defined by restricting connections 
from paths within $\gamma$ to paths within $\gamma'$. 
Now the Hilbert space ${\cal H}_0$ is in fact
the inductive limit of the Hilbert spaces ${\cal H}_{\gamma,0}$, that is,
there exist isometric monomorphisms 
\be \label{2.31}
\hat{U}_{\gamma'\gamma}:\;{\cal H}_{\gamma',0}\to {\cal H}_{\gamma,0};\;
f_{\gamma'}\mapsto p_{\gamma\gamma'}^\ast f_{\gamma'}
\ee
for all $\gamma'\subset\gamma$. These maps satisfy 
the consistency condition
\be \label{2.32}
\hat{U}_{\tilde{\gamma}\gamma}\hat{U}_{\gamma'\tilde{\gamma}}
=\hat{U}_{\gamma'\gamma}
\ee
for all $\gamma'\subset\tilde{\gamma}\subset\gamma$. An operator $\hat{O}$
on ${\cal H}_0$ can be thought of as the inductive limit of a family of 
operators $\{\hat{O}_\gamma\}_{\gamma\in\Gamma}$, that is,  
$\hat{O}_\gamma$ is densely defined on ${\cal H}_\gamma$ subject to the 
consistency condition 
\be \label{2.33}
\hat{O}_\gamma\hat{U}_{\gamma'\gamma}=
\hat{U}_{\gamma\gamma'}\hat{O}_{\gamma'}
\ee
for all $\gamma'\subset\gamma$ (there is also a condition for the domains of
definition which we skip here). Thus, in particular, the complexifier
is a consistently defined operator family all of whose members 
are self-adjoint and positive on the respective ${\cal H}_{0,\gamma}$. 
Therefore, if $f_{\gamma'}$
depends only on connections restricted to paths within $\gamma'$ we have 
\ba \label{2.34}
&&\int_{\ag_\gamma} d\mu_{\gamma,m} [p_{\gamma\gamma'}^\ast f_{\gamma'}]
=(\int_{\ag_\gamma} d\mu_{\gamma,0} \delta_{A',\gamma}
[e^{-\hat{C}_\gamma/\hbar} 
\hat{U}_{\gamma'\gamma}f_{\gamma'}])_{A\to A^{(\Cl)}}
\nonumber\\
&=& (\int_{\ag_\gamma} d\mu_{\gamma,0} \delta_{A',\gamma}
[\hat{U}_{\gamma'\gamma}e^{-\hat{C}_{\gamma'}/\hbar} f_{\gamma'}])_{A\to 
A^{(\Cl)}} 
= (\int_{\ag_{\gamma'}} d\mu_{\gamma',0} \delta_{A',\gamma'}
[e^{-\hat{C}_{\gamma'}/\hbar} f_{\gamma'}])_{A\to A^{(\Cl)}} 
\nonumber\\
&=&\int_{\ag_{\gamma'}} d\mu_{\gamma',m} f_{\gamma'}
\ea
The projective limit of these measures coincides with the measure 
$\psi_m d\mu_0$. The notation is abusing because it suggests that
$\mu_m$ is absolutely continuous with respect to $\mu_0$ which certainly
is not the case because $\psi_m\not\in L_1(\ab,d\mu_0)$.\\
\\
Let us close this subsection by comparing with the concrete set 
of states constructed in \cite{10,11}. The states considered 
there do not define a complex probability measure because we wanted 
to keep the states background independent. However, if one gives 
up background independence then one can trivially 
repair this by making the parameter $a$, used there in analogy to 
(\ref{2.24}), edge dependent. Roughly, one replaces $E_j(S_e)/a^2$
by $l_e E_j(S_e)/a^2$ where $S_e$ is a surface, intersected by $e$,
in a cell complex dual to the graph and the dimensionless numbers
$l_e$ satisfy $l_e+l_{e'}=l_{e\circ e'},\;l_{e^{-1}}=l_e$, see section
\ref{s4}. The associated complexifier operator $\hat{C}_\gamma$ then 
becomes cylindrically consistent and defines an operator $\hat{C}$ 
and a complex probability 
measure if we just {\it define} abstract elements $g_e\in SL(2,\Cl)$ 
subject to
$g_e g_{e'}=g_{e\circ e'},\;g_{e^{-1}}=g_e^{-1}$. 
The trouble, however, is that $\hat{C}$ does 
not have a classical limit, see \cite{9,10} and the remark at the end of 
section \ref{s2.1}. Therefore we do not have a graph independent map
map $m\mapsto g_e(m)$ and no classical interpretation of the
distributional states a priori.
This is precisely the reason why the dual cell complex has been introduced 
in the first place in \cite{9,10} because it enables us to explicitly
construct a map $m\mapsto g^\gamma_e(m)$ {\it graph by graph} and 
therefore we have a semiclassical interpretation of the cut-off states.

In this paper we will propose a new set of states 
which {\it do} come from a classical complexifier whose corresponding 
operator is densely defined 
with explicitly known spectrum and which are defined for arbitrary compact 
gauge 
groups, not only Abelean. Moreover, they do not require the additional 
structure of the cell complex. The analysis of their semiclassical 
properties is beyond the scope of the present paper but since their 
mathematical structure is similar to those defined in \cite{10,11},
most proofs will we be easily adaptable.

\subsection{Varadarajan States}
\label{s2.4}

We will sketch the main ideas of Varadarajan's elegant construction 
\cite{17} for the 
case of the gauge group $U(1)$ (Maxwell theory). The case of $U(1)^3$ 
(linearized gravity) is similar.\\
\\
We consider classical Maxwell theory on Minkowski space, that is, 
$\sigma=\Rl^3$ and the spatial metric is simply $\delta_{ab}$, we do not 
need to worry about positions of tensor indices. By $A_a,E^a$ we denote the 
magnetic potential 
and electric field respectively. Due to our choice that $A$ has dimension
cm$^{-1}$, the Feinstruktur constant is given by $\alpha=\hbar q^2$
where $\kappa_M=q^2$ is the Maxwell coupling constant (electric charge 
squared). Then $E$ has dimension cm$^{-2}$. Let $f$ be an arbitrary but 
fixed test function of rapid decrease which is even under reflection, 
i.e. 
$f(x)=f(-x)$ and which has dimension cm$^{-3}$. We define $f-$smeared 
connections and electric fields respectively by
\be \label{2.35}
E^a_f(x)=\int_\sigma d^3y f(x-y) E^a(y) \mbox{ and }
A_a^f(x)=\int_\sigma d^3y f(x-y) A^a(y) 
\ee
Let us also define the distributional and smeared form factor respectively 
by
\be \label{2.36}
X^a_p(x):=\int_p dy^a \delta^{(3)}(x,y) \mbox{ and }
X^a_{p_f}(x):=\int_\sigma d^3y f(x-y) X^a_p(y)=\int_p dy^a f(x-y)
\ee
where $p$ is any piecewise analytic path. Both objects have dimension
cm$^{-2}$.  
Consider the following two sets of elementary variables
\be \label{2.37}
(E^a(x),A(p_f)=e^{i\int d^3x X^a_{p,f} A_a}) 
\mbox{ and }
(E^a_f(x),A(p)=e^{i\int d^3x X^a_p A_a}) 
\ee
where $A(p),A(p_f)$ denote the 
holonomies of the unsmeared and smeared connection respectively.

It is a remarkable feature of Abelean gauge groups that the smeared
holonomy $A(p_f)$ still transforms covariantly under local 
gauge transformations, that is, if $A^g=-dg g^{-1}/i+A$ then
with $g=e^{i\lambda}$ 
\ba \label{2.38}
A^g(p_f) &=& A(p_f)e^{-i\int d^3x\lambda_{,a} X^a_{p_f}}
=A(p_f)e^{i\int d^3x \lambda \partial_a X^a_{p_f}}
\nonumber\\
&=& A(p_f)e^{i\int d^3x \lambda(x) \int_p dy^a \partial_{x^a} f(x-y)} 
=A(p_f)e^{-i\int d^3x \lambda(x) \int_p dy^a \partial_{y^a} f(x-y)} 
\nonumber\\
&=& A(p_f)e^{-i\int d^3x \lambda(x) \int_p df(x-y)} 
=A(p_f)e^{-i\int d^3x \lambda(x) [f(x-f(p))-f(x-b(p))]}
\nonumber\\
&=& g_f(b(p))A(p_f)g_f(f(p))^{-1}
\ea 
where the integration by parts in the second step is justified by
the temeredness of $f$ and as before $b(p),f(p)$ denote the beginning and 
final point of $p$ respectively.
Here $g_f(x)=e^{i \int d^3y \lambda(y) f(x-y)}$ is the {\it smeared 
gauge transformation} which still takes values in $U(1)$.
In particular, if $p$ is a closed path then the smeared holonomies
are still gauge invariant.
Moreover, since already $E^a$ is gauge invariant, so is $E^a_f$.
{\it These facts no longer hold in the Non-Abelean context}, the analogs
of $E_f,A(p_f)$ are no longer gauge covariant ! 

It is trivial to check that again, {\it precisely because $U(1)$ is 
Abelean}, both Poisson algebras are closed. Moreover, they are {\it 
isomorphic}
\ba \label{2.39}
&& \{A(p_f),A(p'_f)\}=\{E^a(x),E^b(y)\}=0,\;
\{E^a(x),A(p_f)\}=i q^2 X^a_{p_f}(x)
\nonumber\\
&& \{A(p),A(p)\}=\{E^a_f(x),E^b_f(y)\}=0,\;
\{E^a(x),A(p)\}=i q^2 X^a_{p_f}(x)
\ea
and the reality conditions on smeared and unsmeared objects are identical.
Thus, at a purely Poisson $^\ast-$algebraic level we cannot distingiush 
between the Poisson algebras generated by these two pairs of variables 
although their values at a given point $m=(A_a,E^a)\in{\cal M}$ in the 
classical phase space are quite different. (To be precise, one should 
also prove that any algebraic identities in the subalgebra generated by
the $A(p)$ is a corresponding identity in the subalgebra generated by the 
$A(p_f)$ for, say connections of rapid decrease. Please refer to 
the detailed analysis in the first reference of \cite{17} where this proof 
can be found 
for the specific choice of $f$ in (\ref{3.15}). In what follows, we will
assume this to hold for the function $f$ under consideration.)

Consider the resulting abstract $^\ast-$algebras ${\cal A}_F$ and 
${\cal A}_P$ respectively, to which we will refer as the 
{\it Fock algebra and polymer algebra} respectively, generated by 
requiring that Poisson brackets become commutators divided by $i\hbar$
and that reality conditions become $^\ast-$relations. That is,
\ba \label{2.39a}
&& [\hat{A}(p_f),\hat{A}(p'_f)]=\{\hat{E}^a(x),\hat{E}^b(y)]=0,\;
[\hat{E}^a(x),\hat{A}(p_f)\}=i\alpha X^a_{p_f},
\nonumber\\ &&
\hat{A}(p_f)^\ast-\hat{A}(p_f)^{-1}=\hat{E}^a(x)^\ast-\hat{E}^b(x)=0,
\nonumber\\
&& [\hat{A}(p),\hat{A}(p')]=\{\hat{E}^a_f(x),\hat{E}^b_f(y)]=0,\;
[\hat{E}^a_f(x),\hat{A}(p)\}=i\alpha X^a_{p_f},
\nonumber\\ &&
\hat{A}(p)^\ast-\hat{A}(p)^{-1}=\hat{E}^a_f(x)^\ast-\hat{E}^b_f(x)=0
\ea
Then we have an isomorphism of $^\ast$algebras
\be \label{2.40}
{\cal I}:\; {\cal A}_F\to {\cal A}_P;\;(\hat{A}(p_f),\hat{E}^a(x))\mapsto 
(\hat{A}(p),\hat{E}^a_f(x))
\ee

Let $\omega_F$ be a ``Fock state", that
is, a normalized, positive linear functional on the Fock algebra. Then the 
GNS construction provides a cyclic vector $\Omega_F$ on a Hilbert space
${\cal H}_F$ and a representation $\pi_F$ of ${\cal A}_F$ such that
\be \label{2.41}
<\Omega_F,\pi_F(b)\Omega_F>_F=\omega_F(b) 
\ee
for each $b\in {\cal A}_F$ and $<.,.>_F$ denotes the inner product on
${\cal H}_F$. 
Using the isomorphism 
(\ref{2.40}) we trivially obtain a positive linear functional $\omega_P$ 
on ${\cal A}_P$ and corresponding GNS data via
\be \label{2.42}
\omega_P({\cal I}(b)):=\omega_F(b),\;
<\Omega_P,\pi_P({\cal I}(b))\Omega_P>_P:=\omega_P({\cal I}(b)) 
\ee
  
This works for any $\omega_F$. Let us now consider the particular
Fock representation $\omega_F^H$ induced by the Maxwell Hamiltonian on 
Minkowski space
\be \label{2.43}
H=\frac{1}{2q^2}\int d^3x \delta_{ab}[E^a E^b+B^a B^b]
\ee
where $B^a=\epsilon^{abc}\partial_b A_c$ is the magnetic field.
Using the transversal projection operator
\be \label{2.44}
(P\cdot A)_a(x)=A_a(x)-\partial_a[\Delta^{-1}\partial^b A_b](x)
\ee
with the Laplacian $\Delta=\delta^{ab}\partial_a\partial_b$
and using the Gauss constraint $E^a_{,a}=0$ we may rewrite $H$ in the form
\be \label{2.44a}
H=\int d^3x \delta^{ab} \bar{z}_a(x)[\hbar\sqrt{-\Delta}] (P\cdot z)_b(x)
\ee
where we have defined the quantity ($m=(A,E)\in{\cal M}$ a phase space 
point) 
\be \label{2.45}
z_a=z_a(m)=\frac{1}{\sqrt{2\alpha}}
[(\root[4]\of{-\Delta})A_a-i(\root[4]\of{-\Delta})^{-1}E^a]
\ee
which has dimension cm$^{-3/2}$. They obey the standard canonical brackets
and reality conditions
\be \label{2.46}
\{z_a(x),z_b(y)\}=\{\bar{z}_a(x),\bar{z}_b(y)\}=0,\;
\{z_a(x),\bar{z}_b(y)\}=-iq^2/\alpha\delta_{ab}\delta(x,y),\; 
\overline{z_a(x)}=\bar{z}_a(x) 
\ee

The positive linear functional $\omega^H_F$ can be formally derived by the 
following steps:\\
1) We want to represent $\hat{A}_a(x)$ and $\hat{E}_a(x)$ respectively
as multiplication operator $\pi_F^H(\hat{A}_a(x)):=A_a(x)$ and functional 
derivative $\pi_F^H(\hat{E}^a(x)):=i\alpha\delta/\delta A_a(x)$ on a 
Hilbert space ${\cal H}_F^H$ of square integrable functions of 
$P\cdot A$ (they are gauge invariant and all operators should be 
projected by $P\cdot$).\\
2) The vacuum state $\Omega_F^H$ should be selected by the condition  
that $\pi_F^H(P\cdot\hat{z_a})$ annihilates $\Omega_F^H$.\\
3)Canonical commutation relations and adjointness conditions should be 
implemented according to (\ref{2.47}).

Condition 3) is formally satisfied if we choose 
${\cal H}_F^H=L_2(\ag',[dA]/||\Omega^H_F||^2)$
where $[dA]$ is the formal infinite dimensional Lebesgue measure,
$\ag'$ is a distributional extension of the space of 
smooth $U(1)$ connections modulo gauge transformations yet to be determined 
(it will turn out to be the 
space of transversal vector valued tempered distributions on $\Rl^3$) 
and the infinite
constant $||\Omega_F^H||$ denotes the norm of the ground state.  

Conditions 1) and 2) are satisfied if we define the ground state formally
as 
\be \label{2.47}
\Omega_F^H(A)=e^{-\frac{1}{2\alpha}\int d^3x A_a\sqrt{-\Delta}(P\cdot 
A)_b}
\ee
Notice that none of the objects $[dA],\Omega_F^H,||\Omega_F^H||^2$
makes sense separately, however, the combination
\be \label{2.48}
d\mu_F^H:="[dA] \frac{|\Omega_F^H|^2}{||\Omega_F^H||^2}"
\ee
that appears in expectation values will make rigorous sense as 
a Gaussian measure. For instance, if $p$ is a closed loop, then
$A(p_f)$ only depends on $P\cdot A$ and we find by formally completing 
the square in the appearing Gaussian integral that
\be \label{2.49}
\omega_F^H(\hat{A}(p_f)):=<\Omega_F^H,\pi_F^H(\hat{A}(p_f))\Omega_F^H>
=e^{-\frac{\alpha}{4}\int d^3x \delta_{ab} X^a_{p_f} 
\sqrt{-\Delta}^{-1}X^b_{p_f}}
\ee
There is no transversal projector needed in (\ref{2.49}) because for a 
closed path $p$ the form factor $X^a_{p_f}$ is already transversal.
Similarly we can compute expectation values of polynomials
involving $\hat{E}^a(x)$ and $\hat{A}(p_f)$ by using the rules of 
Gaussian integrals. It should be noted, however, that $\hat{E}^a(x)$
is not an operator but rather an operator valued distribution. In order
to turn it into an operator we must smear it with a test function, i.e.
we must consider the operators 
\be \label{2.50}
\hat{E}(l)=\int d^3x l_a(x) \hat{E}^a(x)
\ee
where $l_a$ is a transversal test function of rapid decrease of dimension
cm$^{-1}$.
We compute
\be \label{2.51}
\omega_F^H(\hat{E}(l))=0,\;
\omega_F^H(\hat{E}(l)^2)=\frac{\alpha}{2}\int d^3x (l^a\sqrt{-\Delta} l_a)
\ee
%
%
%
%
From this calculation we observe that if we would replace $l_a$ by
a singular expression of the form
\be \label{2.51a}
Y^S_a(x)=\frac{1}{2}\int_S \epsilon_{abc} dy^b\wedge dy^b 
\delta^{(3)}(x,y)
\ee
which would be appropriate for an electric flux operator $\hat{E}(Y^S)$
then its fluctuation diverges since e.g. for $S=[-L,L]^2\times\{0\}$
\ba \label{2.52}
&&\int d^3x (Y^S_a \sqrt{-\Delta} Y^S_a)
=\int_{[-L,L]^2} d^2y \int_{[-L,L]^2} d^2y' 
[\sqrt{-\Delta}_x  \delta(x,(y_1',y_2',0))]_{x=(y_1,y_2,0)}
\nonumber\\
&=& \int \frac{d^3k}{(2\pi)^3} ||k|| \int_{[-L,L]^2} d^2y \int_{[-L,L]^2} 
e^{i[k_1(y_1-y_1')+k_2(y_2-y_2')}
\ea
and at least the integral over $k_3$ blows up. Thus, electric flux
operators are not well-defined in the standard Fock representation
since e.g. 
$||\pi_F^H(\hat{E}(Y^S)\Omega_F^H||^2=\omega_F^H(\hat{E}(Y^S)^2)=\infty$.

Having obtained $\omega_F^H$ we now obtain $\omega_P^H$ by the above 
prescription. In order to compare the resulting GNS Hilbert space
${\cal H}_P^H$ with the standard Hilbert space ${\cal H}_0$ for 
diffeomorphism invariant theories of $U(1)$ connections we recall
that a suitable orthonormal basis for ${\cal H}_0$ is given by the charge 
network states $T_s$ with $s=(\gamma(s),\vec{n}(s))$ where $\gamma$ is a 
closed piecewise analytic graphs and the integers $\vec{n}$ labelling its 
edges are subject to the gauge invariance constraint
\be \label{2.53}
\sum_{b(e)=v} n_e-\sum_{f(e)=v} n_e=0
\ee
for each vertex $v$ of $\gamma$. It is easy to see that 
\be \label{2.54}
T_s(A)=\prod_e A(e)^{n_e}=:A(s)=e^{i \int d^3x A_a X^a_s} \mbox{ where }
X^a_s=\sum_e n_e X^a_e
\ee
If we define 
\be \label{2.55}
X^a_{s_f}(x)=\int d^3y f(x-y) X^a_s(y)
\ee
then by the isomorphism
\be \label{2.56}
\omega_P^H(\hat{A}(s))=\omega_F^H(\hat{A}(s_f))=e^{-\frac{\alpha}{4}
\int d^3x X^a_{s_f}\sqrt{-\Delta}^{-1} X^b_{s_f} \delta_{ab}}
\ee
defines a positive linear functional on the $C^\ast$algebra generated 
by the holonomies and therefore the GNS Hilbert space is of the 
form $L_2(\ab,d\mu_f)$ where the measure is uniquely defined via the 
Riesz representation theorem (on $C^\ast$ algebras, positive linear
functionals are automatically continuous) and 
certainly depends on $f$. On the other hand, the Hilbert space 
${\cal H}_0$ is the GNS Hilbert space arising from the positive linear
functional on the $C^\ast$ algebra defined by 
$\omega_0(\hat{A}(s))=\delta_{s,(\emptyset,\vec{0})}$.

The measures $\mu_f,\mu_0$ on $\ab$ are mutually singular with 
respect to each other, see \cite{18} or appendix \ref{sa}.

Our next task will be to construct, next to the vacuum state 
$\Omega_F^H$, also coherent states $\Omega_{F,m}^H$ labelled by a point
$m\in {\cal M}$. Using the definition of being eigenstates of the 
annihilation operators $P\cdot \hat{z}_a$, we obtain the formal 
expression 
\be \label{2.59}  
\Omega_{F,m}^H=
e^{-\frac{1}{2}\int d^3x \bar{z}^a P\cdot z_a}\;
e^{\int d^3x z^a(m)P\cdot \hat{z}_a^\dagger}\Omega_F^H
\ee
where $z=z(m)$ as in (\ref{2.45}).
In order to compute the expectation value of the operator
$\hat{A}(s_f)$ for instance, we make use of the identity
\be \label{2.60}
\int d^3x X^a_{s_f} \hat{A}_a= 
\sqrt{\alpha/2}\int d^3x [(\root[4]\of{-\Delta})^{-1}X^a_{s_f}]
[\hat{z}_a+\hat{z}_a^\dagger]
\ee
and use the Baker-Campbell-Hausdorff (BCH) formula 
%
%
in order to normal order $\hat{A}(s_f)$.
\ba \label{2.61}
<\Omega_{F,m}^H,\hat{A}(s_f)\Omega_{F,m}^H>& = &
A(s_f)
e^{-\frac{1}{2}[i\sqrt{\alpha/2}\int d^3x 
((\root[4]\of{-\Delta})^{-1}X^a_{s_f})\hat{z}_a^\dagger,
i\sqrt{\alpha/2}\int d^3x 
((\root[4]\of{-\Delta})^{-1}X^a_{s_f}\hat{z}_a]}
\nonumber\\
&=& A(s_f)
e^{-\frac{\alpha}{4}[\sqrt{\alpha/2}\int d^3x 
X^a_{s_f}\sqrt{-\Delta}^{-1} X^a_{s_f}]}
\ea
so it is just the classical value times a fluctuation correction.
Similarly one proceeds with expectation values involving $\hat{E}^a$,
one just expresses them in terms of annihilation and creation operators.\\
\\
We now ask the crucial question: Given the states $\omega_{F,m}^H$
we can construct, again using the isomorphism, a state $\omega_{P,m}^H$,
and corresponding GNS data $(\Omega_{P,m},{\cal H}_{P,m}^H,\pi_{P,m}^H)$.
Is it possible to relate these GNS data to those of $\omega_0$, that is,
$(\Omega_0=1,{\cal H}_0=L_2(\ab,d\mu_0),\pi_0)$ ?

Since the states $T_s$ form an orthonormal basis of ${\cal H}_0$ we will
make an ansatz of the form 
\be \label{2.62}
\Omega_{P,m}^H=\sum_s c_{s,m} \overline{T_s}
\ee
where the coefficients $c_{s,m}$ are to be determined. Notice that
we do not expect (\ref{2.62}) to be normalizable as it stands with respect 
to the inner product on ${\cal H}_0$, just as it is the case for 
$\Omega^H_{F,m}$ with respect to $L_2(\a',[dA])$. In order
to arrive at the expecation value functional $\omega_{P,m}^H$ we must
divide by the (infinite) norm $||\Omega_{P,m}^H||_0^2$ just as we 
have to divide $[dA]$ by the norm squared of $||\Omega_{F,m}^H||^2$.
This is also the reason for additional
complex conjugation involved in (\ref{2.62}) as compared to \cite{17} 
which is due to the fact that we will not treat $\Omega_{P,m}^H$ as an 
element of a distributional extension of ${\cal H}_0$ but formally as a 
(non-normalizable) element of ${\cal H}_0$. It will become normalizable  
by dividing by its (infinite) norm later on, thereby inducing a new inner
product. This is similar in spirit to the group averaging proposal, see
\cite{6} and references therein. It is also similar in spirit to 
constructive QFT methods where one first regularizes a formally defined 
interacting measure by introducing IR and UV cut-offs, thereby obtaining
a well-defined measure which is absolutely continuous with respect to a 
free one, and then takes the limit to arrive at a well-defined 
interacting measure.

In order to determine the $c_{s,m}$ we will ask that $\Omega_{P,m}^H$
is an annihilation operator eigenstate, but the question is, what 
annihilation operator should be chosen ? The idea of \cite{17} 
is to use the basic
generators $\hat{A}(p_f),\hat{E}^a$ of ${\cal A}_P$, to relate them to
to the annihilation operators $\hat{z}_a$ and then to use the isomorphism
$\cal I$. Thus we write 
\be \label{2.63}
\int d^3x X^a_{s_f} A_a
=\int d^3x X^a_{s_f} 
[\sqrt{2\alpha}(\root[4]\of{-\Delta})^{-1} z_a+i\sqrt{-\Delta}^{-1} E^a]
\ee
so that, by purely algebraic manipulations, we obtain the operator 
identity
\ba \label{2.64}
\hat{A}(s_f) 
&=& e^{i\int d^3x  
[(\sqrt{2\alpha}(\root[4]\of{-\Delta})^{-1} X^a_{s_f}) \hat{z}_a
+i(\sqrt{-\Delta}^{-1}X^a_{s_f}) \hat{E}^a]}
\nonumber\\
&=& 
e^{-\frac{\alpha}{2}\int d^3x X^a_{p_f} \sqrt{-\Delta}^{-1} X^a_{s_f}}\;
e^{-\int d^3x (\sqrt{-\Delta}^{-1}X^a_{s_f}) \hat{E}^a}\;
e^{i \sqrt{2\alpha}\int d^3x  
(\root[4]\of{-\Delta})^{-1} X^a_{s_f}) \hat{z}_a}
\ea
where we have used again the BCH formula. Thus, our coherent state
$\Omega_{F,m}^H$ can be characterized by the requirement that
\be \label{2.64a}
\hat{A}(s_f)\Omega_{F,m}^H=
e^{-\frac{\alpha}{2}\int d^3x X^a_{s_f} \sqrt{-\Delta}^{-1} X^a_{s_f}}\;
e^{-\int d^3x (\sqrt{-\Delta}^{-1}X^a_{s_f}) \hat{E}^a}\;
e^{i \sqrt{2\alpha}\int d^3x  
(\root[4]\of{-\Delta})^{-1} X^a_{s_f}) z_a(m)}\Omega_{F,m}^H
\ee
Notice that a similar relation like (\ref{2.63}) in the language 
of the algebra ${\cal A}_P$ is not possible because the distributional
form factor $X_p$ is not square integrable. Also the operator valued
distribution $\hat{z}_a$, while well-defined on ${\cal H}_{F,m}^H$, has no 
analog on ${\cal H}_0$ on which we would like to define our annihilation 
operator. Thus, a definition of an annihilation operator along these lines 
fails. Therefore, as a substitute the authors of \cite{17,19} chose to 
translate (\ref{2.64}), guided by the isomorphism ${\cal I}$, as follows 
\be \label{2.65}
\hat{A}(s)\Omega_{P,m}^H=
e^{-\frac{\alpha}{2}\int d^3x X^a_{s_f} \sqrt{-\Delta}^{-1} X^a_{s_f}}\;
e^{-\int d^3x (\sqrt{-\Delta}^{-1}X^a_{s_f}) \hat{E}^a_f}\;
e^{i \sqrt{2\alpha}\int d^3x  
(\root[4]\of{-\Delta})^{-1} X^a_{s_f}) z_a(m)}\Omega_{P,m}^H
\ee
It is crucial to remark, however, that there is a certain amount of 
arbitrariness in this choice: For instance, should we not reexpress
$z(m)$ in terms of $A(p_f),E^a$ in (\ref{2.64}) and then substitute by
$(A(p),E_f^a)$ in (\ref{2.65}) ? Only for the vacuum state with $m=0$
there is no such ambiguity. We will come back to this question in the next 
section where we derive a precise relation between coherent states from
the general complexifier principle: It turns out that (\ref{2.65}) is 
not quite the correct classical correspondence because the expectation 
values for $\hat{A}(p),\hat{E}_f^a$ do not come out the right way. This is 
automatically fixed by the complexifier approach to coherent states.

Let us then solve (\ref{2.65}). It is a remarkable feature, {\it again 
true for Abelean gauge theories only}, 
that the $T_s$ are eigenfunctions of 
3D smeared electric field operators, specifically
\be \label{2.66}
\hat{E}_f^a(x) T_s=-\alpha X^a_{s_f} T_s
\ee
Now $\overline{T_s}=T_{-s}$ where $-s=(\gamma(s),-\vec{n}(s))$ 
and $T_s T_{s'}=T_{s+s'}$ where 
$s+s'=(\gamma(s)\cup\gamma(s'),\vec{n}(s)+\vec{n}(s'))$. The notation 
means that if $e$ is an edge of $\gamma(s)$
then there is a unique decomposition 
$e=\prod_{\tilde{e}}\tilde{e}^{\sigma(e,\tilde{e})}$ where 
$\sigma(e,\tilde{e})=1,-1,0$ if an edge $\tilde{e}$ of 
$\tilde{\gamma}=\gamma\cup \gamma'=\gamma(s+s')$ is overlapped by $e$ 
with equal (1) or (-1) opposite orientation or not overlapped at all (0)
(in the Abelean case ordering of that composition is irrelevant).
We then define $n_{\tilde{e}}(s)=\sum_e n_e\sigma(e,\tilde{e})$ 
and similar for the edges of $\gamma(s')$. Thus, (\ref{2.65}) 
becomes 
\ba \label{2.67}
&&\hat{A}(s^0)\Omega_{P,m}^H=
\sum_s c_{s,m} \overline{T_{s-s^0}}=\sum_s c_{s+s^0,m} \overline{T_s}
\\
&=& 
e^{-\frac{\alpha}{2}\int d^3x X^a_{s^0_f} \sqrt{-\Delta}^{-1} 
X^a_{s^0_f}}\;
e^{i \sqrt{2\alpha}\int d^3x  
(\root[4]\of{-\Delta})^{-1} X^a_{s^0_f}) z_a(m)}
\sum_s  c_{s,m} e^{-\alpha \int d^3x (\sqrt{-\Delta}^{-1}X^a_{s^0_f}) 
X^a_{s_f}}
\overline{T_s} \nonumber
\ea
Since the $\overline{T_s}$ provide a basis we conclude
\be \label{2.68}
c_{s+s^0,m}= c_{s,m}
e^{-\frac{\alpha}{2}\int d^3x X^a_{s^0_f} \sqrt{-\Delta}^{-1} 
X^a_{s^0_f}}\;\;\;
e^{i \sqrt{2\alpha}\int d^3x  
(\root[4]\of{-\Delta})^{-1} X^a_{s^0_f}) z_a(m)}
e^{-\alpha \int d^3x (\sqrt{-\Delta}^{-1}X^a_{s^0_f}) X^a_{s_f}}
\ee
Set $c_m=c_{(\emptyset,\vec{0}),m}$ then we find by setting $s=0$ in
(\ref{2.68}) that
\be \label{2.69}
c_{s^0,m}= c_m
e^{-\frac{\alpha}{2}\int d^3x X^a_{s^0_f} \sqrt{-\Delta}^{-1} 
X^a_{s^0_f}}\;\;\;
e^{i \sqrt{2\alpha}\int d^3x  
(\root[4]\of{-\Delta})^{-1} X^a_{s^0_f}) z_a(m)}
\ee
which now, upon reinsertion into (\ref{2.68}) leads to the following 
consistency requirement
\ba \label{2.70}
&& 
e^{-\frac{\alpha}{2}\int d^3x X^a_{s+s^0_f} \sqrt{-\Delta}^{-1} 
X^a_{s+s^0_f}}\;\;\;
e^{i \sqrt{2\alpha}\int d^3x  
(\root[4]\of{-\Delta})^{-1} X^a_{s+s^0_f}) z_a(m)}
\\
&=&
[e^{-\frac{\alpha}{2}\int d^3x X^a_{s_f} \sqrt{-\Delta}^{-1} 
X^a_{s_f}}\;\;\;
e^{i \sqrt{2\alpha}\int d^3x  
(\root[4]\of{-\Delta})^{-1} X^a_{s_f}) z_a(m)}]\times\nonumber\\
&& \times
[e^{-\frac{\alpha}{2}\int d^3x X^a_{s^0_f} \sqrt{-\Delta}^{-1} 
X^a_{s^0_f}}\;\;\;
e^{i \sqrt{2\alpha}\int d^3x  
(\root[4]\of{-\Delta})^{-1} X^a_{s^0_f}) z_a(m)}\;\;\;
 e^{-\alpha \int d^3x (\sqrt{-\Delta}^{-1}X^a_{s^0_f}) X^a_{s_f}}]
\nonumber
\ea
which is an identity since $X_{s_f}+X_{s^0_f}=X_{(s+s_0)_f}$. From the 
present point of view the remarkable identity (\ref{2.70}) looks like 
a miracle. We will see however in the next section that it follows 
trivially from the consistency of the complexifier operator on ${\cal 
H}_0$.

Concluding, we find the unique solution (up to a free function
$m\mapsto c_m$ which we set equal to unity since this is the value that
is automatically selected by the complexifier method) 
\be \label{2.71}
\Omega_{F,m}^H=\sum_s 
e^{-\frac{\alpha}{2}\int d^3x X^a_{s_f} \sqrt{-\Delta}^{-1} 
X^a_{s_f}}\;
e^{i \sqrt{2\alpha}\int d^3x  
(\root[4]\of{-\Delta})^{-1} X^a_{s_f}) z_a(m)} \overline{T_s}
\ee

In \cite{19} the authors made the following observation: Define
for edges $e,e'$ of a graph $\gamma$
\be \label{2.72}
G^{e e'}_f:=\int d^3x X^a_{e_f} \sqrt{-\Delta}^{-1} X^a_{e'_f}
\mbox{ and }
z^e_f(m):=\int d^3x  (\root[4]\of{-\Delta})^{-1} X^a_{e_f}) z_a(m)
\ee
then (\ref{2.71}) can be rewritten as 
\be \label{2.73}
\Omega_{F,m}^H=\sum_s 
e^{-\frac{\alpha}{2}\sum_{e,e'} G^{ee'}_f n_e n_{e'}}
e^{i \sqrt{2\alpha}\sum_e z^e_f(m) n_e} \overline{T_s}
\ee
where the sums over edges at given $s$ are over those of $\gamma(s)$.
If we denote by $Y_e$ the right invariant vector field on $U(1)$ acting
on the degree of freedom $A(e)$ then we may rewrite (\ref{2.73}) further 
in the form 
\be \label{2.74}
\Omega_{F,m}^H=\sum_s 
e^{\frac{\alpha}{2}\sum_{e,e'} G^{ee'}_f Y_e Y_{e'}}
e^{-\sqrt{2\alpha}\sum_e z^e_f(m) Y_e} \overline{T_s}
\ee
This formula suggests an immediate generalization to $SU(2)$ by 
replacing the charge network states $T_s$ by spin network states,
replacing $Y_e Y_{e'}$ by $Y^j_e Y^j_{e'}$ and $z^e(m) Y_e$ by
$z^e_{j,f}(m) Y^j_e$ where $Y^j_e$ are right invariant vector fields on 
$SU(2)$. Thus, they propose
\be \label{2.75}
\Omega_{F,m}^H=\sum_s 
e^{\frac{\alpha}{2}\sum_{e,e'} G^{ee'}_f Y^j_e Y^j_{e'}}
e^{-\sqrt{2\alpha}\sum_e z^e_{j,f}(m) Y^f_e} \overline{T_s}
\ee
However, several remarks are in order:\\
1)\\
Since the ``edge metric" $G^{ee'}_f$ is not diagonal, the operator
$Y^j_e Y^j_{e'}$ is not gauge invariant for $e\not=e'$. In \cite{19} the 
authors propose to repair this by averaging the states over the $SU(2)$ 
gauge group. Another, maybe more direct method is to work instead with 
operators of the form 
\be \label{2.76}
\mbox{Tr}([\mbox{Ad}_{\hat{A}(\rho_{b(e)})} Y_e]
\;[\mbox{Ad}_{\hat{A}(\rho_{b(e')})} Y_{e'}])
\ee
where $x\mapsto \rho_x$ defines a system of paths between the point $x$
and an arbitrary but fixed point $x_0$. Again, it is {\it only for
Abelean groups} not necessary to do this. However, while the operator
in the exponent of (\ref{2.75})
is negative definite when restricted to functions over the graph in question,
\be \label{2.77}
\hat{G}^f_\gamma:=\sum_{e,e'} G^{e e'}_f Y^j_e Y^j_{e'}
=-\int d^3x [\sum_e X^a_{e_f} Y^j_e]^\dagger(x)
(\sqrt{-\Delta}^{-1} [\sum_e X^a_{e_f} Y^j_e])(x)
\ee
the path holonomies in (\ref{2.76}) spoil positivity. They also
spoil the convenient fact that (\ref{2.77}) leaves each of the Hilbert
spaces ${\cal H}_{0,\gamma}$ separately invariant. Thus, not only the 
spectral analysis of (\ref{2.76}) is unclear, also if we are thinking of 
cut-off 
states (shadows), they are no longer granted to be normalizable.\\
2.)\\
The operators $\hat{G}^f_\gamma$ do not form a consistent system of 
operators and therefore no inductive limit operator $\hat{G}^f$ exists.
This is closely related to the fact that, {\it unless the gauge group is 
Abelean}, there is no continuum function, i.e. a complexifier, which 
gives rise to the states as will be demonstrated in the next section.\\
3.)\\
Since there is no classical complexifier function available, it is
unclear what the function $z^e_{j,m}$ should be where 
$m=(A_a^j,E^a_j)$. Abstracting from (\ref{2.72}) one might guess, e.g. in 
the case of general relativity the formula
\ba \label{2.78}
z^e_{j,f}(m) &:=& \int d^3x  (\root[4]\of{-\Delta})^{-1} X^a_{e_f}) 
z^j_a(m)
\nonumber\\
z^j_a(m)&=&\frac{1}{\sqrt{2\alpha}}[\root[4]\of{-\Delta}A_a^j-\frac{i}{a}
(\root[4]\of{-\Delta})^{-1} E^a_j]
\ea
where $a$ is a is a constant with unit cm. Also, the Feinstruktur constant
$\alpha$ must be exchanged against some dimensionless constant of the 
form $[\ell_p/a]^n$ for some power of $n$. 
Notice that $P\cdot z_a^j$ is not gauge 
covariant, moreover, since there is no physically interesting non-Abelean 
Hamiltonian which is bilinear in $z_a^j,\bar{z}_a^j$ there is no good 
motivation for the proposal (\ref{2.78}).\\
4.)\\
Since there is no underlying complexifier, we have 
no guarantee that the expectation values of the operators $\hat{A}(e),
\hat{E}^a_{f,j}$ have any close relation with their classical values at 
the phase space point $m$. What is more, the operator $\hat{E}^a_{j,f}$
fails to be densely defined in the representation ${\cal H}_0$ so 
that it is impossible to define it (or its dual) on the state
$\Omega_{P,m}^H$. Therefore, the commutator of these operators is 
ill-defined so that the expectation value of the commutators 
cannot have anything to do with the classical Poisson brackets. 
From the point of view of the isomorphism ${\cal I}$ this is a consequence 
of the fact that for non-Abelean gauge theories there is no 
gauge invariant state
$\omega_{F,m}^H$ from which we could derive $\Omega_{P,m}^H$ by similar
methods, moreover, there is no isomorphism of Poisson algebras because 
smeared holonomies do not transform gauge covariantly any longer and 
the Poisson algebra between 3D smeared electric fields and unsmeared
holonomies no longer closes.\\
5.\\
It is unclear what the definition of a suitable annihilation operator
should be.\\
\\
We conclude this section by showing that in the Abelean case 
the $\hat{G}_\gamma$ actually form a consistent family. 
Responsible for this are:\\
1) The smeared form factors satisfy the following identities
\be \label{2.79}
X_{(e^{-1})_f}=-X_{e_f},\;X_{(e\circ e')_f}=X_{e_f}+X_{e'_f}
\ee
which implies corresponding identities for $z^e_f(m)$.\\
2) The vector fields $Y_e$ form an Abelean algebra {\it only in the 
Ablean case} and satisfy 
\be \label{2.80}
Y_{e^{-1}}=-Y_e,\;(Y_{e})_{|e\circ e'}=(Y_{e'})_{|e\circ e'}=Y_{e\circ e'}
\ee
where the notation in the latter identity means that the vector fields are 
restricted to functions of $A(e\circ e')$.\\
3) We have 
\ba \label{2.81}
\hat{G}^f_\gamma&:=&\sum_{e,e'} G^{e e'}_f Y_e Y_{e'}
=\int d^3x [\sum_e X^a_{e_f} Y_e](x)
(\sqrt{-\Delta}^{-1} [\sum_e X^a_{e_f} Y_e])(x)
\nonumber\\
&=:&\int d^3x Y^f_\gamma(x)(\sqrt{-\Delta}^{-1} Y^f_\gamma)(x)
\ea
Now consistency means that $\hat{U}_{\gamma'\gamma}\hat{G}_{\gamma'}
=\hat{G}_\gamma\hat{U}_{\gamma'\gamma}$ for any $\gamma'\subset\gamma$
where we can confine ourselves to the case that for the sets of 
oriented edges of a graph holds a) $E(\gamma')=E(\gamma)-\{e\}$
does not contain some edge $e$, b) 
$E(\gamma')=(E(\gamma)-\{e\})\cup\{e^{-1}\}$ contains 
some edge inverted or c) 
$E(\gamma')=(E(\gamma)-\{e_1,e_2\})\cup\{e_1\circ e_2\}$ contains 
two edges only as their composition. It is straightforward to check,
using properties 1),2),3), that consistency already holds for the 
vector fields $Y^f_\gamma$ so that consistency for $\hat{G}_\gamma$ 
follows
easily from the Abelean nature of the $Y_e$. For the same reason also
the vector fields
\be \label{2.82}
Z^f_\gamma=\sum_e z^e_f Y_e
\ee
form a consistent family. All of this will be spoiled in the non-Abelean
context where (\ref{2.80}) is replaced by 
\be \label{2.83}
Y^j_{e^{-1}}=-O_{jk}(A(e))Y^k_e,\;(Y^j_{e})_{|e\circ e'}=Y^j_{e\circ e'},\;
(Y^j_{e'})_{|e\circ e'}=O_{jk}(A(e))Y^k_{e\circ e'}
\ee
where $O_{jk}(h)\tau_k=\mbox{Ad}_h(\tau_j)$ is an orthogonal matrix and 
$\tau_j$ is a basis of $Lie(G)$, unless $G^{e e'}_f$ is diagonal as in
\cite{10,11} which of course is not the case here.\\
\\
Finally, notice the similarity between (\ref{2.73}) which we write 
in the more suggestive form
\be \label{2.84}
\Omega_{F,m}^H=\sum_s e^{-\lambda_s/\hbar} 
T_{s_f}(A^\Cl)  \overline{T_s}
\ee
and the complexifier coherent states of the previous subsection.
Here $\lambda_s=\frac{\alpha\hbar}{2}\sum_{e,e'} G^{ee'}_f n_e n_{e'}$
and $A^\Cl=A-i\sqrt{-\Delta}^{-1} E$. In the next
section we will show the complexifier theoretic origin for all 
of these facts, the only thing that is disturbing is the 
appearance of smeared holonomies in (\ref{2.84}) which suggests
that $\Omega_{F,m}^H$ is peaked at $m_f=(A_f,E_f)$ rather than $m$ as we 
would expect. In fact, this is due to the fact that the translation of the 
annihilation condition (\ref{2.64a}) into (\ref{2.65}) is ambiguous 
without any additional structural principle which will be automatically 
provided by the complexifier construction.

\section{Complexifier Theoretic Origin of Varadarajan States}
\label{s3}

In order to motivate the complexifier underlying the Varadarajan states
we begin again with the usual Fock states $\Omega^H_{F,m}$ but this time 
we will not derive them by asking to be eigenstates of an 
annihilation operator, rather we would like to derive them from a 
complexifier. The corresponding, exponentiated, annihilator will be 
automatically produced as a byproduct. This line of derivation makes
the similarity between the Fock representation and the polymer 
representation even more transparent.

We claim that the usual Maxwell coherent states follow from the 
complexifier
\be \label{3.1}
C^H_F=\frac{1}{2q^2} \int d^3x E^a(x)(\sqrt{-\Delta}^{-1} E^a)(x)    
\ee
In order to verify this, it is sufficient to check that 
\be \label{3.2}
\sum \frac{i^n}{n!} \{A_a,C^H_F\}_{(n)}=A_a-i\sqrt{-\Delta}^{-1} E^a    
=\sqrt{2\alpha}\root[4]\of{-\Delta} z_a=A_a^\Cl(m)
\ee
is indeed the complexified connection that is induced by the Maxwell
Hamiltonian. It follows from the rules of canonical quantization that
\be \label{3.3}
\hat{C}^H_F/\hbar=-\frac{\alpha}{2} \int d^3x 
\frac{\delta}{\delta A_a(x)}
(\sqrt{-\Delta}^{-1} \frac{\delta}{\delta A_a})(x)    
\ee
which is formally self-adjoint and positive on the formal Hilbert 
space ${\cal H}^H_F=L_2(\ag',[dA])$.
Thus we can compute the complexifier coherent states once we have a 
suitable $\delta$-distribution with respect to the Lebesgue measure
$[dA]$ at our disposal. A suitable representation can be formally
given as the formal functional integral over transversal momenta $k^a$
\be \label{3.4}
\delta^F_{A'}(A)=
\int [\frac{d^2k}{(2\pi)^2}] e^{i\int d^3x k^a(x)[A'_a(x)-A_a(x)]}
\ee
Notice the formal similarity between (\ref{3.4}) and the 
$\delta$-distribution on ${\cal H}_0=L_2(\agb,d\mu_0)$ defined in section
\ref{s2.3} and which specializes for $U(1)$ to
\be \label{3.5}
\delta^P_{A'}(A)=\sum_s T_s(A')\overline{T_s(A})
\ee
where the sum is over gauge invariant charge networks.
Thus we get correspondences between objects on the Fock side 
on the one hand and objects on the Polymer side on the other,
summarized in the following table 
\be \label{3.6}
\left. \begin{array}{ccc}
\mbox{Lebesgue measure }[dA] & \leftrightarrow & \mbox{Uniform measure }
d\mu_0 \\ 
\mbox{Tempered distributions }\ag' & \leftrightarrow & 
\mbox{Ashtekar-Isham spectrum } \agb \\
\mbox{Functional integral } \int [\frac{d^2k}{(2\pi)^2}] 
& \leftrightarrow & 
\mbox{Discrete sum } \sum_s \\
\mbox{3D smeared ``holonomy" } e^{i\int d^3x k^a(x) A_a(x)} 
& \leftrightarrow & 
\mbox{holonomy } A(s)=T_s(A)=e^{i\int d^3x X^a_s A_a} 
\end{array} \right.
\ee
Interestingly, while the objects on the Fock side are only formal, the 
polymer objects are {\it rigorously defined} !

Our complexifier coherent states on the Fock side become 
\ba \label{3.7}
\Omega_{F,m}^H(A)&=& 
[(e^{-\hat{C}^H_F/\hbar}\delta_{A'})(A)]_{A'\to A^\Cl(m)}
\nonumber\\
&=&
(\int [\frac{d^2k}{(2\pi)^2}] 
e^{-\frac{\alpha}{2} \int d^3x k^a \sqrt{-\Delta} k_a}\; e^{i\int d^3x 
k^a[A'_a-A_a]})_{A'\to A^\Cl(m)}
\nonumber\\
&=& {\cal N}(A^\Cl)
e^{-\frac{1}{2\alpha}\int d^3x 
[A_a^\Cl-A_a]P\cdot \sqrt{-\Delta}[A_a^\Cl-A_a]}
\ea
where ${\cal N}(A^\Cl)$ is the usual infinite constant coming from the 
functional determinant in a Gaussian integral. It is easy to check that
(\ref{3.7}) divided by its (infinite) norm coincides with our earlier 
definition up to a finite phase, similar to the case of a harmonic 
oscillator. 

An interesting object is the exponentiated annihilation operator 
corresponding to a smeared holonomy
\ba \label{3.8}
\hat{g}_{p_f} &:=& 
e^{-\hat{C}^H_F/\hbar}\hat{A}(p_f)e^{\hat{C}^H_F/\hbar}
=e^{e^{-\hat{C}^H_F/\hbar}\; i\int d^3x X^a_{p_f} \hat{A}_a\; 
e^{\hat{C}^H_F/\hbar}}
\nonumber\\
&=&
e^{i\sum \frac{1}{n!}\int d^3x X^a_{p_f} [\hat{A}_a,\hat{C}^H_F/\hbar]_{(n)}}
=e^{i\int d^3x X^a_{p_f} [\hat{A}_a-i\sqrt{-\Delta}^{-1}\hat{E}^a]}
\nonumber\\
&=&
\hat{A}(p_f) e^{\int d^3x (\sqrt{-\Delta}^{-1} X^a_{p_f})\hat{E}^a]}
e^{-\frac{\alpha}{2}\int d^3x X^a_{p_f} \sqrt{-\Delta}^{-1} X^a_{p_f}}
\nonumber\\
&=&
e^{\int d^3x (\sqrt{-\Delta}^{-1} X^a_{p_f})\hat{E}^a]} \hat{A}(p_f)
e^{\frac{\alpha}{2}\int d^3x X^a_{p_f} \sqrt{-\Delta}^{-1} X^a_{p_f}}
\ea
which can be easily checked to have eigenvalue 
$e^{i\int d^3x X^a_{pf} A_a^\Cl(m)}$ on $\Omega_{F,m}^H$.

We now turn to the polymer representation. It is easy to see that the 
operator $\hat{C}^H_F$ is not well-defined on charge network states,
specifically, $\hat{C}^H_F/\hbar T_s=[\frac{\alpha}{2}\int d^3x
X^a_s \sqrt{-\Delta}^{-1} X_s] T_s$ which blows up due to the too singular 
behaviour of the form factors. The isomorphism $\cal I$ suggests to try 
the smeared {\bf Varadarajan complexifier}
\be \label{3.9} 
C^H_P=\frac{1}{2q^2} \int d^3x E^a_f (\sqrt{-\Delta}^{-1} E^a_f)
\ee
which corresponds to a slightly different complexified connection
\be \label{3.10}
A_{a,f}^\Cl(m)=A_a-i\int d^3x f(\sqrt{-\Delta}^{-1} E^a_f)
\ee
in which the electric field appears automatically {\it doubly smeared}.
The operator corresponding to (\ref{3.9}) is in fact a densely defined,
positive, essentially self-adjoint operator which is diagonalized by the 
charge network states, that is, its spectrum is pure point ! Specifically
\be \label{3.11}
\hat{C}^H_P/\hbar T_s=[\frac{\alpha}{2}\int d^3x X_{s_f}^a 
(\sqrt{-\Delta}^{-1} X^a_{s_f})]T_s=:\lambda_s/\hbar T_s
\ee
The reader will immediately recognize the eigenvalue $\lambda_s$ 
{\it precisely} as the eigenvalue that appeared in Varadarajan's 
derivation of the coherent states $\Omega_{P,m}^H$. 
Our complexifier method now produces the coherent states
\ba \label{3.12}
\Omega_{P,m}^H(A)
&:=& [(e^{-\hat{C}^H_P/\hbar}\delta_{A'})(A)]_{A'\to A^\Cl_f(m)}
\nonumber\\
&=& \sum_s e^{-\lambda_s/\hbar} T_s(A^\Cl_f(m)) \overline{T_s(A)}
\ea
Our (exponentiated) annihilation operators are defined in the charge 
network basis as 
(notice that $\hat{E}^a_f T_s=-\alpha X^a_{s_f} T_s$)
\ba \label{3.13}
\hat{g}_{s^0} T_s&:=& 
e^{-\hat{C}^H_P/\hbar}\hat{A}(s^0)e^{\hat{C}^H_P/\hbar} T_s
=e^{-\lambda_{s+s^0}/\hbar} e^{\lambda_s/\hbar} T_{s+s^0}
\nonumber\\
&=& e^{-\lambda_{s^0}/\hbar-\alpha\int d^3x X^a_{s_f} 
\sqrt{-\Delta}^{-1} X^a_{s^0_f}} T_{s+s^0}
\nonumber\\
&=& 
e^{-\frac{\alpha}{2}\int d^3x X^a_{s^0_f}(\sqrt{-\Delta})^{-1} 
X^a_{s^0_f})}\;\hat{A}(s^0)\;
e^{\int d^3x (\sqrt{-\Delta}^{-1} X^a_{s^0_f})\hat{E}^a_f}\; T_s
\nonumber\\
&=& 
e^{\frac{\alpha}{2}\int d^3x X^a_{s^0_f}(\sqrt{-\Delta})^{-1} 
X^a_{s^0_f})}\;
e^{\int d^3x (\sqrt{-\Delta}^{-1} X^a_{s^0_f})\hat{E}^a_f} 
\;\hat{A}(s^0)\; T_s
\ea
from which 
\ba \label{3.14}
\hat{g}_{s^0} &=& 
e^{-\frac{\alpha}{2}\int d^3x X^a_{s^0_f}(\sqrt{-\Delta})^{-1} 
X^a_{s^0_f})}\;\hat{A}(s^0)\;
e^{\int d^3x (\sqrt{-\Delta}^{-1} X^a_{s^0_f})\hat{E}^a_f}
\nonumber\\
&=& 
e^{\frac{\alpha}{2}\int d^3x X^a_{s^0_f}(\sqrt{-\Delta})^{-1} 
X^a_{s^0_f})}\;
e^{\int d^3x (\sqrt{-\Delta}^{-1} X^a_{s^0_f})\hat{E}^a_f} 
\;\hat{A}(s^0)
\ea
which should be compared with formula (\ref{3.8}). It is easy to check
that the $\Omega_{P,m}^H$ are eigenstates for the $\hat{g}_s$
with eigenvalue $T_s(A^\Cl_f(m))$ which by comparison with (\ref{3.14})
is the correct classical value corresponding to that operator. This is an
important difference with the states (\ref{2.71}) for which the operators
$\hat{g}_s$ have eigenvalue $T_{s_f}(A^\Cl(m)$ which is the correct
eigenvalue for the operators (\ref{3.8}) but not for (\ref{3.14}).
Thus we see that the complexifier method, being based on an underlying
canonical transformation that complexifies a configuration space,
is automatically produces coherent states with the correct classical 
correspondence. Likewise one can explicitly check that the commutator
among the $\hat{g}_s,\hat{g}_s^\dagger$ precisely corresponds to the 
Poisson brackets between the $g_s,\bar{g}_s$.

Of course, the states (\ref{3.12}), even in their cut-off form,
have not been demonstrated to have all the desired semiclassical
properties. In their cut-off form, the methods of \cite{10,11}
could be of help provided the function $f$ is stronly peaked at the origin 
of the coordinate system, so that the off-diagonal elements of the 
edge metric are suppressed. For instance, following \cite{34}, Varadarajan 
proposed 
\be \label{3.15}
f(x)=\frac{e^{-\frac{||x||^2}{2r^2}}}{(\sqrt{2\pi}r)^3}
\ee
which becomes a $\delta-$distribution in the limit $r\to 0$. If 
we parameterize edges as the image of the closed interval in $[0,1]$
then it is easy to see that the edge metric components become
\be \label{3.16}
G^{ee'}_f=\frac{1}{2\pi^2 r} \int_0^1 dt\int_0^1 dt' 
\frac{<\dot{e}(t),\dot{e}'(t')>}{||e(t)-e'(t')||}
\int_0^\infty dk\; e^{-k^2}\sin(k\frac{||e(t)-e'(t')||}{r})
\ee
where $<.,.>$ is the Euclidean inner product on $\Rl^3$.
Let $y\not=0$, then using the Taylor expansion of the sine function 
we find    
$$
\int_0^\infty dk e^{-k^2}\frac{\sin(ky)}{y}=\frac{1}{2}
\sum_{n=0}^\infty \frac{(-y^2)^n\;n!}{(2n+1)!}\approx e^{-y^2/4}
$$
where we have used Stirling's approximation for the factorials. Thus 
\be \label{3.17}
G^{ee'}_f\approx\frac{1}{(2\pi r)^2} \int_0^1 dt\int_0^1 dt' 
<\dot{e}(t),\dot{e}'(t')> e^{-\frac{||e(t)-e'(t')||^2}{4r^2}}
\ee
As an example, let us compute (\ref{3.17}) for $e(t)=u\epsilon t,;\
e'(t)=N\epsilon w+v\epsilon t$ where $u,v,w$ are unit vectors, $\epsilon$ 
is the 
typical edge length of a given graph and $N$ denotes the distance between
the edges in lattice units. Then
\be \label{3.18}
G^{ee'}_f\approx\frac{\epsilon^2 <u,v>}{(2\pi r)^2} \int_0^1 dt\int_0^1 dt' 
e^{-\epsilon^2\frac{||Nw+vt'-ut||^2}{4r^2}} \mbox{ and }
G^{ee}_f\approx\frac{\epsilon^2}{(2\pi r)^2} \int_0^1 dt\int_0^1 dt' 
e^{-\epsilon^2\frac{(t-t')^2}{4r^2}} 
\ee
So the off-diagonal elements are suppressed by an order of 
$e^{-(\frac{N\epsilon}{2r})^2}$ which is small for $N\epsilon\gg 2r$.
Thus, since $r$ is fixed, for sufficiently fine graphs an order of 
$N\approx r/\epsilon$ nearest neighbour elements cannot be neglected
which becomes arbitrarily large as $\epsilon \to 0$ so that one will tie
$\epsilon\approx r$ in order to control this.
Concluding, the Poisson transformation that was the important tool in 
\cite{10,11} to arrive at peakedness estimates and Ehrenfest
theorems can still be applied but due to the non-diagonal nature 
of $G^{ee'}_f$ and since for Maxwell theory $\alpha\approx 1/137$ is not
extremely small, the estimates will become worse.

Let us now compute the fluctuation of the basic operators 
$\hat{A}(p),\hat{E}(S)$, on which the polymer Hilbert space ${\cal H}_0$
is based, in the state $\Omega_{P,m}^H$ (in the cut-off version these are
fine of course since the states are normalizable). In order to compute 
those the strategy will be to try to express them in terms of the 
operators $\hat{g}_e,\hat{g}_e^\dagger$. Using (\ref{3.14})
we find 
\ba \label{3.19}
\hat{g}_p^\dagger \hat{g}_p &=&
e^{-\alpha\int d^3x X^a_{p_f}(\sqrt{-\Delta})^{-1} X^a_{p_f})}
e^{2\int d^3x (\sqrt{-\Delta}^{-1} X^a_{p_f})\hat{E}^a_f}
\nonumber\\
\hat{A}(p)& = & 
e^{-\frac{\alpha}{2}\int d^3x X^a_{p_f}(\sqrt{-\Delta})^{-1} 
X^a_{p_f})}\;[\sqrt{\hat{g}_{p^{-1}}^\dagger\hat{g}_{p^{-1}}}]\;\hat{g_p}
\ea
The first line represents a function of $\hat{E}$ only from which we 
would like to
extract $\hat{E}(S)$ through a limiting procedure using small loops 
$p$.

{\it Notice that all
calculations will be performed without recourse to the Fock state
$\omega_{F,m}^H$, all calculations are just using techniques from the Hilbert
space ${\cal H}_0$} ! This is an important difference to \cite{17,19} 
because it shows that {\it we do not need to first define a completely 
new representation} (the GNS representation derived from $\omega_{P,m}^H$)
which reconfirms {\it ${\cal H}_0$ as the fundamental representation}
(at the kinematical level) from which everything else is to be derived.

We begin with the expectation value of $\hat{A}(p)$. In order to get rid 
of the square root in (\ref{3.19}) we notice the basic estimate, valid
for any $y\ge 0$,
\be \label{3.20}
\frac{1}{2}[1+y-(y-1)^2]\le\sqrt{y}\le \frac{1}{2}[y+1]
\ee
which could be sharpened further by taking more powers of $y$ into account,
however, (\ref{3.20}) will be sufficient for our purposes. The upper and 
lower bound in (\ref{3.20}) coincide the better with the actual value 
the smaller $|y-1|$. We will exploit this fact in the subsequent estimate.
Let us define the normal ordered, positive operator (recall 
$\hat{g}_p^{-1}=\hat{g}_{p^{-1}}$) 
\be \label{3.21}
\hat{y}_{p^{-1}}:(m)=|g_p(m)|^2\hat{g}_{p^{-1}}^\dagger \hat{g}_{p^{-1}}
\ee
whose expectation value in the state $\Omega_{P,m}^H$ equals unity. Using 
the commutation relations it is easy to verify that
\be \label{3.22}
\hat{y}_{p^{-1}}(m)^2=|g_p(m)|^4(\hat{g}_{p^{-1}}^\dagger)^2 
(\hat{g}_{p^{-1}})^2
e^{2\alpha\int d^3x X^a_{p_f} \sqrt{-\Delta}^{-1} X^a_{p_f}}
\ee
Writing the holonomy operator in the form 
\be \label{3.23}
\hat{A}(p)=e^{-\frac{\alpha}{2}\int d^3x X^a_{p_f}(\sqrt{-\Delta})^{-1} 
X^a_{p_f})}|g_p(m)|^{-1} \sqrt{\hat{y}_{p^{-1}}}\hat{g_p}
\ee
we find for its expectation value
\be \label{3.24}
\frac{<\Omega_{P,m}^H,\hat{A}(p)\Omega_{P,m}^H>}{||\Omega_{P,m}^H||^2}
=e^{-\frac{\alpha}{2}\int d^3x X^a_{p_f}(\sqrt{-\Delta})^{-1} X^a_{p_f}} 
[A(m)](p)
\frac{<\Omega_{P,m}^H,\sqrt{\hat{y}}\Omega_{P,m}^H>}{||\Omega_{P,m}^H||^2}
\ee
Thus, putting everything together
\be \label{3.25}
1-\frac{e^{2\alpha\int d^3x X^a_{p_f}(\sqrt{-\Delta})^{-1}X^a_{p_f}}-1}{2}
\le 
e^{\frac{\alpha}{2}\int d^3x X^a_{p_f}(\sqrt{-\Delta})^{-1}X^a_{p_f}}
\frac{<\hat{A}(p)>_m}{[A(m)](p)}\le 1
\ee
Thus, for sufficiently short paths $p$ the expectation value is rather close
to its classical value and the quantum corrections are a power series in 
$\alpha$. Notice that the trick (\ref{3.20}) can also be applied to other
important operators in quantum general relativity which involve 
square roots, such as area and volume operators. Fluctuations of the 
holonomy operator for loops $p$ can be reduced to expectation values 
since $\hat{A}(p)^2=\hat{A}(p^2)$ so formula (\ref{3.25}) can be applied.
We will leave this task to the interested reader.

Next we turn to the expectation value of $\hat{E}(S)$. From 
(\ref{3.19}) we find the explicit formula
\ba \label{3.26}
&& \int d^3x (\sqrt{-\Delta}^{-1} X^a_{p_f})\hat{E}^a_f
=\frac{\alpha}{2}\int d^3x X^a_{p_f}(\sqrt{-\Delta})^{-1} X^a_{p_f}
+\ln(|g_p(m)|)+\frac{1}{2}\ln(\hat{y}_p)
\nonumber\\
&=& -\frac{\alpha}{2}\int d^3x X^a_{p_f}(\sqrt{-\Delta})^{-1} X^a_{p_f}
+\ln(|g_p(m)|)-\frac{1}{2}\ln(\hat{y}_{p^{-1}})
\ea
Using the elementary estimate $1-y^{-1}\le\ln(y)\le y-1$, valid for all 
$y>0$, we find 
\be \label{3.27}
|\int d^3x (\sqrt{-\Delta}^{-1} X^a_{p_f})<\hat{E}^a_f>_m-\ln(|g_p(m)|)|\le
\frac{\alpha}{2}\int d^3x X^a_{p_f}(\sqrt{-\Delta})^{-1} X^a_{p_f}
\ee
In order to deduce from this estimate information about $\hat{E}_f$ and 
then the electric flux itself we must remove the smeared form 
factors. To do this, let $S_p$ be any surface with $\partial S_p=p$
then upon performing a couple of integrations by parts we obtain
\be \label{3.28}
E(p_f):=\int d^3x (\sqrt{-\Delta}^{-1} X^a_{p_f}) E^a_f
=\int_{S_p} dS_a(x) ((\sqrt{-\Delta}^{-1}(\nabla\times E)_f)_f)^a(x)
\ee
where the subscript denotes smearing with $f$ as before.  
Consider now the special circular loops $p^a_\epsilon(x)$ with radius 
$\epsilon$ in the $x^a=const.$ plane with center $x$ and let
$S^a_\epsilon(x)$ be the surface in the $x^a=const.$ plane bounded by it.
Then 
\be \label{3.29}
\lim_{\epsilon\to 0} \frac{E([p^a_\epsilon(x)]_f)}{\pi\epsilon^2}=
((\sqrt{-\Delta}^{-1}(\nabla\times E)_f)_f)^a(x)
\ee
In order to extract $E$ itself from this formula, we must specify 
the smearing function $f$. Let us assume, as in \cite{17} that 
$f(x,y)=e^{-||x-y||^2/(2r^2)}/(\sqrt{2\pi}r)^3$. Then in fact $f$ is
the kernel of the smoothening operator $e^{\Delta r^2}$ which has 
an inverse precisely on functions of the form $e^{\Delta r^2} F$
where $F$ is smooth, say. Thus 
\be \label{3.30}
e^{-\Delta_x r^2} \sqrt{-\Delta_x} e^{-\Delta_x r^2} \lim_{\epsilon\to 0} 
\frac{E([p^a_\epsilon(x)]_f)}{\pi\epsilon^2}=(\nabla \times E)(x)
\ee
Finally, observing the identity $P=\Delta^{-1}\nabla\times\nabla\times$
for the transversal projection operator we obtain
\be \label{3.31}
\Delta_x^{-1} \epsilon^{abc} \partial_{x^b}
e^{-\Delta_x r^2} \sqrt{-\Delta_x} e^{-\Delta_x r^2} \lim_{\epsilon\to 0} 
\frac{E([p^c_\epsilon(x)]_f)}{\pi\epsilon^2}= (P\cdot E)^a(x)
\ee
The fact that only gauge invariant holonomies $A^\Cl_f(p)$ are allowed
implies that we can only extract the tranversal piece of the electric 
field which is precisely what we are interested in. 

We can now study the expectation value of $(P\cdot\hat{E})^a_f(x)$
which from the intuition of the isomorphism ${\cal I}$ should be 
a well-defined operator valued distribution, or in other words,
for any tranversal smearing field $l_a$ of dimension cm$^{-1}$ the doubly 
smeared 
object $\hat{E}_f(l)=\int d^3x l_a \hat{E}^a_f$
should be a well-defined operator with finite fluctuations in the state
$\Omega^H_{P,m}$. Already at this point it is intuitively clear that
the electric flux operator can be at most an operator valued distribution
whose fluctuations therefore should diverge in the coherent states. Let us 
verify these statements explicitly: We have 
\be \label{3.32}
\hat{E}_f(l)=\int d^3x (\Delta^{-1} \nabla\times l)_a(x)   
\sqrt{-\Delta_x} e^{-\Delta_x r^2} \lim_{\epsilon\to 0} 
\frac{\hat{E}([p^a_\epsilon(x)]_f)}{\pi\epsilon^2}
\ee
From (\ref{3.27}) we find  
\ba \label{3.33}
0&\le& |\int d^3y (\sqrt{-\Delta}^{-1} 
X^b_{[p^a_\epsilon(x)]_f})(y)<\hat{E}^b_f(y)>_m-\ln(|g_{p^a_\epsilon(x)}(m)|)|
\nonumber\\
&\le&
\frac{\alpha}{2}\int d^3y X^b_{[p^a_\epsilon(x)_f}(y)(\sqrt{-\Delta})^{-1} 
X^b_{[p^a_\epsilon(x)]_f}(y)
\nonumber\\
&=& (\pi\epsilon^2)^2\sum_{b\not= a} \int d^3y 
f_{,b}(u-x)(\sqrt{-\Delta}^{-1} f_{,b})(u-x)+O(\epsilon^5)
\ea
Thus 
\be \label{3.34}
<\hat{E}_f(l)>_m=[E_f(l)](m)
\ee
has no quantum corrections as expected since in the Fock representation
$\hat{E}(l)$ is already normal ordered. In order to compute the 
fluctuation of $\hat{E}_f(l)$ we could again use formula 
(\ref{3.26})
and use estimates for $\ln(y)^2$. It is, however, technically more 
instructive to use a different, more direct method. We observe that 
\be \label{3.35}
\lim_{\epsilon\to 0} \frac{|g_{p^a_\epsilon(x)}(m)|^2-1}{2\pi\epsilon^2}
=((\sqrt{-\Delta}^{-1}(\nabla\times E)_f)_f)^a(x)
\ee
so that 
\be \label{3.36}
\hat{E}_f(l)
=\int d^3x [\Delta^{-1}\nabla\times l]_a
[\sqrt{-\Delta_x} e^{-\Delta_x r^2} \lim_{\epsilon\to 0} 
\frac{\hat{g}_{p^a_\epsilon(x)}^\dagger
\hat{g}_{p^a_\epsilon(x)}-1}{2\pi\epsilon^2}]
\ee
where we have judiciously normal ordered the operator. It can be verified 
once more with (\ref{3.36}) that (\ref{3.35}) holds.
In computing the square of (\ref{3.36}) we encounter
\ba \label{3.37}
&&
[\hat{g}_{p^a_\epsilon(x)}^\dagger\hat{g}_{p^a_\epsilon(x)}-1]
[\hat{g}_{p^b_\sigma(y)}^\dagger\hat{g}_{p^b_\sigma(y)}-1]
\nonumber\\
&=&
\hat{g}_{p^a_\epsilon(x)}^\dagger\hat{g}_{p^b_\sigma(y)}^\dagger
\hat{g}_{p^a_\epsilon(x)}\hat{g}_{p^b_\sigma(y)}
e^{2\alpha\int d^3z X^c_{[p^a_\epsilon(x)]_f}(z)
[\sqrt{-\Delta}^{-1} X^c_{[p^b_\sigma(y)]_f}](z)}
\nonumber\\
&&
-\hat{g}_{p^a_\epsilon(x)}^\dagger\hat{g}_{p^a_\epsilon(x)}
-\hat{g}_{p^b_\sigma(y)}^\dagger\hat{g}_{p^b_\sigma(y)}+1
\ea
which we have written in normal ordered form. The expectation value 
of (\ref{3.37}) is thus given by
\ba \label{3.38}
&&
[|g_{p^a_\epsilon(x)}|^2-1][ |g_{p^b_\sigma(y)}|^2-1]
\\
&& +(e^{2\alpha q}-1)\{
[|g_{p^a_\epsilon(x)}|^2-1][ |g_{p^b_\sigma(y)}|^2-1]
+[|g_{p^a_\epsilon(x)}|^2-1]+[ |g_{p^b_\sigma(y)}|^2-1]+1\}
\nonumber
\ea
where 
\ba \label{3.39}
q&=&\int d^3z X^c_{[p^a_\epsilon(x)]_f}(z)
[\sqrt{-\Delta}^{-1} X^c_{[p^b_\sigma(y)]_f}](z)
\nonumber\\
&=&
\int_{S^a_\epsilon(x)} dS_c(u)
\int_{S^b_\sigma(y)} dS_d(v)\int d^3w
(\delta^{cd} \delta^{ef}-\delta^{cf} \delta^{de})
f_{,e}(w-u)[\sqrt{-\Delta}^{-1} f_{,f}](w-v)
\nonumber\\
&=& (\pi\epsilon\sigma)^2
\int d^3z (\delta^{ab} \delta^{cd}-\delta^{ad} \delta^{bc})
f_{,c}(z-x)[\sqrt{-\Delta}^{-1} f_{,d}](z-y) 
+O(\epsilon^3\sigma^2,\epsilon^2\sigma^3)
\ea
We conclude that
\ba \label{3.40}
&&<\hat{E}_f(l)^2>_m-[E_f(l)](m)^2
\nonumber\\
&=& \frac{\alpha}{2}
\int d^3x [\Delta^{-1}\nabla\times l]_a(x)
\sqrt{-\Delta_x} e^{-\Delta_x r^2} 
\int d^3y [\Delta^{-1}\nabla\times l]_b(y)
\times\nonumber\\ && \times
\sqrt{-\Delta_y} e^{-\Delta_y r^2} 
\int d^3z (\delta^{ab} \delta^{cd}-\delta^{ad} \delta^{bc})
f_{,c}(z-x)[\sqrt{-\Delta}^{-1} f_{,d}](z-y) 
\nonumber\\
&=& \frac{\alpha}{2}\int d^3x l_a(x)[\sqrt{-\Delta} l_a](x)
\ea
where we used transversality of $l$. From this formula it is easy to 
compute the expectation value and fluctuation of the electric flux 
operator $P\cdot\hat{E}(S)$ by using the distributional smearing field
\be \label{3.41}
l_a(x)=P_x\cdot e^{-r^2\Delta_x}\int_S dS_a(y) \delta(x-y)
\ee
Upon inserting (\ref{3.41}) into (\ref{3.40}) the integral blows up even 
much worse than on the Fock space side. This is to be expected from
the isomorphism $\cal I$ because the 2D smeared electric field in the 
polymer algebra has no preimage in the Fock algebra.

What is remarkable about these calculations, however, is that we can perform
strong operator limits of the form $\epsilon\to 0$ as in (\ref{3.36}),
as observed in \cite{17}, that is, with respect to the new 
representations defined by coherent states and their excitations.
Such limits were not possible with respect to the normalizable states 
of ${\cal H}_0$ which signals that we are dealing here with a new 
representation upon passing to non-normalizable (distributional) states.
In \cite{27} such states were given a home, the algebraic dual
Cyl$^\ast$ of the dense subspace of smooth cylindrical functions
Cyl$\subset{\cal H}_0$. Of course, this is possible only because we are 
allowed to use the Minkowski background metric all the time which makes 
it possible that the coefficients of the spin-network expansion 
of $\Omega_{P,m}^H$ depend smoothly on the graph rather than being
diffeomorphism invariant. On the other hand, in order to be of use 
for the quantization of diffeomorphism invariant operators on ${\cal H}_0$
which use 2D smeared electric fluxes in an essential way we must
use a complexifier which controls the fluctuations of $\hat{E}(S)$
and $\hat{A}(p)$ simultaneously in such new representations. This brings 
us to the next section.

\section{Electric Flux Operator}
\label{s4}

In the previous section we have found that the fluctuation of the 
electric flux operator is ill-defined in the representation 
${\cal H}_{P,m}^H$. This indicates an obstruction to using states of a 
similar kind, suitably generalized to $SU(2)$, in order to analyze the 
semiclassical behaviour of diffeomorphism covariant operators in quantum 
general relativity such as the Hamiltonian constraint operator. On the 
other hand, there was no problem with the fluctuation of the holonomy.
Could it be that the complexifier is too much adapted to the
holonomies but unsufficiently so to electric fluxes ? Can we distribute
the fluctuations more democratically so that both become well-defined ?

In this section we examine, still in the context of $U(1)$ to keep
things simple, whether a modification of the complexifier can fix this.
Specifically we will consider all complexifiers $C$ which are homogeneous
bilinear polynomials in the electric fields and which satisfy the 
following conditions:\\
1) $A^\Cl_a(m):=A_a-iq^2 \delta C/\delta A_a$ or at least 
$g_p(m)=\exp(i\int_p A^\Cl)$ is a well-defined classical function.
Distributional Poisson brackets among the $g_p,\overline{g_p}$ will be 
allowed. This will ensure a suitable correspondence between the 
coherent states $\psi_m$ and $m\in {\cal M}$.\\
2) $\hat{C}$ is well-defined on ${\cal H}_0$. This will make sure 
that $\psi_m$ is a well-defined distribution in Cyl$^\ast$.\\
3) The fluctutions of both $\hat{A}(p),\hat{E}(S)$ are finite. This 
will ensure that we can compute expectation values of (not 
necessarily normal ordered) operators more 
complicated than linear in these two basic operators as it is important
for quantum general relativity.

We will see shortly that the set of complexifiers satisfying 1),2),3)
is empty. Actually, for $U(1)$ conditions 2) automatically holds 
due to our restricted ansatz for $C$ and for the same reason 1) holds
for all gauge theories. So we just need to consider 3).\\
Due to our quadratic ansatz for $C$, the only freedom left is in the 
choice of a symmetric, positive kernel $K_{ab}(x,y)=K_{ba}(y,x)$ of 
dimension cm$^{-2}$, which we require to have an inverse on a suitable 
function space. Also, for simplicity, we assume that it is an even 
function of $x-y$ (the analysis for non-translational invariant kernels 
is similar but more complicated). Then the complexifier becomes 
\be\label{4.1}
C=\frac{1}{2q^2}\int d^3x E^a(x)[K\cdot E]_a(x):=
\frac{1}{2q^2}\int d^3x\int d^3y E^a(x)K_{ab}(x,y) E^b(y)
\ee
In the two previous sections we had $K=f\cdot\sqrt{-\Delta}^{-1}\cdot f$.
The complexified connection is 
\be \label{4.2}
A^\Cl_a(m)=A_a-i (K\cdot E)_a
\ee
The operator $\hat{C}/\hbar$ is again a positive, essentially self-adjoint
operator with pure point spectrum, eigenstates being given by charge 
network states
\be \label{4.3}
\hat{C}/\hbar T_s=[\frac{\alpha}{2}\int d^3x X^a_s [K\cdot X_s]_a]T_s
=:\frac{\alpha}{2}<X_s,K\cdot X_s>\; T_s
\ee
The annihilation operator becomes
\be \label{4.4}
\hat{g}_s=e^{-\hat{C}/\hbar}\hat{A}(s) e^{\hat{C}/\hbar}
=e^{\alpha<X_s,K\cdot X_s>} e^{<\hat{E},K\cdot X_s>}\hat{A}(s)
=e^{-\alpha<X_s,K\cdot X_s>} \hat{A}(s)e^{<\hat{E},K\cdot X_s>}
\ee
corresponding to the classical holonomy
\be \label{4.5}
g_s(m)=A(s)e^{<E,K\cdot X_s>}
\ee
Let $p^a_\epsilon(x)$ be again the loops of the previous section,
then for any surface $S$ we have  
\ba \label{4.6}
(P\cdot E)(S) &=&\int_S dS_a(x) 
[K^{-1}\cdot\Delta^{-1}\nabla\times\nabla\times
\lim_{\epsilon\to 0}\frac{[|g_{p_\epsilon}|^2-1]}{2\pi\epsilon^2}]^a(x)
\nonumber\\
A(p)&=& |g_{p^-1}|g_p
\ea
where the $x,a$ dependence of the fraction is the one of 
$p^a_\epsilon(x)$.
We then define 
\ba \label{4.7}
(P\cdot \hat{E})(S) &:=&
\int_S dS_a(x) 
[K^{-1}\cdot P\cdot
\lim_{\epsilon\to 0}
\frac{[\hat{g}_{p_\epsilon}^\dagger
\hat{g}_{p_e\epsilon}-1]}{2\pi\epsilon^2}]^a(x)
\nonumber\\
\hat{A}(p)&=& [\sqrt{\hat{g}_{p^{-1}}^\dagger\hat{g}_{p^{-1}}}]\hat{g}_p
\ea
Consider the operator 
$$
\hat{y}_{p^{-1}}(m)=|g_p(m)|^2\hat{g}_{p^{-1}}^\dagger\hat{g}_{p^{-1}}
$$
which has the square
\be \label{4.8}
\hat{y}_{p^{-1}}(m)^2=
|g_p(m)|^4(\hat{g}_{p^{-1}}^\dagger)^2(\hat{g}_{p^{-1}})^2
e^{2\alpha<X_p,K\cdot X_p>}
\ee
Then we compute, using our estimates for $\sqrt{y}$ from the previous 
section,
\ba\label{4.8a}
&& 1-\frac{1}{2}[e^{2\alpha<X_p,K\cdot X_p>}-1]\le
\frac{<\hat{A}(p)>_m}{[A(p)](m)}=<\sqrt{\hat{y}_{p^{-1}}(m)}>_m
\le 1 \mbox{ and} 
\nonumber\\
&& <[P\cdot \hat{E}](S)^2>_m-[[P\cdot E](S)](m)^2
\nonumber\\
&=&\int_S dS_a(x)\int_S dS_b(y) 
\lim_{\epsilon\to 0}\lim_{\sigma\to 0}
\int d^3u \int d^3v [K^{-1}\cdot P]^a_{a'}(x,u) [K^{-1}\cdot P]^b_{b'}(y,v)
\times\nonumber\\ &&\times
\frac{e^{2\alpha<X_{p^{a'}_\epsilon(u)},K\cdot X_{p^{b'}_\sigma(v)}>}
-1}{(2\pi\epsilon\sigma)^2}
\nonumber\\
&=&\frac{\alpha}{2}
\int_S dS_a(x)\int_S dS_b(y) \epsilon^{a'cd}\epsilon^{b'ef}
\int d^3u \int d^3v [K^{-1}\cdot P]^a_{a'}(x,u) 
[K^{-1}\cdot P]^b_{b'}(y,v)
\times\nonumber\\ &&\times
\partial^2 K_{df}(u,v)/\partial u^c\partial v^e
\nonumber\\
&=&\frac{\alpha}{2}
\int_S dS_a(x)\int_S dS_b(y) 
\int d^3u \int d^3v [K^{-1}\cdot P\cdot\nabla\times]^{ac}(x,u) 
[K^{-1}\cdot P\cdot\nabla\times]^{bd}(y,v) K_{cd}(u,v)
\nonumber\\
&=&\frac{\alpha}{2}
\int_S dS_a(x)\int_S dS_b(x) \epsilon^{acd}\epsilon^{bef}
\partial^2 [P\cdot K^{-1}\cdot P]_{df}(x,y)/\partial x^c\partial y^e
\ea
Thus, the necessary and sufficient condition for the kernel $K$ such
that the fluctuations of our elementary variables are well-defined 
is that
\ba \label{4.9}
&& \oint_p dx^a \oint_p dx^b K_{ab}(x,y) \le \infty \mbox{ and}
\\
&&
\int_S dS_a(x)\int_S dS_b(y) \epsilon^{acd}\epsilon^{bef}
\partial^2 [P\cdot K^{-1}\cdot P]_{df}(u,v)/\partial x^c\partial y^e
\nonumber\\ 
&=& \oint_{\partial S} dx^a \oint_{\partial S} dy^b 
[P\cdot K^{-1}\cdot P]_{ab}(x,y)
\le \infty \nonumber
\ea
for any surface $S$ and any loop $p$. In terms of form factors we may 
rewrite (\ref{4.9}) as
\be \label{4.10}
<X_p,K\cdot X_p>\le \infty \mbox{ and}
<X_{\partial S},P\cdot K^{-1} \cdot P\cdot X_{\partial S}>\le \infty 
\ee
Since for a loop $p$ the form factor is transversal, specifically
\be \label{4.11}
\partial_a X^a_p(x)=\int_p dy^a \partial_{x^a}\delta(x,y)
=-\int_p [d\delta_x](y)=-[\delta_x]_{\partial p}=0
\ee
we simply find the compact condition
\be \label{4.12}
<X_p,K\cdot X_p>\le \infty \mbox{ and}
<X_p,K^{-1}\cdot X_p>\le \infty 
\ee
for any closed path $p$. Notice that $X_p$ is a distribution, thus 
both $K$ and $K^{-1}$ must be non-distributional. We can rewrite 
(\ref{4.12}) in terms of its Fourier transform $\tilde{K}_{ab}(k)$
as 
\be \label{4.13}
\int d^3k \tilde{K}_{ab}(k) F^a_p(k) \overline{F^b_p(k)} \le \infty
\mbox{ and }
\int d^3k \tilde{K^{-1}}_{ab}(k) F^a_p(k) \overline{F^b_p(k)} \le \infty
\ee
where $F^a_p(k)=\int_p dx^a e^{ik_b x^b}$. To see why (\ref{4.13}) is 
impossible, consider first the simplified case $\tilde{K}_{ab}(k)=
\delta_{ab}\rho(k)$ with a positive function $\rho(k)$ whence (\ref{4.13}) 
turns into
\be \label{4.14}
\int d^3k \rho(k) \sum_a |F^a_p(k)|^2 \le \infty
\mbox{ and }
\int d^3k \frac{1}{\rho(k)} \sum_a |F^a_p(k)|^2 \le \infty
\ee
Now for a loop $t\mapsto p(t)=x_0+\epsilon n(t)$ where $n(t)$ is a unit
vector and $||x_0||\gg \epsilon$ we have $F^a_p(k)\approx 
e^{ik_b x^b_0} \epsilon \int_n dx^a$ so that for such loops
(\ref{4.14}) essentially turns into
\be \label{4.14a}
\int d^3k \rho(k) \le \infty
\mbox{ and }
\int d^3k \frac{1}{\rho(k)} \le \infty
\ee
which is impossible since the first condition requires $\rho(k)$ to die 
off at infinity at least as $1/||k||^{3+\epsilon},\;\epsilon>0$ so that
the second integral diverges. The case with non-diagonal $\tilde{K}_{ab}$
is similar and will be left to the reader.\\
\\
This concludes this section. We have shown that the representations 
induced by complexifiers of the form (\ref{4.1}), where 
$K_{ab}(x,y)=K_{ab}(x-y)$ is a symmetric positive kernel, cannot 
implement both, the holonomy and electric flux functions, as well
defined operators. This indicates an obstruction to using those 
representations for the semiclassical analysis of quantum general relativity.  
Thus, although it would be much preferred to use the distributional 
states (due to their graph independence), at least at the kinematical level
it seems that we are forced to work with graph dependent ones.

\section{Generalization to Non-Abelean Gauge Groups}
\label{s5}

So far, Varadarajan type of complexifiers have been defined only for Abelean
gauge groups (the proposal in \cite{19} for $SU(2)$ is not based on a 
complexifier as we will show below).
The underlying reason is the following: The associated 
complexifiers are of the type (\ref{4.1}). If we want to generalize 
this to a Non-Abelean compact gauge group then the most general,
homogeneous, bilinear ansatz is given by
\be \label{5.1}
C=\frac{1}{a\kappa}\int d^3x \int d^3y E^a_j(x) K_{ab}^{jk}(x,y) E^b_k(y)
\ee
where $K_{ab}^{jk}(x,y)=K_{ba}^{kj}(y,x)$ is a dimensionfree, symmetric, 
positive kernel
and $a$ is a constant of the appropriate dimension.
The first observation is that, unless $K_{ab}^{jk}(x,y)$ is proportional
to $\delta_{jk}\delta(x,y)$, then $C$ will not be gauge invariant.
This can be fixed only if we allow the kernel to depend non-trivially
on the connection as well. For instance, we could fix once and for all 
a point $x_0\in\sigma$ and for each $x\in\sigma$ we choose once and 
for all a path $\rho_x$ with $b(\rho_x)=x_0,f(\rho_x)=x$. Then 
\be \label{5.2}
C=\frac{1}{\kappa}\int d^3x \int d^3y 
\mbox{Tr}(\tau_j\mbox{Ad}_{A(\rho_x)}(E^a(x)))
K_{ab}(x,y) \mbox{Tr}(\tau_k\mbox{Ad}_{A(\rho_y)}(E^b(x)))
\ee
is gauge invariant, however, the corresponding operator will be very 
complicated, certainly it does not leave every subspace 
${\cal H}_{0,\gamma}$ separately invariant so that its spectral resolution 
will be very complicated, it is not even clear that it will be positive.
Even worse, it can not be defined on ${\cal H}_0$ {\it at all}:\\
To see this, choose $G=SU(2)$ for quantum general reativity. Then
$E^a_j$ is dimensionfree so that $a$ has dimension cm$^4$. Let us 
apply  the naive quantization of (\ref{5.1}) with 
$\hat{E}^a_j=\ell_p^2\delta/\delta A_a^j $
to a spin network function $T_s$. The result, after proper regularization, 
is 
\be \label{5.3}
\hat{C}/\hbar T_s=-\frac{\ell_p^2}{a}\Delta_{\gamma(s)} T_s
\ee
where 
\ba \label{5.4}
\Delta_\gamma &=&
\sum_{e,e'} \int_0^1 dt \int_0^1 dt' K_{ab}(e(t),e'(t'))\dot{e}^a(t)
\dot{e}^{b\prime}(t') O_{jk}(A(e_t)) O_{jl}(A(e'_{t'})) Y^k_e Y^l_{e'}
\nonumber\\
&& -2   
\sum_e \int_0^1 dt \int_0^1 dt' K_{ab}(e(t),e(t'))
\dot{e}^a(t) \dot{e}^b(t') \mbox{Tr}(A(e_t) A(e_{t',t}) A(e_{t'})^{-1}
\tau_k) Y^k_e
\nonumber\\
&=:& \sum_{e,e'} G^{e,e'}_{kl}(A) Y^k_e Y^l_e-2\sum_e G^e_k(A) Y^k_e   
\ea
Here the sums are over the edges of $\gamma$,
$Y^j_e=Y^j(A(e))$ is the right invariant vector field acting on 
the group degree of freedom $A(e)$, $e_{t,t'}$ denotes the segment
of $e$ between the points $e(t),e(t')$ with $e_t=e_{0,t}$ and 
$O_{jk}(h)\tau_k:=\mbox{Ad}_h(\tau_j)$ where $\tau_j$ is a basis
of $su(2)$. Actually, the result (\ref{5.4}) holds for any $G$.
Notice that the ``white noise kernel" $K_{ab}(x,y)=\delta_{ab}\delta(x,y)$
is not allowed due to the only one dimensional integrals.

Since we have started from a well-defined classical expression (\ref{5.1})
it is not unexpected, but nevertheless {\it absolutely non-trivial} to 
check that the family of operators $\Delta_\gamma$ is cylindrically
consistent. In particular, the term linear in $X^k_e$ cannot be discarded,
it is very essential in order to have a consistent operator family. The 
corresponding calculations are straightforward but fill pages and will not 
displayed here. Notice that by cylindrical consistency we mean here
that $\Delta_\gamma$ arestricted to functions over a smaller graph 
$\gamma'$ coincides with $\Delta_{\gamma'}$, of course, due to its 
non-trivial $A-$dependence residing in $G^{ee'}_{kl},G^e_k$ the 
operator $\Delta_\gamma$ does not preserve the space Cyl$_\gamma$ of 
functions over $\gamma$.  

This sounds promising, but we want more than just a 
consistent operator, it should be positive definite. Now the explicit
form of (\ref{5.4}) suggests that a formula like 
\be \label{5.5}
\Delta_\gamma=\sum_{e,e'} Y^k_e G^{e,e'}_{kl}(A) Y^l_{e'}  
\ee
should hold so that the term linear in $X^k_e$ in (\ref{5.4}) simply is 
due to pulling the operator $Y^k_e$ through $G^{e,e'}_{kl}$. In fact,
this is precisely the reason for its occurence in the regularized 
calculation that led to (\ref{5.4}). If that were the case 
then positivity could follow because $G^{e,e'}_{kl}(A)$ is a 
Hermitean matrix. Unfortunately, (\ref{5.5}) is already ill-defined as 
it stands because the operator $X^k_e$ only knows how to act on functions 
over graphs which depend only on the holonomy $A(e)$ but not on the 
segments $A(e_t)$. Thus, (\ref{5.5}) {\it is false}. However, even if
it was valid, there is an even worse obstacle: {\it The distributional 
nature of $\ab$ is such that none of the integrals involved in the 
definition of $G^{e,e'}_{kl},G^e_k$ make sense} ! Namely, as shown   
in \cite{28}, the map $t\mapsto A(e_t)$ is not $dt$ measurable.
{\it It is only in the Abelean case that the $A$-dependence 
of $G^{e,e'}_{kl}$ disappears and $G^e_k=0$}.\\
\\
{\it We conclude that, for Non-Abelean gauge groups, a bilinear ansatz of 
the form (\ref{5.1}) does not lead to a well-defined operator on 
${\cal H}_0$}.\\
\\
The question then arises, whether there exist replacements
for $G^{ee'}_{kl},G^e_k$ in (\ref{5.4}) which depend only on the edges 
of $\gamma$, such that $\Delta_\gamma$ becomes cylindrically defined, 
consistent and negative definite (i.e. $\hat{C}$ positive definite). 
In appendix \ref{sb} we will show that the {\it unique} solution to this 
problem is the operator 
\be \label{5.5a}
\Delta_\gamma=\sum_{e\in E(\gamma)} l_e Y^k_e Y^k_e
\ee
already found in \cite{30} where the positive function $e\mapsto l_e$
satisfies $l_{e\circ e'}=l_e+l_{e'},\;l_{e^{-1}}=l_e$.

Why do we then not use simply (\ref{5.5a}) as the complexifier for 
non-Abelean gauge groups ? The answer is that, while the operator is 
consistent and positive definite, it has no classical limit as shown in
\cite{9} and thus there is no map $m\mapsto A^\Cl(m)$ available. This is
precisely the reason for why we have struggled in \cite{10} to define 
a map $m\mapsto A^\Cl_\gamma(m)$ for each $\gamma$ separately so that
at least the cut-off coherent states $\psi_{\gamma,m}$ are useful for 
semiclassical analysis. (In \cite{10,11} we did not bother with 
the length functions $l_e$ for simplicity but it is trivial to incorporate 
them into the map $m\mapsto A^\Cl_\gamma(m)$). However, precisely because 
the continuum operator underlying the family
(\ref{5.5a}) has no classical limit the maps $m\mapsto A^\Cl_\gamma(m)$
do not come from a single map $m\mapsto A^\Cl(m)$. Thus, we may say that
as far as the class of complexifiers (\ref{5.6}) is concerned, 
\cite{10,11} is the best one can possibly do.\\
\\
In what follows we define a new class of complexifiers which are free from 
these drawbacks, namely each of them has the following properties:\\
1) it has a classical limit, so we obtain a well-defined map (canonical
transformation) $m\mapsto A^\Cl(m)$ and thus make immediate 
correspondence between classical and quantum theory.\\
2) it is gauge invariant.\\
3) it is positive definite.\\
4) it is cylindrically defined.\\
5) it has an explicitly known pure point spectrum.\\
6) it is almost of the type (\ref{5.1}).\\
\\
The clue for how to construct a complexifier with all of these 
properties comes from the observation that for non-Abelean gauge theories 
whose Hilbert space is based on holonomies the only known, 
well-defined and cylindrical momentum operators come from electric fluxes
\be \label{5.25}
E_j(S)=\int_S dS_a(x) E^a_j(x)
\ee
These objects are not gauge invariant, however, there are precisely two
basic invariants that one can build from those, namely 
$E_j(S) E_k(S')\delta^{jk}$ and
$E_j(S) E_k(S') E_l(S^{\prime})\epsilon^{jkl}$ in the limit as the 
surfaces involved shrink to a single point. The operators on ${\cal H}_0$
for which this shrinking process converges to a well-defined operator 
are precisely the area operator on the one hand and volume -- and 
length operators on the other \cite{30a}. We already have discussed the 
volume operator as a possible complexifier above and, in fact, it seems 
to be the more natural possibility because we do not need to introduce any
other structure, however, since its spectrum is presently only poorly 
understood, we will turn to the area operator. By definition, the 
area operator is only supported on a given surface but we must obtain a 
complexifiers which is supported everywhere in order that a damping factor 
is produced {\it for every graph}. Moreover, as we have shown in section
\ref{2.1}, we must use a power of the area operator which is greater than
one in order to arrive at an entire analytic function (convergence) and 
since with an embedding $X:\;\check{S}\subset \Rl^2\to S$
\be \label{5.26}
\mbox{Ar}(S)=\int_{X^{-1}(S)} d^2u \sqrt{\det(X^\ast q)}(u)
=\int_{X^{-1}(S)} d^2u \sqrt{[E^a_j(X(u))n_a^S(u)]^2}
\ee
where $n_a^S(u)=\epsilon_{abc}X^b_{,u^1} X^c_{,u^2}$ we see easily that
$\lim_{S\to x}[\mbox{Ar}(S)]^2/[E_j(S) E_j(S)]=1$. Thus, the natural 
power,
from the point of view of \cite{10,11} which was built on a 
gauge invariant version of objects of the type $E_j(S) E_j(S)$, is two.
We will then approximate a Gaussian decay as closely as we can in the 
non-Abelean context.

How should we then construct a complexifier built from objects of the 
kind $[\mbox{Ar}(S)]^2$ which is supported everywhere in $\sigma$ ?
There are many possibilities and we will present a few of them:\\
\\
Version 1: {\it Foliation and Parquet}\\
Let us introduce $D$ linearly independent foliations $X^I_t$ of $\sigma$,
that is, for each $t\in \Rl$ we obtain an embedding of a $D-1$ surface
$X^I_t:\;\check{S}^I_t\subset \Rl^{D-1}\to \sigma$ whose topology may 
vary with $t,I$ and linear independence means that at each point 
$x\in\sigma$ the $D$ ``normal" covectors
\be \label{5.27}
n_a^I(x):=\epsilon_{aa_1..a_{D-1}} [X^{Ia_1}_{,u^1} 
..X^{Ia_{D-1}}_{,u^{D-1}}]_{X^I_t(u)=x}
\ee
or equivalently the $D$ tangents 
$[(\partial X^I_t(u))/\partial t]_{X^I_t(u)=x}$ are linearly independent.
Within each leaf of the foliation $X^I_t$ fix a {\it parquet}
$P^I_t$, that is, 
a partition into smaller $D-1$ surfaces of fixed (say simplicial) topology
and we require that for each $I$ the parquet varies smoothly with $I$.
Notice that all of these structures do not refer to a background metric.
The parquet is quite similar in nature to the polyhedronal decomposition 
dual to a graph defined in \cite{10} but it is diferent because it is 
{\it graph-independently defined} so that the resulting complexifier 
can be defined already classically rather than only in quantum theory 
graph-wise. We then propose
\be \label{5.28}
C=\frac{1}{2a\kappa}\sum_{I=1}^D \int_{\Rl} dt \sum_{\Box\in P^I_t}
[\mbox{Ar}(\Box)]^2 
\ee
where $a$ is again an appropriate dimensionful parameter which we 
could also make dependent on $\Box$. For instance for quantum general
relativity in $D=3$, $a$ would have dimension cm$^2$ if we take the 
parameter $t$ dimension-free. 

The corresponding complexified connection would be 
\be \label{5.29}
A^{\Cl j}_a(x)=A_a^j(x)-i\sum_{I=1}^D 
(\frac{\mbox{Ar}(\Box^I_x)}{|\det(\partial X^I_t/\partial(t,u))|}
\frac{E^b_j(x) n^{\Box^I_x}_b(t,u)}{\sqrt{[E^c_j(x)n^{\Box^I_x}_c(t,u)]^2}}
n^{\Box^I_x}_a(t,u))_{X^I_t(u)=x}
\ee
where $\Box^I_x\in P^I_{t^I(x)},\;X^I_{t^I(x)}(u^I(x))=x$ is the surface 
containing $x$. From (\ref{5.29}) we see why we cannot do without 
the parquet since then we would have to work with the areas of the 
whole leaves which would be to an unsufficiently local object. 
However, even (\ref{5.29}) only allows us to reconstruct $E$
from $A^\Cl$ only with a precision that is defined by how fine the parquet 
is.

The spectrum of the corresponding complexifier operator is essentially 
derived from the known spectrum of the area operator, together with an 
important key observation which is responsible for making this operator
really leave all the Cyl$_\gamma$ separately invariant. Recall that
given an open, analytic, oriented surface $S$ and a graph $\gamma$ we 
can always subdivide its 
edges in such a way that any of them belongs to precisely one of  
the four disjoint subsets $E_{in},E_{out},E_{up},E_{down}$ of edges of
$\gamma$ where $e\in E_{in}\Rightarrow e\cap S=e,\; 
e\in E_{out}\Rightarrow e\cap S=\emptyset,\;
e\in E_{up}\Rightarrow e\cap S=b(e)\mbox{ and e points up},\;
e\in E_{down}\Rightarrow e\cap S=b(e)\mbox{ and e points down}$.
Here ``up,down" denote one of the two half spaces of $\sigma$ that (the 
analytic extension of) $S$ bounds. Let $P(S,\gamma)=\{b(e),\;e\in 
E_{up}\cup E_{down}\}$ and given $p\in P(S,\gamma)$ let
$X^j_{up}(p)=\sum_{e\in E_{up}(p)} X^j_e,
X^j_{down}(p)=\sum_{e\in E_{down}(p)} X^j_e$. The operators
$\Delta_{up}(p)=(X^j_{up}(p))^2,\;\Delta_{down}(p)=(X^j_{down}(p))^2,\;
\Delta_{updown}(p)=(X^j_{up}(p)+X^j_{down}(p))^2$ are simultaneously
diagonazable with ($-2$ times) total angular momentum spectrum.
The area operator is given by
\be \label{5.30}
[\widehat{\mbox{Ar}}(S)]_{\mbox{Cyl}_\gamma}=\frac{\hbar\kappa}{4}
\sum_{p\in P(S,\gamma)} 
\sqrt{-2\Delta_{up}(p)-2\Delta_{down}(p)+\Delta_{updown}(p)}
\ee
and its spectrum for $SU(2)$ reads explicitly
\be \label{5.31}
[\mbox{Spec}(\widehat{\mbox{Ar}}(S))]_{\mbox{Cyl}_\gamma}
=\frac{\hbar\kappa}{2}
\sum_{p\in P(S,\gamma)} 
\sqrt{2 j_u(p)(j_u(p)+1)+2 j_d(p)(j_d(p)+1)-j_{ud}(p)(j_{ud}(p)+1)}
\ee
with $j_u(p)+j_d(p)\ge j_{ud}(p)\ge |j_u(p)-j_d(p)|$

The key point is now that the subdivision of edges of $\gamma$ into the 
classes $E_{in},E_{out},E_{up},E_{down}$ {\it depends on the surface S} !
That is, a given spin-network state $T_s$ is not an eigenstate 
of a given operator $\widehat{\mbox{Ar}(S)}$, rather we must subdivide
the edges of $\gamma(s)$ adapted to $S$ and then decompose the 
intertwiners $I(s)$ in such a way that we get eigenfunctions of 
$\Delta_{up}(p),\Delta_{down}(p),\Delta_{updown}(p)$ for all vertices $p$ 
of $\gamma$ respectively. It follows that the function
$\widehat{\mbox{Ar}(S)} T_s$ depends, in the non-Abelean case, generally
no longer on the edges of $\gamma$ but also on the subdivision of the 
edges of $\gamma$ as adapted to $S$. This is dangerous because we are 
dealing with operators of the form $\int dt [\widehat{\mbox{Ar}(S_t)}]^2$
for a foliation $t\mapsto S_t$ and the function 
$[\widehat{\mbox{Ar}(S_t)}]^2 T_s$ therefore depends on the parameter $t$.
If it would depend on a graph $\gamma_t$ where $\gamma_t$ dpends on a 
subdivision of edges according to $S_t$ then the operator $\hat{C}$ would 
not exist since $[\widehat{\mbox{Ar}(S_t)}]^2 T_s$ is not $dt-$measurable
as we showed above. Fortunately this does not happen:

A point $p\in P(\gamma,S_t)$ falls only into one of the two categories:
Either it is a vertex of $\gamma$ in which case the subdivision of edges
does not change the graph or $p$ is an {\it interior point of a single 
edge}. However, in the latter case a spin-network function is already 
an eigenfunction: If $e=e_u(t)\circ e_d^{-1}(t)$ denotes the adapted 
decomposition of the corresponding edge of $\gamma$ with 
$p:=S_t\cap e=b(e_u(t))=b(e_d(t))$ then from (\ref{5.31}) due to
gauge invariance at $p$  
\be \label{5.32}
\sqrt{-2\Delta_{up}(p)-2\Delta_{down}(p)+\Delta_{updown}(p)} T_s
=\hbar\kappa \sqrt{j_e(j_e+1)}
\ee
is completely independent of $t$. We conclude that spin-network functions 
$T_s$ are simultaneous eigenfunctions of all possible 
$\widehat{\mbox{Ar}}(S_t)$
as long as $S_t$ does not contain a vertex of $\gamma(s)$. However, for 
given $T_s$ the number of vertices of $\gamma$ is finite and set 
$\{t\in \Rl;S_t\cap V(\gamma)\not=\emptyset\}$ is discrete and thus has
$dt-$measure zero. 

The spectrum of our complexifier operator therefore can easily be computed 
as follows:\\
We will assume that the graph $\gamma(s)$ is contained in a region such
that each of the embedded surfaces $t\mapsto X^I_t,\;t\in[a,b]$ has 
topology independent of $t$ with $\gamma\subset \cup_{t\in [a,b]} X^I_t$
for all $I$. The more general case including topology change
just involves introducing more notation and does not lead to new insights 
and thus will be left to the reader. Our assumptions about the parquet 
imply then that, given $I$, we have a corresponding family of surfaces
$S^I_{\Box,t}$ with a discrete label $\Box$.   
Fix $I,\Box$, and a set of intersection numbers $n^{I,\Box}_e=0,1,2,..;\;
e\in E(\gamma)$ one for each edge of $\gamma$ and  
denote by $t^I_{\Box}(\gamma,\vec{n})$ the $dt-$measure of the 
set $\{t\in [a,b];\;|S^I_{\Box,t}\cap e|=n^{\Box,I}_e\}$ (notice that we 
only 
count isolated intersection points). 
Then
\ba \label{5.33}
\frac{\hat{C}}{\hbar} T_s&=&
\frac{\ell_p^2}{a} \{\sum_{I,\Box,\vec{n}^{\Box,I}(s)}   
t^I_{\Box}(\gamma(s),\vec{n}^{\Box,I}(s)) 
[\sum_{e\in E(\gamma)} n^{\Box,I}_e \sqrt{j_e(j_e+1)}]^2\} T_s
\nonumber\\
&=&
\frac{\ell_p^2}{a}
\{\sum_{e,e'\in E(\gamma)} 
\sqrt{j_e(j_e+1)}\sqrt{j_{e'}(j_{e'}+1)}
\sum_{I,\Box,\vec{n}^{\Box,I}(s)}   
t^I_{\Box}(\gamma(s),\vec{n}^{\Box,I}(s)) 
n^{\Box,I}_e n^{\Box,I}_{e'}\} T_s
\nonumber\\ &=:&
\frac{\ell_p^2}{a}
\{ \sum_{e,e'\in E(\gamma)} G^{e,e'}_s
\sqrt{j_e(j_e+1)}\sqrt{j_{e'}(j_{e'}+1)}\}
T_s
\ea
which displays a suitable, non-Abelean generalization of the edge metric
which is automatically consistent because the area operator is. 

Interestingly, if the parquet is much finer than the graph then
each of the surfaces $S^I_\Box$ will typically intersect at most 
one edge $e^I_\Box$ of the graph and if so then only once. Therefore,
$t^I_{\Box}(\gamma(s),\vec{n}^{\Box,I}(s)) 
n^{\Box,I}_e n^{\Box,I}_{e'}$ vanishes unless 
$n^{\Box,I}_e=\delta_{e,e^I_\Box}$ up to small corrections in the vicinity
of vertices. Thus the sum over edges reduces approximately to diagonal
contributions and the sum over surfaces and their intersction numbers 
at given $e$ reduces approximately to $l^I_e$, the $dt$ measure of the set
$\{t\in [a,b];\; |S^I_t\cap e|=1\}$. This means that (\ref{5.33}) is 
approximated by 
\be \label{5.33a}
\frac{\hat{C}}{\hbar} T_s \approx
\frac{\ell_p^2}{a}
\sum_{e\in E(\gamma)} j_e(j_e+1) [\sum_I l^I_e] T_s
=:\frac{\ell_p^2}{a}
\sum_{e\in E(\gamma)} j_e(j_e+1) l_e T_s
\ee
which provides a concrete realization and classical interpretation of the 
numbers $l_e$ of \cite{30}.
In other words, at least for parquets much finer than a given graph,
the function (\ref{5.28}) provides a suitable continuum limit of 
the complexifier used in \cite{10,11} ! Of course, the exact 
operator has a non-diagonal edge metric and one has to repeat all the 
estimates of \cite{10,11} for this more general case, however,
on graphs sufficiently coarse as compared to the parquet the approximation 
given by (\ref{5.33a})
should be already quite good. We will provide analytic estimates in a 
future publication.\\
\\
In appendix \ref{sc} we present more versions of this operator which
depend on a background metric and make it possible to introduce the freedom
of arbitrary covariances and to make the operator Euclidean invariant 
(on a flat background).

These examples show that there is sufficient freedom in working with 
expressions bilinear in the area functionals, in the non-Abelean context,  
as a substitute for the bilinear expressions in the electric field, in the 
Abelean context, in order to produce well-defined operators whose 
expressions actually come arbitrarily close to those of the Abelean case.
We can work with the same kernels $K$ that also work in the Abelean case. 
Properties for various choices of $K,f$ and {\it for all} backgrounds
will be explored in future papers.    

Notice that all of this works directly in the polymer representation 
without recourse to the Fock representation, the only input is the 
complexifier. We have shown that it can be used, modulo functional 
analytic niceties, to generate non-Gaussian measures on the ``would-be" 
Fock side. This can be done either directly or, at least in the Abelean 
case, by 
starting from a complexifier on the polymer side which is a function of 
$E_f$ and then translating it by means of the isomorphism ${\cal I}^{-1}$
to the Fock side.\\
\\
Unfortunately, all these representations that we obtain from 
complexifiers of the form (\ref{5.28}) still do not control the 
fluctuations of the electric flux operator as is obvious from the 
closeness of the expression (\ref{5.33}) to the corresponding one in the 
Abelean case. Thus, while we have no complete proof, we have 
demonstrated strong 
indications that the new representations induced by this type of 
complexifier do not support the operator 
algebra of QGR as presently formulated and are therefore forced to work
with cut-off states or have to use completely different kinds of 
complexifiers.

\section{Averaging of Coherent States: Dirichlet-Voronoi Types of 
Constructions} 
\label{s6}

As we have just seen, the distributional coherent states $\psi_m$ 
corresponding to a given complexifier do not give rise 
to well-defined electric flux operators which in turn are basic building 
blocks for diffeomorphism invariant operators arising in quantum general 
relativity. However, the corresponding cut-off states $\psi_{\gamma,m}$
associated to a given graph are actually well-defined $L_2$ elements of 
${\cal H}_0$ provided that the complexifier satisfies the criteria 
outlined in section (\ref{s2.1}) and thus we can perform semiclassical 
analysis with those. The immediate caveat is that $\psi_{\gamma,m}$
depends on the given graph $\gamma$ and therefore introduces a huge 
amount of ambiguity. Which graphs should be chosen in order to achieve 
good semiclassical behaviour ? Several proposals have been made in order 
to overcome this:\\
1) {\it Averaging}\\
In \cite{20,19} one averages over graphs, based on the so-called 
Dirichlet-Voronoi construction, in order to reduce the graph dependence.\\
2) {\it Random Graphs}\\
In \cite{31} we use random graphs which, while depending on the 
particulars of the graph on the microscopic scale (with respect to a 
given background metric), look isotropic and homogeneous on a macroscopic 
scale (within each close to flat coordinate patch).\\
3) {\it Operator-State Correspondence}\\
Instead of building states that behave semiclassically with respect to 
a given class of operators we can turn things around and build operators 
that behave semiclassically with respect to a given class of states. 
As long as the classical limit of the theory is the desired classical theory
both quantization procedures are equally acceptable. This point of view
has been put forward in the last reference of \cite{11} which, in 
particular, means that 
operators are quantized in a graph dependent way. This still results in
consistently defined families of operators but their expression is 
quite ugly and generically 
different from the ones that have appeared in the literature already whence
this approach is disfavoured.\\
\\
In this section we will show that, unfortunately, an averaging over 
graphs of the Drichlet-Voronoi type does not help in order to achieve 
good semi-classical behaviour in the following sense: While they help to 
improve the 
semi-classical behaviour of the three metric $q_{ab}$ as represented by
area operators, they destroy the 
semi-classical behaviour of the extrinsic curvature $K_{ab}$ as 
represented by holonomy operators and thus, from this point of view,
can be called, at best, weaves. In the next section we show how to 
remove this problem.\\ 
\\
In order to show that the Dirichlet-Voronoi type of averaging over graphs 
destroys the semiclassical behaviour of the holonomy operator it is not 
necessary to go into the details of \cite{20}. All that we need is:\\
a) A point $m\in {\cal M}$.\\
b) A subset $\Gamma_m\subset\Gamma$ of compactly supported, piecewise 
analytic graphs depending on $m$.\\
c) A probability measure $\nu_m$ on $\Gamma_m$ for which discrete 
sets are of measure zero if $\Gamma_m$ is not countable.\\
The concrete Dirichlet-Voronoi construction actually only uses the 
three-metric information contained in $m$ in order to produce a 
continuous set of graphs $\Gamma_m$ and a measure $\nu_m$ thereon by a 
particular, beautiful construction. However, the details will not be 
important in the subsequent argument.

In order to use these structures we assume that we are given a map
\be \label{6.1}
\psi:\;\Gamma\times {\cal M}\to {\cal H}_0; (\gamma,m)\mapsto \psi_{\gamma,m}
\ee
as for instance through the cut-off states of the complexifier coherent state
machinery. We then have the following options:\\
\\
A) {\it Pure State}\\
Consider the formal object
\be \label{6.2}
``\;\psi_m:=\int_{\Gamma_m} d\nu_m(\gamma) \psi_{\gamma,m}\;\;\;"
\ee
which looks like a superposition of states in ${\cal H}_0$. The object 
(\ref{6.2}) should, however, not be confused with the graph independent 
distributions of section \ref{s3} since it still depends on the particular
choice $\Gamma_m$. Let us check whether $\psi_m\in {\cal H}_0$. If that 
were the case then $<T_s,\psi_m>\not=0$ for at most countably many $s$ 
and $||\psi_m||^2:= \sum_s |<T_s,\psi_m>|^2$. Now
\be \label{6.3}
<T_s,\psi_m>:=
\int_{\Gamma_m} d\nu_m(\gamma) <T_s,\psi_{\gamma,m}>
\ee
Making use of the spin-network decomposition of $\psi_{\gamma,m}$ we find\\
$<T_s,\psi_{\gamma,m}>=\sum_{\gamma'\subset\gamma} 
c_{s,\gamma'}(m)\delta_{\gamma',\gamma(s)}$ for certain coefficients 
$c_{s,\gamma'}$. Thus, (\ref{6.3}) vanishes unless $\Gamma_m$ is a 
countable set (which is not the case for 
the Dirichlet-Voronoi construction) since the integrand is supported 
on a $\nu_m$ measure zero subset. Thus, $\psi_m=0$ unless $\Gamma_m$
is a discrete set and $\nu_m$ a corresponding, weighted counting measure.
One can ask whether we still can use (\ref{6.2}) if we formally interchange 
integration over $\Gamma_m$ with scalar products in ${\cal H}_0$. But 
even then it is easy to see that $||\psi_m||^2=0$ unless $\Gamma_m$ is 
countable.\\
\\
B) {\it Mixed State}\\
Consider the object
\be \label{6.4}
\hat{\rho}_m(.):=\int_{\Gamma_m} d\nu_m(\gamma)
\psi_{\gamma,m}<\psi_{\gamma,m},.>
\ee
which looks like a density matrix. Let us check whether (\ref{6.4}) 
defines a trace class operator on ${\cal H}_0$. If that were the case 
then $<T_s,\hat{\rho}_m T_s>\not=0$ for at most finitely many $s$ and 
$\mbox{Tr}(\hat{\rho}_m):=\sum_s <T_s,\hat{\rho}_m T_s>$. But
\be \label{6.5}
<T_s,\hat{\rho}_m T_s>=
\int_{\Gamma_m} d\nu_m(\gamma) |<T_s,\psi_{\gamma,m}>|^2
\ee
vanishes for the same reason as (\ref{6.3}) does, except if $\Gamma_m$ is 
countable.\\
\\
C) {\it Expectation Value Functional}\\
Let us interchange the integral over 
$\Gamma_m$ and the trace operation. This leads to a sensible result since 
$\sum_s |<T_s,\psi_{\gamma,m}>|^2=||\psi_{\gamma,m}||^2=1$. Thus we 
simply define a positive linear functional on the operator algebra over
${\cal H}_0$ by
\be \label{6.6}
\omega_m(\hat{O}):=\int_{\Gamma_m} d\nu_m(\gamma) 
<\psi_{\gamma,m},\hat{O}\psi_{\gamma,m}>
\ee
Thus, while for continuous sets $\Gamma_m$ the operator $\hat{\rho}_m=0$
vanishes, the state $\omega_m$ makes sense.\\
\\
Let now $p$ be a closed path and $\hat{W}_p:=\mbox{Tr}(\hat{A}(p))$ the 
usual, gauge invariant Wilson loop operator. It is clear that
$<\psi_{\gamma,m},\hat{W}_p\psi_{\gamma,m}>=0$ unless $p\subset \gamma$.
Thus, if $\Gamma_m$ is a continuous set 
\be \label{6.7}
\omega_m(\hat{W}_p)=0
\ee
for each $m\in {\cal M}$. Choosing flat initial data $m=(A=0,E=const.)$
for which $W_p(m)=2$ we see that (\ref{6.7}) does not display good 
semiclassical behaviour with respect to the magnetic degrees of freedom
unless the set $\Gamma_m$ is countable.

One might hope that the situation can be rescued by passing to ``thickened
loops", i.e. tubes. But, apart from the fact that it is not thickened 
three-dimensional loops on which ${\cal H}_0$ is built, but rather the 
ususal one-dimensional ones, even this idea does not work: Let 
$(u^1,u^2)\mapsto p_u$ be a two-parameter congruence of loops so that 
$\cup_u p_u$ is topologically a solid torus. Then one might 
consider the altered Wilson operator 
$\hat{W}'_p:=\int d^2u \hat{W}_{p_u}$. The only hope that this operator 
has a non-vanishing expectation value is that the congruence of 
loops is generated by the same process that generates $\Gamma_m$. 
Suppose that this is even the case. Then
\be \label{6.8}
<\psi_{\gamma,m},\hat{W}'_p\psi_{\gamma,m}>
=\int d^2u <\psi_{\gamma,m},\hat{W}_{p_u}\psi_{\gamma,m}>=0
\ee
since $\{u\in \Rl^2;\;p_u\subset \gamma\}$ has $d^2u-$measure zero.

Thus, if averaging should have any chance to produce good semiclassical 
states (rather than distributions) then $\Gamma_m$ must be a countable set 
and the pure state $\psi_m$ or the mixed state $\hat{\rho}_m$ will be 
semiclassical for $\hat{W}_p$ at most if $p\subset \gamma$ for at least 
one $\gamma\in \Gamma_m$. For instance, in one dimension we could fix 
$M$ points in the interval $[-L/2,L/2]$, distributed according to the 
scale of variation of the one-dimensional metric and we could consider 
all possible graphs $\Gamma_m$ that can be formed by using $N\ll M$ 
of these points as vertices. In the end one could let $L,M,N\to\infty$
keeping the average lattice length $l=L/N$ and the sample number 
characteristic $M/N$ 
fixed. If the spatial metric is flat then 
the $M$ points will be equidistantly distributed and the weight for each 
of the approximately
$\left( \begin{array}{c} M\\ N \end{array}\right)$ configurations will be the 
same, that is, approximately $1/\left( \begin{array}{c} M\\ N \end{array} 
\right)$. It is now 
straightforward to estimate $\omega_m(\hat{W}_p)$ for the case that $G=U(1)$
(and $p$ is an open path for simplicity):\\
Even in this one dimensional case, where $p$ and $\gamma$ are 
contained in the same one-dimensional manifold, only those graphs 
$\gamma$ contribute which not only contain $p$ but also contain $b(p),f(p)$
as vertices of its edges (i.e. these points must not be interior points 
of edges). Thus the number of contributing graphs is at most
$\left( \begin{array}{c} M\\ N-2 \end{array} \right)$ since both end points 
have to be 
contained in the configuration. Since for each of the individual 
expectation values holds $|<\psi_m,\hat{W}_p\psi_m>|\le 1$ we easily
find 
\be \label{6.7a}
|\omega_m(\hat{W}_p)|\le 
\frac{\left(\begin{array}{c} M\\ N-2 \end{array} 
\right)}{\left(\begin{array}{c} M\\ N \end{array} \right)}\approx (N/M)^2
\ee
which is a tiny number (in order that we have many configurations), no matter
how the $\psi_m$ are chosen, and is thus grossly off the expected value
with $|W_p(m)|=1$. Results in higher dimensions will be 
even worse because there are relatively even less graphs $\gamma$ which 
contain $p$.\\ 
\\
All of this points us to the fact that averaging simply does not work
as far as holonomy operators are concerned
because the representation ${\cal H}_0$ is such that the functional
$T_s\mapsto <1,T_s>_{{\cal H}_0}$ is continuous only if 
the set $\Gamma$ is assigned the discrete topology.

\section{A Resolution: Diffeomorphism Invariant Operators}
\label{s7}

Up to this point the analysis lets us conclude that for the 
semiclassical analysis of QGR at best normalizable cut-off states are useful 
which are therefore necessarily graph-dependent. The non-averaged ones do 
not approximate area operators well, the unaveraged ones do not 
approximate at all the holonomy operators. More in detail,
graph dependent, unaveraged coherent states 
$\psi_{\gamma,m}$ were criticized due to the staircase problem, see 
the last reference in \cite{11}. That is, given e.g. a regular graph 
$\gamma$ (say of cubic topology) and a surface $S$, unless 
the surface $S$ is adapted to $\gamma$ (is composed of the surfaces 
defined by 
the plaquettes of the cubic graph), the expectation value of the 
area operator should be way off its classical value. While this can possibly
be overcome using a random graph for which no surface is adapted to its 
edges so that none is distinguished, given an arbitrary path the expectation 
value of the corresponding 
holonomy will be zero again unless $p$ is contained in the graph. This is 
an improvement over averaged coherent states which do not give good 
semiclassical results {\it for any path}, however, it is not good enough.

The observation is now that {\it Holonomy and electric flux operators 
for given coordinate paths and surfaces respectively have no invariant
physical meaning}! Physical meaning have only gauge invariant, spatially
diffeomorphism invariant operators which also (weakly) commute with the 
Hamiltonian constraint. At this moment we have little control on the 
Hamiltonian constraint so let us content ourselves with gauge -- and 
spatially diffeomorphism invariant quantities. Then we can define, 
following \cite{33}, for 
instance, something like a diffeomorphism invariant area operator by using
matter degrees of freedom. Thus, in order to measure the area, say of the 
surface of the sheet of paper that you are reading right now, we must
prepare a state which depends on gravitational and electromagnetic degrees
of freedom say and which is peaked on an electromagnetic field whose 
electromagnetic field energy is concentrated on that sheet of paper and 
which is peaked on a flat 
gravitational metric. Now we must construct an operator which does not 
measure a given coordinate surface but rather the surface of any region 
in space within which the electromagnetic field energy is non-zero.
Suppose that we have constructed such an operator $\widehat{\mbox{Ar}}$
on the Hilbert space ${\cal H}_0^E\otimes {\cal H}_0^M$, see \cite{1}.
Notice that instead of depending on a coordinate surface $S$, this 
operator depends on both gravitational and electromagnetic operators.
The key point is now that the semiclassical state on ${\cal H}_0^E\otimes 
{\cal H}_0^M$ which describes the sheet of paper in an ambient flat spacetime
is necessarily of the form $\psi_{\gamma;m_E,m_M}$ \cite{1}, where 
$m=(m_E,m_M)$ is the corresponding point in the combined Einstein-Maxwell 
phase space, {\bf because 
matter can only be located where geometry is excited}. In other words,
{\it the surface whose area is to be measured is dynamically automatically
adapted to the graph on which the coherent state is based}. Similar
statements can be made concerning the holonomy operator.

What we learn from these considerations is that the staircase problem 
is actually never there if we consider invariant quantities rather than 
coordinate dependent ones. For instance, one might think that there 
should be a gravitational semiclassical state peaked on flat space which 
assigns to the area operator for any possible coordinate surface the same 
area expectation value as long as those surfaces are only translated and 
rotated copies of each other. However, physically this is completely 
unnecessary because there exists no matter configuration which describes,
at the same instant of time, an uncountably infinite number of surfaces 
where electromagnetic field energy is concentrated. Thus, every 
measurement of one of those translated or rotatetd surfaces corresponds 
to a different physical situation, a different point $m=(m_E,m_M)$ in phase 
space and a different measurement
and it is therefore physically correct to use a different coherent state 
for each of them corresponding to different points $m\in {\cal M}$ in 
the combined matter -- geometry phase space. We 
will come back to this point in much detail in a future publication,
but what can be said here is that the types of operators which
are very well approximated by graph dependent coherent states, whether 
averaged or not, arise from integrated scalar densities of weight one
which is precisely the class of operators in which the Hamiltonian 
constraint falls. See \cite{31} for first explicit calculations in that 
direction.

\section{Summary}
\label{s8}

In this article we have summarized the present status of semiclassical 
states for quantum general relativity. 

As far as distributional coherent states are 
concerned we have shown that the states defined by Varadarajan in 
\cite{17} for $G=U(1),U(1)^3$ are {\it precisely} of the type of  
the complexifier coherent states defined in \cite{9} which unifies 
the current proposals under a common principle, the complexifier machinery. 
Conceptually, we have made a change of viewpoint as compared to 
\cite{17}
in that we showed that the inner product on the distributional coherent 
states is not an extra ingredient (coming from the isomorphism
$\cal I$ which in turn relies on a Gaussian measure) but {\it 
can be derived} from the 
already existent inner product on the kinematical Hilbert space ${\cal 
H}_0$ for 
quantum general relativity through a limiting procedure. This is interesting
because it confirms the latter Hilbert space as the ``common source" of 
possibly interesting other kinematical representations and thus gives it the 
status of a ``fundamental" representation.

On the other hand, we have shown that those distributional states that 
come from complexifiers which are bilinear in the electric field operator
do not support an electric flux operator. A first reaction might 
be that such kind of complexifiers are simply not well-defined operators 
on ${\cal H}_0$ if $G$ is not Abelean and that for QGR 
for wich electric flux operators are basic building blocks we 
must consider different kinds of complexifiers. We have constructed a 
particularly simple one which is bilinear in area functionals rather than 
electric fields and still is close to the one of the Abelean theory. 
This works because the complexifier machinery equips us with a huge freedom 
for how to adapt the complexifier to the quantum dynamics and is completely
independent of the existence of a corresponding representation on the 
Fock side.

The corresponding distributional states, although now appropriate for 
non-Abelean 
gauge groups, still do not admit well-defined electric flux operators.
Thus, the analysis suggests that the problem with electric flux
fluctuations has nothing to do with $G$ being Abelean or not but rather
lies in the nature of things: The representations induced by the
distributional states constructed are somehow too close to Fock like 
representations for which the preimage of an electric flux operator 
under Varadrajan's isomorphism $\cal I$ is an even more singular object 
than an operator valued distribution. 

Fortunately, at least the cut-off versions of these distributional
states on given graphs, being elements of ${\cal H}_0$, make sense with 
respect to the QGR operator algebra and the associated   
distributional state, which can also be considered as a complex measure, 
should be considered as the ``universal" structure underlying those cut-off 
states. 
Only these normalizable states thus seem appropriate for the semiclassical
analysis of QGR unless we want to redefine the theory from scratch.

In the past the graph 
dependence of these states was considered to be a bad feature and therefore 
averaging 
techniques have been proposed. We have shown here that all of these proposals
and natural variations thereof cannot lead to good coherent states 
either, if coordinate dependent operators are considered as the basic
ones, since 
they assign expectation values to holonomy operators which are (close to) 
equal to zero. However, we have shown that the graph dependence
is actually meaningful once one no longer talks about coordinate 
curves, surfaces and regions but rather about invariantly defined ones 
(for instance by matter). This happens because matter can only be where
geometry is excited and since matter determines the locus of those 
curves, surfaces and areas it follows that the graph underlying a 
coherent state for both gravitational and matter degrees of freedom 
is {\it dynamically} automatically adapted to the curves, areas and regions 
whose holonomy (or length), area or region is to be measured.

Encouraged by the results of \cite{31}, we are optimistic that
graph dependent coherent states defined through 
the complexifier machinery, both in their averaged and unaveraged form,
provide a suitable starting point for the kinematical semiclassical analysis
of QGR, at least as far as providing checks on the Hamiltonian constraint
are concerned.\\
\\
\\
{\large\bf Acknowledgements}\\
\\
We are grateful to Abhay Ashtekar, Luca Bombelli, 
Arundhati Dasgupta, Rodolfo Gambini,
Jurek Lewandowski, Jorge Pullin, Hanno 
Sahlmann and Oliver Winkler for countless discussions about coherent 
states, semiclassical states and their averages. Great thanks also go
to Mahavan Varadarajan first for numerous discussions about distributional
coherent states, kinematical and dynamical semiclassical states and 
the semiclassical limit of QGR and for a careful reading of the 
manuscript.

\begin{appendix}

\section{Review of the Kinematical Structure of Diffeomorphism Invariant 
Theories of Connections}
\label{s2.2}

The complexifier coherent state machinery can be applied, in particular, to 
quantum 
field theories which in their canonical formulation can be described by a 
phase space of the form ${\cal M}=T^\ast \a$ where $\a$ is a space of 
smooth connections for a principal $G-$bundle with compact gauge group 
$G$ over a $D-$dimensional (spatial) manifold $\sigma$ of arbitrary topology.
Moreover, we consider theories which are not only $G-$invariant but also 
Diff$(\sigma)$ invariant, such as general relativity in terms of real 
connection variables \cite{24}. Of course, connections provide only half of 
the canonical degrees of freedom, there is also a Lie$(G)$ valued vector 
density $E$ of weight one. If we denote spatial tensor indices by 
$a,b,c,..=1,..,D$ and 
Lie algebra indices by $j,k,l,..=1,..,\mbox{dim}(G)$ then the fundamental 
Poisson brackets can be written as 
\be \label{2.14}
\{A_a^j(x),A_b^k(y)\}=\{E^a_j(x),E^b_k(y)\}=0,\;\;
\{E^a_j(x),A_b^k(y)\}=\kappa\delta^a_b\delta_j^k\delta^{(D)}(x,y)
\ee
where $\kappa$ is the coupling constant of the theory.

In what follows we will try to give an outline of an explanation for 
why it is natural to consider the representation ${\cal H}_0$ currently
being used in QGR. The basic, fundamental assumption
in this approach, as stated by Gambini et. al. in \cite{25}
and by Jacobson, Rovelli and Smolin in \cite{26}, is that 
holonomies $A(p)$ of connections $A\in \a$ along 
paths $p\subset\sigma$ (which for technical reasons are assumed to be 
piecewise analytic) can be promoted to well-defined 
quantum operators. This is motivated by two different observations:
First of all, holonomies are the simplest functions of $A$ which transform
covariantly under local gauge transformations $g:\;\sigma\to G;\;
x\mapsto g(x)$, that is, $A^g(p)=g(b(p))A(p)g(f(p))^{-1}$ where 
$b(p),f(p)$ respectively denote the beginning and end point of $p$
respectively. Thus, it is easy construct gauge invariant objects 
like the Wilson loop function $\mbox{Tr}(A(p))$ for closed paths $p$
and one can show that such functions capture all gauge invariant 
information about any given $A$. The second observation is that
our assumption implies that wave functions will depend on the $A(p)$ as 
basic configuration degrees of freedom which in turn means that they will be 
labelled by arbitrary (piecewise analytic) paths. Now, since the 
holonomy is (the non-Abelean generalization) of an integral of 
a $1-$form along a curve, it is {\it background independently} defined.
Therfore
the diffeomorphism group acts on the $A(p)$ also covariantly, namely
$A^\varphi(p)=A(\varphi^{-1}(p))$ for any $\varphi\in$Diff$(\sigma)$ it 
follows that diffeomorphism invariant wave functions should only depend
on diffeomorphism classes of graphs, that is, {\it (generalized) knot 
classes}. This is very attractive because it means that one can 
possibly solve the diffeomorphism constraint rather easily and 
establish a link with topological quantum field theory.

Having motivated to consider the $A(p)$ as our basic configuration variables
we need to decide what our basic momentum variables should be. These 
should be chosen in a background independent way as well such that together 
with the holonomies they form a 
closed subalgebra of the Poisson algebra which is non-distributional in 
order that in the quantum theory we will deal with operators and not 
with operator valued distributions. The latter requirement implies that
the $E's$, to which we will refer to as electric fields in what follows,
have to be smeared in at least $D-1$ dimensions in order to absorb the 
$\delta$ distribution in (\ref{2.14}) when computing Poisson brackets.
The only other option is to smear them in $D$ dimensions, but as one can 
easily check, unless $G$ is Abelian, $D-$smeared electric fields do not
lead to a closed Poisson algebra (the Poisson bracket of two elementary
variables is not a polynomial of elementary variables). 
Thus, in the 
most interesting case of non-Abelean $G$ we are forced to work with the 
background independent {\it electric fluxes}
\be \label{2.15a}
E_j(S):=\frac{1}{(D-1)!}\int_S dx^{a_1}\wedge..\wedge dx^{a_{D-1}} 
\epsilon_{a_1..a_D} E^{a_D}_j
\ee
as the basic momentum degrees of freedom, already from a classical
point of view. The $E_j(S)$ transform covariantly under diffeomorphisms,
$E^\varphi_j(S)=E_j(\varphi^{-1}(S))$ but not under gauge transformations.
This can be easily repaired by replacing $E(x)$ under the integral in 
(\ref{2.15a}) by $\mbox{Ad}_{A(p_S(x)}(E(x))$ where for each $x\in S$
we have chosen a path $p_S(x)\subset S$ with $b(p_S(x))=x_S=const.$ 
and $f(p_S(x))=x$ in 
which case $E^g(S)=\mbox{Ad}_{g(x_S)}(E(S))$. This, however, is 
irrelevant for the construction of gauge invariant functions built from 
the $E(S)$ later on so we do not need not worry about these issues for what 
follows. The philosophy is that, while holonomy and electric flux operators 
are not physically interesting observables themselves, the interesting
observables become composite operators built from them in the quantum 
theory which is why it is important that their algebra is supported by
the Hilbert space representation.

Having made the assumption to represent the Poisson $^\ast$ algebra 
generated by 
$A(p),E(S)$ (with $A(p),E_j(S)$ being $G-$ valued and real-valued 
respectively) as 
operators $\hat{A}(p)=\pi(A(p)),\;\hat{E}(S)=\pi(E(S))$ 
on a Hilbert space we are looking at the representation theory of that
Poisson algebra. In  
particular, we must represent the Abelean Poisson-subalgebra generated by 
the $A(p)$. Since $G$ is compact, the $A(p)$ generate an Abelean function 
algebra of bounded functions which therefore is easily completed to an 
Abelean, unital 
$C^\ast$ algebra $\cal B$ under the sup norm. By basic $C^\ast$ algebra 
theorems, the function algebra can always be thought of as the algebra of 
continuous functions on a compact Hausdorff space $\ab$ (the so-called
spectrum of the original function algebra) which turns out to be a certain
distributional extension of $\a$. Our representation $\pi$ restricted 
to that $\cal B$ is non-degenerate (there is no vector in the 
kerenel of all $\pi(A(p))$) because the $C^\ast$ algebra contains the 
identity operator and thus it is an orthogonal sum of cyclic representations.
But each cyclic representation comes from a positive linear functional 
$\omega$ on $\cal B$ via the GNS construction. Finally, since $\ab$ is a 
compact Hausdorff space, each of these cyclic representations is unitarily
equivalent to a Hilbert space $L_2(\ab,d\mu)$ with respect to some measure
$\mu$ on $\ab$ by the Riesz representation theorem. We conclude that 
$\pi_{|{\cal B}}$ is {\it necessarily} 
represented on a (direct sum of) Hilbert space(s) of the form 
$L_2(\ab,d\mu)$ for 
some probability measure $\mu$, up to unitary equivalence, as an algebra 
of multiplication operators. Moreover, each of these Hilbert spaces 
contains a cyclic vector and it is clear that the common cyclic 
vector is the constant state $1(A)=1\in L_2(\ab,d\mu)$ 
since from this we obtain all continuous functions on $\ab$ by acting
with the $\pi(A(p))$) which are dense in the $L_2$ space. 
On physical grounds we want $\pi_{|{\cal B}}$ to be faithful, so the 
(direct sum of) measures is supposed to be faithful as well.

Now from the representation property $[\pi(E(S)),\pi(A(p))]
=i\hbar\pi(\{E(S),A(p)\})$ and the fact that 
$\{E(S),A(p)\}$ is a linear combination of holonomies it is clear that 
$\pi(E(S))$ is necessarily of the form $\pi(E(S))=i\hbar\kappa X_S+F_S$
where $X_S$ is a certain differential operator on $\cal B$ induced from 
$\{E(S),A(p)\}$ while $F_S\in {\cal B}$ is matrix valued and can map 
between the different sectors. Its purpose is to ensure that $\pi(E(S))$
is self-adjoint (the reality conditions on the $\pi(A(p))$ are 
automatically satisfied since they are multiplication operators). 
Since $X_S 1=0$ it is clear that $F_S$ 
must be an $L_2$ function in order that $\pi(E(S))$ is densely defined.
More details will appear in \cite{32}.
%

The simplest case that one can consider is that there is only one cyclic 
representation and that $F_S=0$. The unique solution to this problem 
is the measure Ashtekar -- Lewandowski measure $\mu_0$ of \cite{21}
as has been demonstrated in \cite{27}, in fact, to 
date there are no known other solutions, although a uniqueness proof is 
certainly missing. The most appealing additional features of $\mu_0$
besides providing a representation of the canonical commutation relations 
and the adjointness conditions is that it is $G-$invariant and 
Diff$(\sigma)$ invariant. This provides sufficient support for 
considering the Hilbert space ${\cal H}_0:=L_2(\ab,d\mu_0)$ as a suitable
starting point for further analysis. Moreover, it has been 
demonstrated in \cite{28} that $D-$smeared electric field operators 
are ill-defined on ${\cal H}_0$. Finally, $(D-1)-$smeared electric field 
operators have led to very appealing and well-defined results concerning
the construction of length, area and volume operators, they are 
ideally suited for defining the Gauss-, Diffeomorphism- and Hamiltonian 
constraint operators etc., see e.g. \cite{6} for a close to comprehensive 
list of these results. Finally, the Hilbert space ${\cal H}_0$ has 
an explicitly known basis of spin-network eigenfunctions which span a 
dense subspace Cyl$\subset {\cal H}_0$. The space of algebraic 
distributions Cyl$^\ast$ on Cyl is a natural home for the space 
of solutions to all constraints. \\ 
\\ 
Concluding, there is strong motivation to work with holonomies and 
electric fluxes in diffeomorphism invariant quantum field theories of 
connections and we will keep it as our postulate that the fundamental theory
should be based on them. Kinematical background dependent representations 
other than the background independent ``fundamental" ones of 
the type ${\cal H}_0$ are certainly allowed but there must be a clear 
relation with ${\cal H}_0$
in the sense that one should obtain them in a certain limit (like 
temperature representations are limits of Fock representations).

\section{Mutual Singularity of Uniform and Induced Fock Measures}
\label{sa}

The measures $\mu_f,\mu_0$ on $\ab$ are mutually singular with 
respect to each other as the following argument shows (see \cite{18} for 
more details):\\ The functions 
$T_s$ are certainly $L_2$ for both measures because $|T_s|=1$ and 
thus they are $L_1$ for both since the measures are probability measures.
Let now $t\mapsto \varphi_t,\;\varphi_t(x)=x+vt$ be a one parameter spatial 
translation 
subgroup of the Poincar\'e group. The measure $\mu_0$ is 
spatially diffeomorphism invariant while $\mu_f$ is spatially 
translation and rotationally invariant since the covariance of the 
Gaussian measure depends only on the operator $\Delta$.
Now
$$
X^a_{[\varphi_t^{-1}(p)]_f}(x)=\int_{\varphi_t^{-1}(p)} dy^a f(x-y)
=\int_p dy^a f(x-\varphi_t(y))=X^a_{p_f}(x-tv)
$$
will vanish in the limit $t\to\infty$ because the function $f$ is of rapid 
decrease and so we check that
\ba \label{2.57}
\lim_{t\to\infty}<T_s,T_{\varphi_t^{-1}(s')}>_{\mu_0}&=&
\delta_{s,(\emptyset,\vec{0})}\delta_{s',(\emptyset,\vec{0})}
=<T_s,1>_{\mu_0}<1,T_{s'}>_{\mu_0}  \mbox{ and}
\nonumber\\
 \lim_{t\to\infty}<T_s,T_{\varphi_t^{-1}(s')}>_{\mu_f} &=&
\lim_{t\to\infty} \omega_F^H(\hat{A}([\varphi_t^{-1}(s')-s]_f)
\nonumber\\
&=& <T_s,1>_{\mu_f}<1,T_{s'}>_{\mu_f}  
\lim_{t\to\infty}
e^{\frac{\alpha}{2}\int d^3x X^a_{s'_f}(x-tv)[\sqrt{-\Delta}^{-1} 
X^a_{s_f}](x)}
\nonumber\\
&=&<T_s,1>_{\mu_f}<1,T_{s'}>_{\mu_f}  
\ea
Thus, the spatial translations are mixing transformations for both 
measures (see e.g. section III.5 of \cite{6}), in particular, they
are ergodic for both. It follows from the definition of ergodicity that
\ba \label{2.58}
&& \mu_0(T_s)\cdot 1=\delta_{s,(\emptyset,\vec{0})}\cdot 1
\nonumber\\
&=_{\mu_0-a.e.}&
\lim_{T\to\infty}\frac{1}{2T}\int_{-T}^T dt T_{\varphi_t^{-1}(s)}(A)
\nonumber\\
&=_{\mu_f-a.e.}& e^{-\frac{\alpha}{4}\int d^3x X^a_{sf} \sqrt{-\Delta}^{-1}
X^a_{s_f}}\cdot 1=\mu_f(T_s)\cdot 1
\ea
Since the constants in the first and last line are different, it follows
from the a.e. equality with the middle term, which depends on a specific 
point $A\in \ab$, that the supports of $\mu_f,\mu_0$ are mutually 
singular, in fact, even a one parameter family of $f$'s each of whose 
members yields mutually different values of the constant in the last line 
of (\ref{2.58}) gives rise to a one parameter family of mutually singular 
measures.

\section{Uniqueness of the Laplacian Complexifier}
\label{sb}

Here we answer the question whether there exist replacements
for $G^{ee'}_{kl},G^e_k$ in (\ref{5.4}) which depend only on the edges 
of $\gamma$, such that $\Delta_\gamma$ becomes cylindrically defined, 
consistent and negative definite (i.e. $\hat{C}$ positive definite). 

Definiteness leads us to an ansatz of the form
\ba \label{5.6}
\Delta_\gamma &=& 
\sum_{e,e'\in E(\gamma)} Y^k_e G^{e e'}_{\gamma;kl} Y^l_{e'}
=\sum_{e,e'\in E(\gamma)} G^{e e'}_{kl} Y^k_e Y^l_{e'}
+\sum_{e\in E(\gamma)} [\sum_{e'\in E(\gamma)} 
(Y^l_{e'} G^{e'e}_{\gamma;lk})] Y^k_e
\nonumber\\
&=:& \sum_{e,e'\in E(\gamma)} G^{e e'}_{\gamma;kl} Y^k_e Y^l_{e'}
+\sum_{e\in E(\gamma)} G^e_{\gamma;k} Y^k_e 
\ea 
where the Hermitean matrix $G^{ee'}_{\gamma;kl}$ may depend on all the 
$A(e)$ where $e$ runs through the edges of $\gamma$. Let us now
explore the consequences of cylindrical consistency. Consider
a graph $\gamma'\subset\gamma$ where\\ 
I) $E(\gamma')=E(\gamma)-\{e_0\}$,\\
II) $E(\gamma')=(E(\gamma)-\{e_0\})\cup \{e_0^{-1}\}$,\\
III) $E(\gamma')=(E(\gamma)-\{e_1,e_2\})\cup \{e_0=e_1\circ e_2\}$.\\
I)\\
Condition I) gives on $f_{\gamma'}$
\ba \label{5.7}
\Delta_\gamma f_{\gamma'}
&=& \{\sum_{e,e'\in E(\gamma')} G^{e e'}_{\gamma;kl} Y^k_e Y^l_{e'}
+\sum_{e\in E(\gamma')} G^e_{\gamma;k} Y^k_e\} f_{\gamma'}
\nonumber\\ &=&\Delta_{\gamma'} f_{\gamma'}
=\{\sum_{e,e'\in E(\gamma')} G^{e e'}_{\gamma';kl} Y^k_e Y^l_{e'}
+\sum_{e\in E(\gamma')} G^e_{\gamma';k} Y^k_e\} f_{\gamma'}
\ea
We conclude that $G^{e,e'}_{\gamma;kl}:=G^{e,e'}_{\gamma-\{e_0\},kl}$
for any $e,e'\not=e_0$. Iterating this over all $\gamma'\subset\gamma$
we conclude that $G^{e,e'}_{\gamma;kl}=G^{e,e'}_{kl}$ should only
depend on $A(e),A(e')$. Then for any $e\in E(\gamma')$ 
\be \label{5.8}
G^e_{\gamma;k}
=\sum_{e'\in E(\gamma')} (Y^l_{e'} G^{e' e}_{lk})+
(Y^l_{e_0} G^{e_0 e}_{lk})=G^e_{\gamma';k}
=\sum_{e'\in E(\gamma')} (Y^l_{e'} G^{e' e}_{lk})
\ee
so 
\be \label{5.9}
Y^l_{e'} G^{e' e}_{lk}=0 \mbox{ for any } e'\not=e 
\ee
We conclude that
\be \label{5.10}
G^e_k=(Y^l_e G^{ee}_{lk})
\ee
only depends on $A(e)$. Thus, condition I) has already restricted our 
ansatz to the form
\be \label{5.11}
\Delta_\gamma= 
\sum_{e,e'\in E(\gamma)} G^{e e'}_{kl} Y^k_e Y^l_{e'}
+\sum_{e\in E(\gamma)} G^e_k Y^k_e 
\ee
Notice that there are indeed solutions to (\ref{5.10}), for instance
$G^{e e'}_{kl}=G^{e e'} O_{kl}(A(e) A(e')^{-1})$ which has the attractive
feature that $G^{ee}_{kl}=G^{ee}\delta_{jk}, G^e_k=0$ so that 
we are reduced to  bilinear ansatz which for diagonal elements is
even gauge invariant. Then, if $G^{ee'}$ is small for $e\not=e'$
we have an almost gauge invariant operator.\\
II)\\
We notice $Y^j_{e^{-1}}=-O_{jk}(A(e)^{-1}) Y^k_e$ and find 
\ba \label{5.12}
&&\Delta_\gamma f_{\gamma'}= \{\Delta_{\gamma-\{e_0\}} 
-\sum_{e'\in E(\gamma)-\{e_0\}} 
[G^{e' e}_{kl}  O_{lm}(A(e)^{-1}) 
+G^{e e'}_{lk}  O_{lm}(A(e)^{-1})]Y^k_{e'} Y^m_{e_0^{-1}} 
\nonumber\\
&& +G^{e_0,e_0}_{kl} 
[O_{km}(A(e_0)^{-1}) O_{ln}(A(e_0)^{-1}) Y^m_{e_0^{-1}} Y^n_{e_0^{-1}}
+O_{km}(A(e_0)^{-1}) (Y^m_{e_0^{-1}} O_{ln}(A(e_0)^{-1}))  Y^n_{e_0^{-1}}]
\nonumber\\
&&-G^{e_0}_k O_{kl}(A(e_0)^{-1}) Y^l_{e_0^{-1}}\} f_{\gamma'}
\ea
It is easy to see that  
$O_{km}(A(e_0)^{-1}) (Y^m_{e_0^{-1}} O_{ln}(A(e_0)^{-1}))\propto 
\epsilon_{kln}$ so the fourth term in (\ref{5.12}) vanishes identically
since the diagonal elements $G^{ee}_{kl}$ are symmetric in $k,l$.
From the second term we find the condition
\be \label{5.13}
G^{e' e}_{km} O_{ml}(A(e)^{-1}) =-G^{e'e^{-1}}_{kl}
\mbox{ for any } e'\not=e 
\ee
which also implies that (\ref{5.13}) holds
with $e'\leftrightarrow e,k\leftrightarrow l$ by using Hermiticity
and that $O_{kl}$ is real valued. 
The third term leads to
\be \label{5.14}
G^{e_0,e_0}_{kl} O_{km}(A(e_0)^{-1}) O_{ln}(A(e_0)^{-1})=
G^{e_0^{-1} e_0^{-1}}_{mn}
\ee
which also implies that condition
\be \label{5.15}
G^{e_0}_k O_{kl}(A(e_0)^{-1})=-G^{e_0^{-1}}_l
\ee
required by the fifth term is then already satisfied. Thus (\ref{5.13})
and (\ref{5.14}) is all that follows from II).\\
III)\\
We notice that $Y^j_{e_1} f_{\gamma'}=Y^j_e f_{\gamma'},\;
Y^j_{e_2} f_{\gamma'}=O_{jk}(A(e_1)) Y^j_e f_{\gamma'}$. Then
\ba \label{5.16}
&& \Delta_\gamma f_{\gamma'}
\nonumber\\
&=& \{\Delta_{\gamma-\{e_0\}} 
+\sum_{e'\in E(\gamma)-\{e_1,e_2\}} 
[(G^{e' e_1}_{kl}+G^{e' e_2}_{km} O_{ml}(A(e_1)))
+(G^{e_1 e'}_{lk}+G^{e_2 e'}_{mk} O_{ml}(A(e_1)))]Y^k_{e'} Y^l_{e_0} 
\nonumber\\
&& 
+[G^{e_1,e_1}_{kl}+[G^{e_2 e_1}_{ml} + G^{e_1 e_2}_{lm}] O_{mk}(A(e_1))
+G^{e_2 e_2}_{mn} O_{mk}(A(e_1))O_{nl}(A(e_1))] Y^k_{e_0} Y^l_{e_0}
\nonumber\\
&&
+[G^{e_1}_k+G^{e_2}_l O_{lk}(A(e_1))]Y^k_{e_0}\}f_{\gamma'}
\ea
The second term rquires that
\be \label{5.17}
G^{e' e_1}_{kl}+G^{e' e_2}_{km} O_{ml}(A(e_1))=G^{e' e_0}_{kl}
\mbox{ for any } e'\not=e_1,e_2 
\ee
which again also implies the same identity with labels 
interchanged due to Hermiticity. We cannot extend (\ref{5.17}) to
$e'=e_1,e_2,e_0$ since it is implicit in the definition of $G^{ee'}_{kl}$ 
that $e,e'$ do not overlap. Thus, from the the third term we find the 
additional condition that
\be \label{5.18}
G^{e_1,e_1}_{kl}+[G^{e_2 e_1}_{ml} + G^{e_1 e_2}_{lm}] O_{mk}(A(e_1))
+G^{e_2 e_2}_{mn} O_{mk}(A(e_1))O_{nl}(A(e_1))
=G^{e_0 e_0}_{kl}
\ee
and the fourth term requires 
\be \label{5.19}
G^{e_1}_k+G^{e_2}_l O_{lk}(A(e_1))=G^{e_0}_k
\ee
which does not follow from (\ref{5.17}),(\ref{5.19}) any more.\\
Thus, there are the 
six conditions 
(\ref{5.9}), (\ref{5.13}), (\ref{5.14}), (\ref{5.17}), (\ref{5.18}),
(\ref{5.19}) to be satisfied for cylindrical consistency to hold.  

Let us now look for solutions of this system of six equations. 
There are obviously two classes of solutions, in the first class we take 
$G^{ee'}_{jk}$ as independent of $A$ while in the second class there is 
non-trivial $A-$dependence. Consider first solutions in the first class.
Then equation (\ref{5.9}) is trivially solved. Equation (\ref{5.13}) 
requires obviously that $G^{e' e}_{jk}=0$ since no
$A-$dependence is allowed, thus the metric is necessarily {\it diagonal}.
Then also equation (\ref{5.19}) is already solved since $G^e_k=0$.
Now, equation (\ref{5.14}) becomes
\be \label{5.20}
G^{e_0 e_0}_{kl} O_{km}(A(e_0)^{-1})O_{ln}(A(e_0)^{-1})
=G^{e_0^{-1} e_0^{-1}}_{mn}
\ee
It follows that $G^{ee}_{kl}=l_e\delta_{kl},\;l_e=l_{e^{-1}}$ since the 
$O_{kl}$ are orthogonal matrices and the only way to get rid of them is 
to contract them with the unit matrix (use elementary Clebsch-Gordan 
theory). 
Finally then equation (\ref{5.17}) is automatically 
satisfied while (\ref{5.18}) implies that 
$l_{e_1}+l_{e_2}=l_{e_1\circ e_2}$. Thus, within the first class there is
a unique solution, namely 
\be \label{5.22} 
\Delta_\gamma=\sum_{e\in E(\gamma)} l_e Y^j_e Y^j_e
\ee
which is the one underlying the heat kernel Segal-Barmann transform of 
\cite{30}.\\
Now it turns out that there is in fact no solution in class two. To see 
this, study first equation (\ref{5.17}). We consider the spin-network 
decomposition for $e\not=e'$
\be \label{5.23}
G^{ee'}_{kl}=
a^{e,e'}_{kl}+b^{e,e'}_{kl}(e)+c^{e,e'}_{kl}(e')+d^{e,e'}_{kl}(e,e')
\ee
where round brackets indicate non-trivial spin dependence on the 
respective edge. Inserting into (\ref{5.17}) shows that 
$a^{e,e'}_{kl}=a_{kl},b^{e,e'}_{kl}(e)=b_{kl}(e)$ are universal functions 
and from here we see by coefficient comparison that $G^{ee'}_{kl}=0$
for $e\not=e'$ so that again the metric is diagonal. By the same token we 
conclude from (\ref{5.19}) that $G^e_k=0$. Now we set
\be \label{5.24}
G^{ee}_{kl}=l_e\delta_{kl}+\alpha^e_{kl}+\beta^e_{kl}(e)
\ee
where $\alpha^e_{kl}\not\propto\delta_{kl}$ and find by the same token from
equation (\ref{5.18}) that $\alpha^e_{kl}=\beta^e_{kl}(e)=0$.\\
\\
Thus, we have shown that the only positive, cylindrically consistent operator
within the class (\ref{5.6}) 
is the one defined in \cite{30} already with a diagonal edge metric.
In other words, the distributional states for $SU(2)$ 
proposed in \cite{19} are not of the complexifier type.

\section{More Non-Abelean Complexifiers}
\label{sc}

Here we display two more versions of the Non-Abelean complexifier.\\
\\
Version 2: {\it Foliation and background dependent smearing}\\
If we allow ourselves to use a background metric, then we can discard 
the parquet as follows: Given $x\in\sigma$, determine for each $I$
the leaf $S^I_x:=S^I_{t^I(x)}$ of the foliation $X^I$ from the identity
$X^I_{t^I(x)}(u^I(x))=x$. Let $f^I_u$ be a smearing function on
$\check{S}^I:=(X^I)^{-1}(S^I_t)$ concentrated (with respect to the 
background metric) at $u$ where we have assumed 
for simplicity that $\check{S}^I$ is independent of $t$ (again, topology 
change can easily dealt with by suitably subdividing the range of $t$
according to the topology type). Then define
\be \label{5.34}
C:=\frac{1}{2a\kappa}\int_\sigma d^3x\int_\sigma d^3y 
\sum_{I,J} K_{IJ}(x,y)
\mbox{Ar}_f(S^I_x)\mbox{Ar}_f(S^J_x)
\ee
where $K_{IJ}(x,y)$ is a background dependent, symmetric kernel, $a$ a 
constant of appropriate dimensionlity and 
\be \label{5.35}
\mbox{Ar}_f(S^I_x)=\int_{(X^I){-1}(S^I_x)} d^2u f^I_{u^I(x)}(u)
\sqrt{[E^a_j(X^I_{t^I(x)}(u))n_a^{S^I_x}(u)]^2}
\ee
For instance, if $\sigma=\Rl^2$ and we consider a flat background 
metric then we may take $\check{S}^I=\Rl^2$
and $f^I_u(u')=e^{-||u-u'||^2/(2r^2)}$ where the foliations are the 
natural ones given by a Cartesian frame. Since $\int d^2 u f_{u'}(u)
=2\pi r^2$ we see that (\ref{5.35}) is concentrated on a disk $\Box^I_x$ 
with center $x$ in the $x^I=const.$ plane of Euclidean area $\approx \pi 
r^2$ so that in this case  $\mbox{Ar}_f(S^I_x)\approx 
\mbox{Ar}(\Box^I_x)$. In order to get close to the previous formula
(\ref{5.28}) we could choose the kernel $K_{IJ}(x,y)=\delta_{IJ}
\delta(x,y)$ where the sum over surfaces $\Box$ inside a leaf $S^I_t$
is then replaced by an integral over the points within the leaf.

The complexified connection is given by 
\be \label{5.36}
A_a^{j\Cl}(x)=A_a^j(x)-i\kappa\delta C/\delta E^a_j(x)
\ee
and this time allows to reconstruct $E$ from $A^\Cl$ given suitable 
assumtions about the kernel $K$. The explicit form of the functional 
derivative in (\ref{5.36}) is similar to the one of (\ref{5.29}) and 
will be left to the reader.

Similar remarks with respect to the spectrum of the area operator
apply as before so that we get 
\ba \label{5.37}
\frac{\hat{C}}{\hbar} T_s &=& \frac{\ell_p^2}{2a} 
\{\int d^3x \int d^3y K_{IJ}(x,y)
\times \nonumber\\
&& \times
[\sum_{e\in E(\gamma)} 
\sum_{p\in S^I_x\cap e} f^I_{u^I(x)}((X^I_{t^I(x)})^{-1}(p)) 
\sqrt{j_e(j_e+1)}]
\times \nonumber\\ && \times
[\sum_{e'\in E(\gamma)} 
\sum_{p\in S^I_y\cap e} f^J_{u^I(y)}((X^I_{t^I(y)})^{-1}(p)) 
\sqrt{j_{e'}(j_{e'}+1)}]\} T_s \nonumber\\
&=:& \frac{\ell_p^2}{2a}  \{\sum_{e,e'} G_{K,f}^{e,e'} 
\sqrt{j_e(j_e+1)}\sqrt{j_{e'}(j_{e'}+1)}\} T_s 
\ea
Again, this gives a suitable well-defined generalization of the edge 
metric to the non-Abelean regime which depends on a foliation and a 
background metric. Using the fact that the edges $e$ have compact 
support one can check that the matrix $G^{ee'}$ is finite and positive
definite and gives rise to a cylindrically consistent operator.\\
\\
Version 3: {\it Background dependent smearing}\\
The idea is to to introduce, for each point $x\in \sigma$ a two parameter
family of surfaces $S_x(\varphi,\theta)$ through $x$ and then to average 
over these surfaces over the sphere $S^2$ with respect to some 
background dependent measure 
$d\nu(\Omega)$ where $\Omega=(\varphi,\theta)$. For instance, if the 
background is flat, $d\nu$ would be naturally chosen to be the rotation
invariant measure on $S^2$. Then, for given $\Omega$ one would let 
$n_a(\Omega)$ be the standard unit vector defined in $\Rl^3$ pointing
into the direction $\Omega$ defined by the natural Cartesian frame and 
$S_x(\Omega)$ would be the plane throgh $x$ orthogonal to $n(\Omega)$. 
Thus the proposal would be in general 
\be \label{5.38}
C:=\frac{1}{2a\kappa}\int_\sigma d^3x\int_\sigma d^3y 
\sum_{I,J} K(x,y) 
\mbox{Ar}_f(x)\mbox{Ar}_f(y)
\ee
where we have again introduced appropriate smearing functions $f$ on the 
$S_x(\Omega)$ (for simplicity we assume that they can be chosen to be 
independent of $x,\Omega$) and    
\be \label{5.39}
\mbox{Ar}_f(x)=\int_{S^2} d\nu(\Omega) 
\mbox{Ar}_f(S_x(\Omega))
\ee
as in (\ref{5.35}). Expression (\ref{5.38}) has the advantage that it is
{\it Euclidean invariant} in the case of a flat background if we choose a 
Euclidean invariant kernel.

The complexified connection is again given by formula (\ref{5.36})
while the spectrum of the complexifier becomes
\ba \label{5.40}
\frac{\hat{C}}{\hbar} T_s &=& \frac{\ell_p^2}{2a} 
\{\int d^3x \int d^3y K(x,y)
\times \nonumber \\
&& \times
[\sum_{e\in E(\gamma)}\int d\nu(\Omega) 
\sum_{p\in S_x(\Omega)\cap e} 
f_{u^\Omega(x)}((X^\Omega_x)^{-1}(p)) 
\sqrt{j_e(j_e+1)}]
\times\nonumber\\ && \times
[\sum_{e'\in E(\gamma)}\int d\nu(\Omega) 
\sum_{p'\in S_y(\Omega)\cap e'} 
f_{u^\Omega(y)}((X^\Omega_y)^{-1}(p')) 
\sqrt{j_{e'}(j_{e'}+1)}]\} T_s \nonumber\\
&=:& \frac{\ell_p^2}{2a}  \{\sum_{e,e'} G_{K,f}^{e,e'} 
\sqrt{j_e(j_e+1)}\sqrt{j_{e'}(j_{e'}+1)}\} T_s 
\ea
where $X^\Omega_x:\;\check{S}\subset\Rl^2\to S_x(\Omega)$ is an embedding
and $X^\Omega_x(u^\Omega(x)):=x$.

\end{appendix}

\end{document}